\documentclass[a4paper]{article}
\pdfoutput=1
\usepackage{ifpdf}
\ifpdf
\usepackage{etex}
\usepackage[pdftex]{graphicx}
\else
\usepackage[dvips]{graphicx}
\fi
\usepackage{a4wide}
\usepackage{amsbsy,amssymb,amsgen,amsfonts}
\usepackage{amsmath,amsfonts,color,latexsym}
\usepackage{pictex}
\usepackage[round]{natbib}
\newcommand{\ben} {\begin{equation}}
\newcommand{\een} {\end{equation}}
\newcommand{\be} [1] {\begin{equation} \label{#1}}
\newcommand{\ee} {\end{equation}}
\newcommand{\bse} [1] {\begin{subequations} \label{#1}}
\newcommand{\ese} {\end{subequations}}
\newcommand{\ban} {\begin{eqnarray*} }
\newcommand{\ean} {\end{eqnarray*} }
\newcommand{\bea} {\begin{eqnarray}}
\newcommand{\eea} {\end{eqnarray}}

\newcommand{\solid}{---$\!$---$\!$--}
\newcommand{\solidcirc}{---$\circ$---}

\newcommand{\dashed}{\hbox{{--}\,{--}\,{--}\,{--}}}
\newcommand{\dashedcirc}{\hbox{{--}\,{--}\,{$\circ$}\,{--}\,{--}}}
\newcommand{\chndot}{---\,$\cdot$\,---}
\def\drawline#1#2{\raisebox{2.5pt}{\rule{#1pt}{#2pt}}}
\def\spacce#1{\hskip #1pt}
\def\bdot{\protect\hbox{\protect\drawline{1}{.5}\spacce{2}}}
\def\dotted{\protect\hbox{\leaders\bdot\hskip 24pt}\nobreak\ }

\def\solidcirc{\protect\rule[2.5pt]{10.pt}{.5pt}$\circ$\hspace*{-1ex}
  \protect\rule[2.5pt]{10.pt}{.5pt}\hspace*{1ex}}
\newsavebox{\mybox}
\sbox{\mybox}{\dashed}

\newcommand{\revision}[2]{{#2}}
\begin{document}
\title{%
Interface-resolved DNS of vertical particulate channel flow in the
turbulent regime
} 
\author{
Markus Uhlmann\footnote{%
Email: {\tt markus.uhlmann@kit.edu}
}
\\
Modeling and Numerical Simulation Unit\\
CIEMAT, 28040 Madrid, Spain
}
\date{9 May 2008}

\maketitle
\begin{abstract}
We have conducted a direct numerical simulation (DNS) study of dilute
turbulent particulate flow in a 
vertical plane channel, considering thousands of finite-size rigid 
particles with resolved phase interfaces. 
The particle diameter corresponds to approximately $11$ wall units and
their terminal Reynolds number is set to $136$. 
The fluid flow with bulk Reynolds number $2700$ is directed upward,
which maintains the particles suspended upon average. 
Two density ratios were simulated, differing by a factor of
$4.5$. The corresponding Stokes numbers of the two flow cases 
were ${\cal O}(10)$ in the near-wall region and ${\cal O}(1)$ in the
outer flow.  
We have observed the formation of large-scale elongated
streak-like structures with streamwise dimensions of the order of
$8$ channel half-widths and cross-stream dimensions of the order of
one half-width. At the same time, we have found no evidence of
significant formation of particle clusters, which suggests that the
large structures are due to an intrinsic instability of the flow,
triggered by the presence of the particles.
It was found that the mean fluid velocity profile tends towards a
concave shape, and the turbulence intensity as well as the normal
stress anisotropy are strongly increased. 
The effect of varying the Stokes number while maintaining the buoyancy,
particle size and volume fraction constant was relatively weak. 
\end{abstract}

\section{Introduction}
The flow of a suspension of solid particles through
vertically-oriented channels, pipes or ducts occurs in a large number
of industrial applications, such as fluidized beds, riser flows and
pneumatic transport lines. 
%
One of the central questions raised by previous studies of these flows
concerns the nature of the mutual interaction between the turbulent
flow field and the motion of the dispersed particles. 

Experimental investigations of vertical wall-bounded flow 
have revealed that the intensity of fluid velocity fluctuations 
can be substantially modified by the addition of even a relatively small
volume fraction of heavy particles 
\citep*[e.g.][]{tsuji:84,rogers:90,kulick:94,suzuki:00,sato:00,hadinoto:05}. 
Depending on the values 
{of} 
the various parameters, either enhancement
or attenuation of the turbulence intensity is observed. 
In general the turbulence modulation effect can be attributed to a
competition between two opposing mechanisms: 
dissipation in the vicinity of particles, 
and generation of velocity fluctuations due to wake shedding
\citep*{hetsroni:89}. 
In addition to these direct consequences, particles can also
indirectly affect the overall turbulent kinetic
energy budget by interacting with specific coherent flow structures,
thereby changing the very structure of the turbulent flow field.
Indeed, it has been experimentally observed that small particles
exhibit so-called 'preferential concentration', e.g.\ in vertical
channel flow \citep*{fessler:94} and homogeneous isotropic turbulence 
\citep*{aliseda:02,wood:05}. \cite*{maxey:87} has shown 
that the accumulation generally occurs in regions of low vorticity and
high strain rate. This feature was later confirmed by various DNS
studies using the point-particle approach
\citep*{squires:90,wang:93,pan:96,rouson:01,ahmed:01,marchioli:02}. 

In 'two-way coupled' point particle simulations the hydrodynamic force
acting upon the particles is fed back to the fluid, thereby allowing
for the description of the influence of dispersed particles upon the
carrier phase.  
Using this methodology, some of the mechanisms of turbulence
modulation through local accumulations of particles has been
elucidated in wall-bounded shear flow \citep*{pan:96}, homogeneous
shear flow \citep*{ahmed:00} and homogeneous isotropic turbulence 
\citep*{squires:90,sundaram:99,ferrante:03}. 
%
It should be underlined, however, that most of the aforementioned results on
'preferential concentration' effects were obtained for particles whose
diameter is smaller than the smallest flow scales, and for small
values of the Reynolds number of the flow around individual particles
\citep*[see also the review by][]{eaton:94}. 

A different type of mechanism can lead to the formation of particle
agglomerations when the particle Reynolds number 
attains values of ${\cal O}(100)$.
In this regime, where the point-particle approach loses its
validity, resolved DNS of the sedimentation of finite-size particles
in ambient fluid \citep*{kajishima:02,kajishima:04} has revealed that
above a critical particle Reynolds number of approximately 300,  
particles tend to form clusters which are elongated in the direction of
gravity. These clusters have a collective settling velocity exceeding
that of an isolated particle, thereby inducing flow perturbations at
length scales much larger than the particle diameter. 
\cite{kajishima:04} has also shown that angular particle motion plays
an important role in the regeneration cycle of the agglomerations. 
Elongated particle accumulations were also reported by
\cite*{nishino:04} in experiments on grid-generated turbulence with
particle diameters corresponding to 2-3 Kolmogorov length scales and
particle Reynolds numbers in the range of 140-210.  
The formation of these clusters has been attributed to the effect of
wake attraction, which is the result of a reduced drag force acting on
a sphere located inside another sphere's wake. 
The variation of drag and lift experienced by pairs of spheres 
in different arrangements has been thoroughly investigated while
keeping the solid objects fixed
\citep*{tsuji:82,tsuji:85,kim:93b,zhu:94}.  
When considering a pair of mobile particles which are settling under
the effect of gravity after initially being vertically separated, 
the lower drag of the trailing sphere leads to an approach
of the particles ('drafting'), a subsequent particle encounter
('kissing') and finally a lateral separation ('tumbling'), yielding a
side-by-side arrangement \citep*[cf.][]{wu:98}. Similar particle
behavior due to wake effects has been identified in fluidization
experiments involving many spheres \citep*{fortes:87}.   

%

Despite considerable progress in the study of dispersed two-phase
flows, a full understanding of the mechanisms driving the
interaction between the fluid and solid phases is still lacking at the
present date, especially in the regime where the smallest length
scales of the turbulent flow are comparable to the particle
diameter. In part this is due to the difficulties which are
encountered in both laboratory and numerical experiments involving
suspension flow, leading to a relative scarcity of detailed data. 
Concerning fully-resolved direct numerical simulations, the challenge
lies in the fact that all relevant flow scales (pertaining to the flow
around the particles and the background turbulence) need to be
accurately represented, while at the same time imposing the
appropriate boundary conditions at the moving phase interfaces.  

High-resolution numerical studies of finite-size particles interacting
with turbulence have 
often focussed on 
individual spheres.
\cite*{kim:98} have analyzed a mobile spherical particle in synthetic
unsteady flow; \cite*{bagchi:03,bagchi:04} have considered a fixed
particle swept by a homogeneous-isotropic turbulent field;
\cite*{burton:05} have studied the interaction between a fixed
particle and homogeneous-isotropic turbulence; \cite*{zeng:07} have 
determined the effect of wall-bounded turbulence upon the flow around
a fixed particle.  
These investigations have provided highly 
detailed data on the time-dependent hydrodynamic forces acting upon
a single particle in unsteady or turbulent flow conditions; in the
case of \cite{burton:05} the effect of the presence of the particle
upon the carrier phase was also evaluated. However, true particle
mobility while allowing for collective effects could not be simulated 
with these approaches. 

To our knowledge, the only previous direct numerical simulation 
study of turbulent vertical 
channel flow with finite-size mobile particles has been performed by
\cite*{kajishima:01}. However, these authors considered only a small
number ($36$) of relatively large particles (with a diameter
corresponding to $32$ wall units) and the angular particle
motion was neglected. 
It should be mentioned that \cite*{pan:97} have conducted resolved DNS
of turbulent particulate flow in a horizontal open channel. They
considered up to $160$ stationary and mobile particles, the particle
Reynolds number being ${\cal O}(10)$ in the latter case. 
\cite*{tencate:04} have performed resolved DNS including a dense
suspension of finite-size particles (solid volume fraction up to
$10^{-1}$) in forced homogeneous-isotropic turbulence. The particle
Reynolds number in their case was ${\cal O}(10)$, precluding
significant wake effects.  

The aim of the present study is first and foremost to generate a body
of detailed data describing the structure of the carrier phase as well
as the motion of the dispersed phase in turbulent wall-bounded
flow. 
For this purpose we have performed interface-resolved DNS of turbulent
flow in a vertical plane channel configuration involving several
thousand spherical particles and integrating the equations of motion
over ${\cal O}(100)$ bulk flow time units.
The parameters are chosen such that the mean particle Reynolds number
is $136$, which means that significant wake effects can be expected. 
The specific questions which we will address concern on the one hand
the occurrence of preferential concentration of the solid phase, and
on the other hand the possible formation of particle-induced flow
structures. 

The outline of this article is as follows. 
In the following section the numerical method employed in our
simulations is explained. 
We report the various validation checks and grid convergence tests
which have been performed, specify the conditions of the simulated
flow cases and describe the initialization procedure. 
In \S~\ref{sec-results-euler-stats} we focus on the Eulerian statistics. 
The procedure for computing the statistical quantities and their
convergence is described in detail.  
We discuss the results for Eulerian one-point moments and probability
density functions of both phases before turning to the Lagrangian
statistics in \S~\ref{sec-results-lag}. 
The spatial structure of the dispersed phase is analyzed in
\S~\ref{sec-results-particle-structure} and the occurrence of new
coherent flow structures is investigated in
\S~\ref{sec-results-fluid-structures}. 
The paper closes with a short summary and discussion. 

\section{Numerical method}\label{sec-numa}
It has been recognized for some time that DNS of particulate flows can
be efficiently performed on computational meshes which are fixed in
space and time, while solving a single set of equations in the entire
domain, i.e.\ including the space occupied by the solid
bodies~\citep*{hoefler:00,glowinski:01,kajishima:02}. 
In this framework, various techniques have been used by different
authors in order to impose the constraint of rigid body motion upon
the fluid at the particle locations.

Our present simulations have been carried out with the aid of a
variant of the immersed boundary technique \citep{peskin:72,peskin:02}
proposed by \cite{uhlmann:04}.
%
This method employs a direct forcing approach, where a localized
volume force term is added to the momentum equations. The additional
forcing term is explicitly computed at each time step as a function of
the desired particle positions and velocities, without recurring to a
feed-back procedure; thereby, the stability characteristics of the
underlying Navier-Stokes solver are maintained in the presence of
particles, allowing for relatively large time steps.  
The necessary interpolation of variable values from 
Eulerian grid positions to particle-related Lagrangian positions (and
the inverse operation of spreading the computed force terms back to
the Eulerian grid) are performed by means of the regularized delta
function approach of~\cite*{peskin:72,peskin:02}. This procedure
yields a smooth temporal variation of the hydrodynamic forces acting
on individual particles while these are in arbitrary motion with
respect to the fixed grid. 

\begin{figure}
  \begin{center}
    \begin{minipage}{.7\linewidth}
      \includegraphics[height=\linewidth]{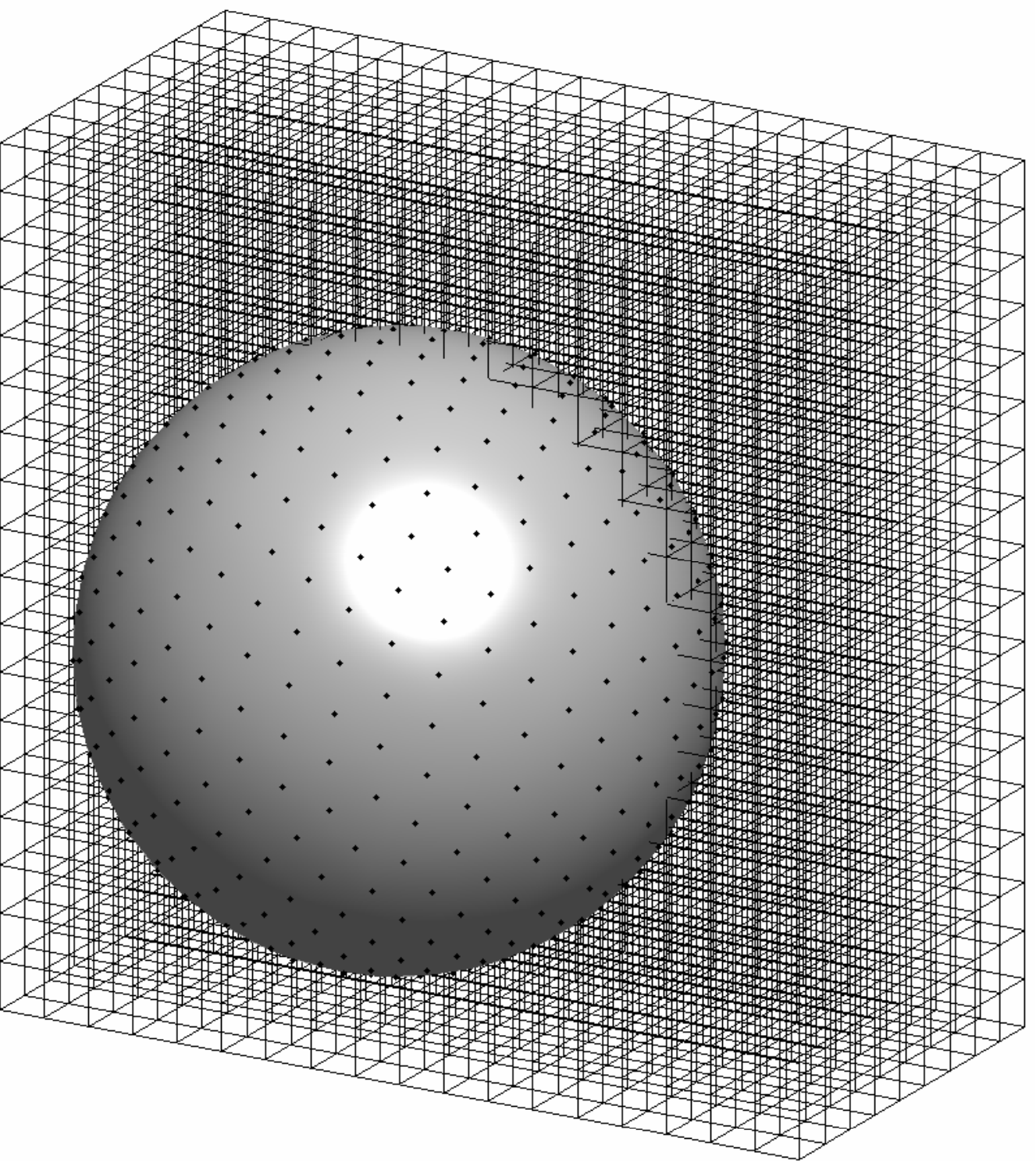}
    \end{minipage}
  \end{center}
  \caption{The Cartesian grid in the vicinity of a sphere with
    diameter $D/\Delta x=12.8$, also showing the Lagrangian force points 
    attached to its surface ($N_L=515$).}
  \label{fig-particles-numa-form-grid}
\end{figure}

Since particles are free to visit any point in the computational
domain and in order to ensure that the regularized delta function
verifies important identities (such as the conservation of the total
force and torque during interpolation and spreading), 
a Cartesian grid 
with uniform isotropic mesh width $\Delta x=\Delta y=\Delta z$ is employed. 
%
For reasons of efficiency, forcing is only applied to the surface of
the spheres, leaving the flow field inside the particles to develop
freely. \cite*{fadlun:00} have demonstrated that the external flow is
essentially unchanged by this procedure; 
in \cite{uhlmann:05a} it was confirmed that the impact upon the
particle motion is indeed small. 
Figure~\ref{fig-particles-numa-form-grid} shows the distribution
of forcing points on the surface of a sphere with diameter $D/\Delta
x=12.8$ and the corresponding Eulerian grid in the vicinity. 

The immersed boundary technique of \cite{uhlmann:04} is implemented in a
standard fractional-step method for incompressible flow. 
The temporal discretization is semi-implicit, based on the
Crank-Nicholson scheme for the viscous terms and a low-storage 
three-step Runge-Kutta procedure for the non-linear part
\citep*{verzicco:96}. The spatial operators are evaluated by central
finite-differences on a staggered grid. The temporal and spatial
accuracy of this scheme are of second order. 

The particle motion is determined by the Runge-Kutta-discretized
Newton equations for translational and rotational rigid-body motion,
which are explicitly coupled to the fluid equations. 
The hydrodynamic forces acting upon a particle are readily obtained
by summing the additional volume forcing term 
over all discrete forcing points and multiplying the result by a
factor of $-1$. 
%
Thereby, the exchange of momentum between the two phases cancels
out identically and no spurious contributions are generated. 
The analogue procedure is applied for the computation of the
hydrodynamic torque driving the angular particle motion. 

During the course of a simulation, particles can approach each other
closely. However, very thin inter-particle films cannot be
resolved by a typical grid and therefore the correct build-up of
repulsive pressure is not captured which in turn can lead to possible 
partial `overlap' of the particle positions in the numerical
computation. In practice, we use the artificial repulsion potential of
\cite*{glowinski:99}, relying upon a short-range repulsion force (with
a range of $2\Delta x$), in order to prevent such non-physical
situations. Essentially the same method is used for the treatment of
particles approaching solid walls.
The stiffness parameter appearing in the above mentioned authors'
definition of the repulsion force has been set to
$\epsilon_p=8\cdot10^{-4}D/(\rho_fu_{c,\infty}^2)$ (where
$u_{c,\infty}$ is the terminal particle settling velocity in ambient
fluid), based on calibration in simulations of two sedimenting
particles and particle-wall interactions.  
It should be noted that the use of an approximate treatment of
particle 'collisions' somehow taints an otherwise 'direct' numerical
simulation method. It is also true that the details of individual
particle encounters sometimes depend on the method for the collision
treatment, e.g.\ in the highly unstable case of
'drafting-kissing-tumbling' of two particles
\citep{glowinski:01}. However, when considering dilute suspensions
particle collisions are expected to occur so infrequently that
representing their exact dynamics does not seem vital for a faithful
description of the flow.  
Previous simulations incorporating the method of \cite{glowinski:99}
have demonstrated that a very good agreement with experimental
measurements can be achieved even in a relatively dense system
\citep*{pan:02}. 
%

\subsection{Validation and grid convergence}
\label{sec-validation}
The present numerical method has been submitted to exhaustive
validation tests
\citep{uhlmann:04c,uhlmann:04,uhlmann:05a,uhlmann:06b}. 
These tests have established the following features:
(i) second-order accuracy of the interpolation scheme and
low sensitivity of the error with respect to the position of the
immersed object relative to the grid;
(ii) favorable smoothness properties of the scheme in computations
involving oscillating objects.

In addition, we have performed a detailed grid convergence study of
the motion of particles in pressure-driven vertical plane channel flow
\citep{uhlmann:06c}. Two regimes were considered: laminar flow with a
single heavy particle of diameter $D/h=1/20$ ($h$ being the channel half
width) at $Re_{D,\infty}$=$100$ and
turbulent flow with four identical particles at
$Re_{D,\infty}$=$136$ (the terminal Reynolds number $Re_{D,\infty}$ is
based upon the particle diameter and the terminal settling velocity in
ambient fluid).  
In the former case, four different grids  
were considered, corresponding to a particle resolution of $D/\Delta
x=\{12.8,19.2,25.6,38.4\}$, while the $CFL$ number was kept constant.
Second order convergence of the
particle velocities was observed in this test. In the turbulent case,
the flow conditions were chosen equal to those of the 
configuration of interest in the following section (cf.\ case B of
table~\ref{tab-particles-channelp-params-phys} and
\ref{tab-particles-channelp-params-num}).   
Three spatial resolutions, $D/\Delta x=\{12.8,19.2,25.6\}$, were
analyzed in a computational domain 
which was sufficiently large to allow for sustenance of the turbulent
state. 
{During the observation interval of approximately $3.6$ bulk
flow time units, all three grids yield the same flow features. The
instantaneous particle velocities obtained from the coarsest and the
finest grid match to within $8.3\%$ of the terminal velocity;
the second-moment fluid statistics exhibit a maximum relative
discrepancy of $7\%$.
It should be noted that particle paths computed with two
different spatial resolutions will invariably deviate with time (even
when both simulations are extremely well resolved) since
perturbations can be amplified under supercritical (i.e.\ turbulent)
conditions.} 
{From these considerations} 
and the above mentioned
comparison with experimental data for sedimentation of a single
sphere, it was concluded that a resolution of $D/\Delta x=12.8$ is
sufficient for our present investigation of turbulent particulate
channel flow. Figure~\ref{fig-particles-numa-form-grid} shows this
grid in the vicinity of a sphere.

In separate simulations it was verified that our DNS code reproduces
single-phase turbulence results for plane channel flow faithfully. In
particular, using a mesh width of $\Delta x/h=1/256$ (which
corresponds to the value chosen in the following section) gives
results in very good agreement with the reference data of \cite*{kim:87}. 
{The mean velocity profile from our single-phase
computation has been included in figure~\ref{fig-results-um}$(b)$.} 
\subsection{Flow configuration}
\label{sec-config}
\begin{figure}
  \begin{center}
    \input{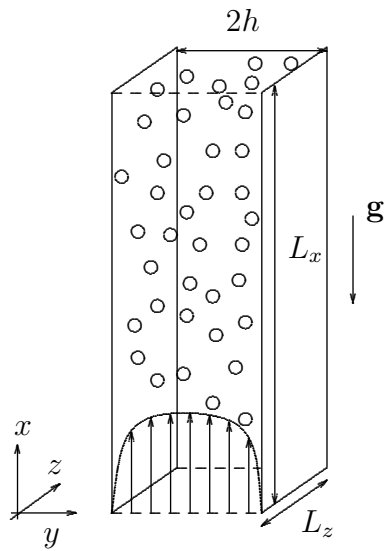}
  \end{center}
  \caption{The geometrical configuration for vertical plane channel
    flow with suspended particles. The fluid motion is induced by a
    negative mean pressure gradient in the direction of the $x$
    coordinate. The computational domain is periodic in $x$ and $z$.} 
\label{fig-particles-channelp-schema}
\end{figure}

We are considering particulate flow in a plane channel which is
aligned along the gravitational ($x$) direction ($y$ is the
wall-normal and $z$ the spanwise coordinate, cf.\
figure~\ref{fig-particles-channelp-schema}). 
The fluid is driven upwards by a mean pressure
gradient $\langle p\rangle_{,x}<0$. 
The computational domain is periodic in the streamwise and spanwise
directions (with periods $L_x$ and $L_z$), i.e.\ particles leaving the
domain on any of the four 
periodic end-planes re-enter at the opposite side. At the solid walls,
the no-slip condition is imposed upon the fluid, while the particles
rebound due to the short-range repulsion force, as specified in
\S\ref{sec-numa}. 
The bulk flow Reynolds 
number is maintained constant at a value of $Re_b=u_b\,h/\nu=2700$
($u_b$ being the bulk velocity and $\nu$ the kinematic viscosity), which
generates a turbulent flow with $Re_\tau=u_\tau\,h/\nu\approx172$
($u_\tau$ being the wall shear velocity) in the absence of 
particles. 
{The particle diameter} is chosen as $D/h=1/20$, corresponding to
$D^+=Du_\tau/\nu=8.6$ in wall units of the unladen flow ($11.25$ wall
units of the laden flow). 
{Physically, this regime of particle sizes is interesting, since the
  particle diameter is comparable to the cross-stream length scales of
  coherent structures typically found in the buffer layer of
  wall-bounded turbulent flow; a mutual interaction can therefore be
  expected. 
  Numerically, this particle size is the smallest value for which we
  can accurately represent the local flow at a particle Reynolds
  number of ${\cal O}(100)$ on the grid required for an accurate
  computation of the single-phase turbulence.}  
As mentioned in the previous section, we choose a spatial
resolution of $D/\Delta x=12.8$, which corresponds to a mesh width of
$\Delta x/h=1/256$, i.e.\ $\Delta x^+=0.67$ in wall units. The time
step for case A (the two flow cases are further specified below) was
set to $\Delta t=0.5761\Delta x/u_b$, which 
corresponds to a maximum $CFL$ number of approximately $1$; for case B
the step was reduced to $\Delta t=0.4148\Delta x/u_b$, i.e.\ a maximum
$CFL$ number of approximately $0.75$.

We have set the bulk fluid velocity ($u_b$) equal to the terminal
particle velocity ($u_{c,\infty}$).  
This condition means that the average vertical velocity of the
particles will be close to zero, supposing that their average wall-normal
spatial distribution is near uniform and neglecting wall effects,
collective effects and possible effects arising from the interaction
with turbulence. 
With these settings it turns out that the vertical velocity averaged
over all particles takes a slightly positive value as we will see
below. From the condition $u_{c,\infty}$=$u_b$ it follows that the terminal
particle Reynolds number 
measures $Re_{D,\infty}\approx136$. From the equality between buoyancy
and drag forces acting upon an isolated particle, we can form the
Archimedes number, which takes the following value: 
\begin{equation}\label{equ-particles-channelp-ar}
  Ar=(D/h)^3\,Re_b^2\,\frac{|{g}_x|h}{u_b^2}
  \left(\frac{\rho_p}{\rho_f}-1\right)
  =13328\,,
\end{equation}
where $\rho_p/\rho_f$ is the density ratio between particle and fluid
and $\mathbf{g}=(g_x,0,0)$ the vector of gravitational acceleration
(with $g_x<0$).  
This leaves us with one parameter free to choose
($\rho_p/\rho_f$, say), the other one 
($|{g}_x|h/u_b^2$) being
fixed by~(\ref{equ-particles-channelp-ar}).    

We have analyzed two cases with parameter values as given in
table~\ref{tab-particles-channelp-params-phys}. 
\revision{}{The current range of density ratios is
  representative of the motion of glass, ceramics or sand in water
  and is often investigated in laboratory experiments
  \cite[e.g.][]{parthasarathy:90a,suzuki:00,kiger:02}. 
  Apart from its relevance for the study of fundamental
  questions in turbulence-particle interaction, this regime has
  important applications in environmental flows and hydrology.}
The relative particle
density in case A is approximately $4.5$ times higher than in case
B, and the Stokes number (defined as the ratio between
particle and fluid time scales) varies accordingly. 
In particular, we
determine the particle time scale from the usual Stokes drag law, 
$\tau_{SD}=D^2\rho_p/(18\nu\rho_f)$, and consider two time scales for 
the fluid motion: the near-wall scale $\tau^+=\nu/u_\tau^2$ and the
bulk flow scale $\tau_{b}=h/u_b$. Consequently, we can form two Stokes
numbers, characterizing the ratio of time scales in the near-wall
region, $St^+=\tau_{SD}/\tau^+$, and in the outer flow
$St_b=\tau_{SD}/\tau_{b}$.
Table~\ref{tab-particles-channelp-params-phys} shows that the present
particles have a relatively large response time with respect to the
flow in the near-wall region, where $St^+>10$, but the scales of both
phases are comparable in the bulk of the flow, i.e.\ $St_b\sim{\cal
  O}(1)$. 

The global solid volume fraction has been set to
$\Phi_s=0.0042$, which can be considered at the upper limit of
the dilute regime. 

The two cases were run in computational domains of different
wall-parallel dimensions,
the volume in case B being eight times larger ($4h\times2h\times
1h$ versus $8h\times2h\times 4h$). The number of particles 
in cases A, B is therefore $N_p=\{512,4096\}$, respectively. The
domain used in case A evidently constitutes a smaller sample size per
instantaneous flow field. However, it allows for simulations over
longer time intervals which is of special interest for the computation
of slowly evolving statistical quantities acquired along the particle
paths. Table~\ref{tab-particles-channelp-params-num} shows the global
grid size of the simulations, which exceeds $10^9$ Eulerian nodes and
$2\cdot10^6$ Lagrangian force points in case B, where the work is
typically distributed over $512$ PowerPC processors with
fiberoptical interconnect (MareNostrum at BSC-CNS,
Barcelona). It should be mentioned that the simulation of case B has
required roughly $10^6$ CPU hours of execution time. 
%
%

\subsection{Initialization of the simulations}
\label{sec-init}
\begin{figure}
  \centering
  \begin{minipage}{5ex}
    $\displaystyle\frac{\langle{u}_c\rangle_p}{u_b}$
  \end{minipage}
  \begin{minipage}{.5\linewidth}
    \centerline{$(a)$}
    \includegraphics[width=\linewidth]{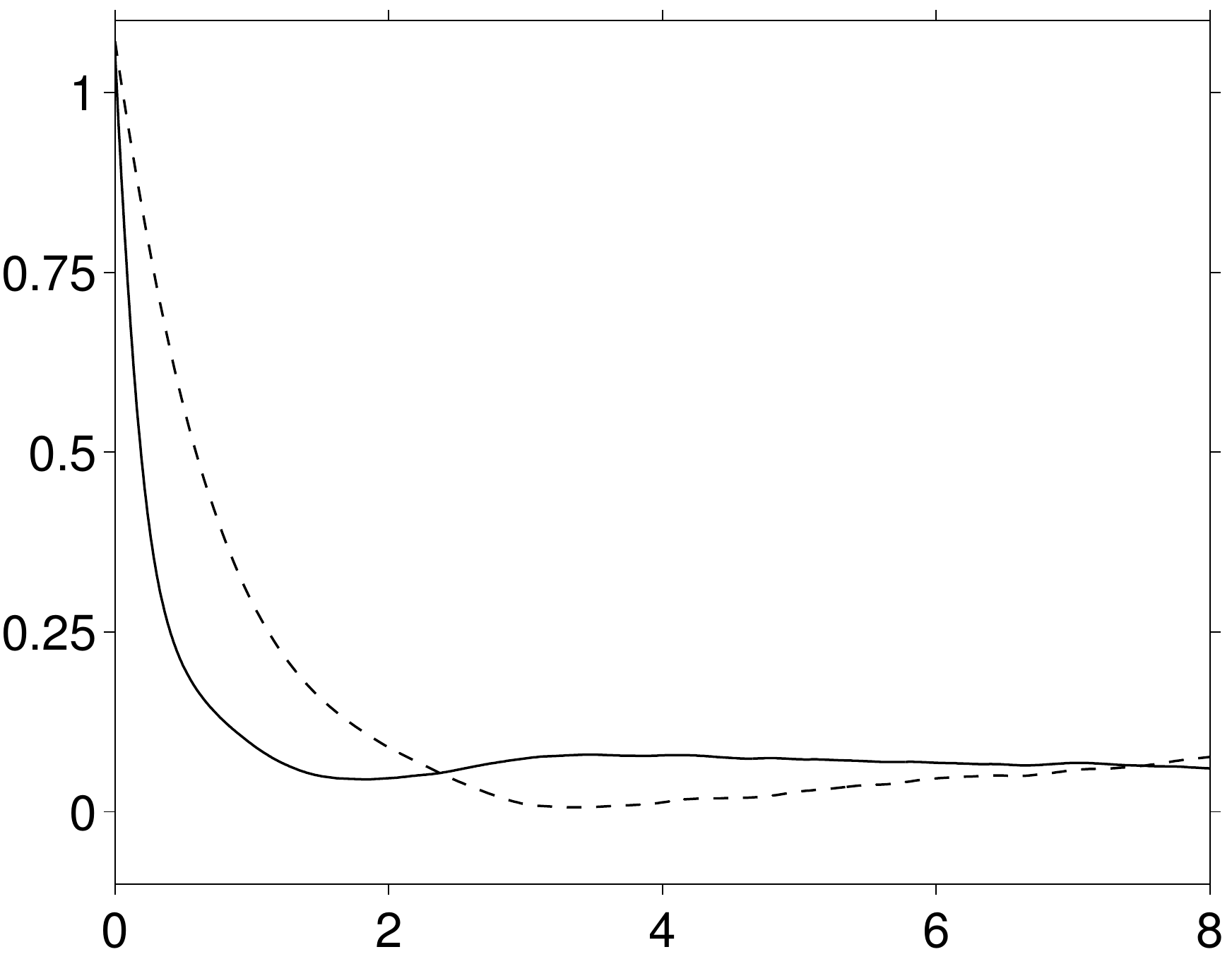}
      \centerline{$(t-t_0)u_b/h$}
  \end{minipage}\\[1ex]
  \begin{minipage}{5ex}
    $Re_\tau$
  \end{minipage}
  \begin{minipage}{.5\linewidth}
    \centerline{$(b)$}
    \includegraphics[width=\linewidth]{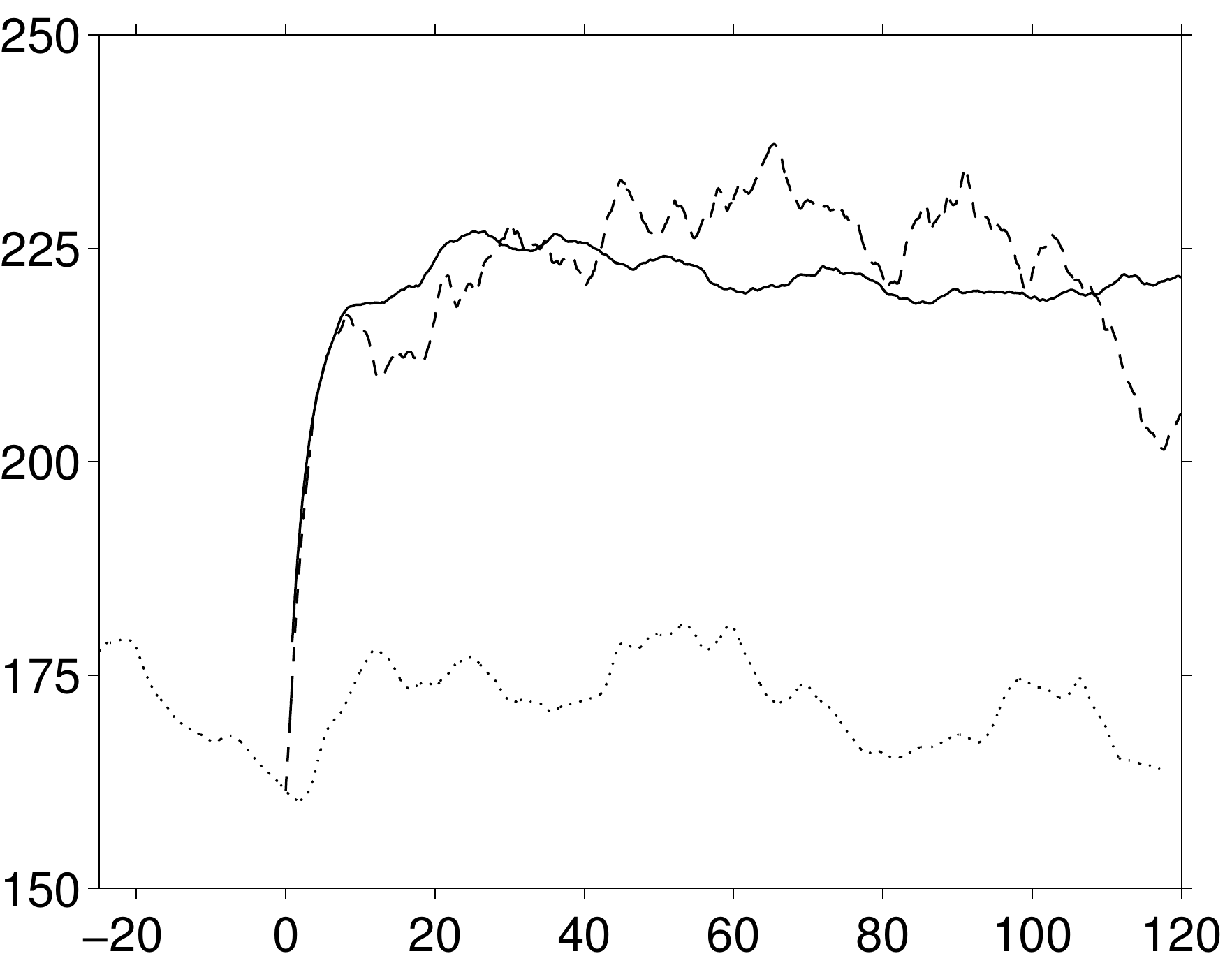}
      \centerline{$(t-t_0)u_b/h$}
  \end{minipage}
  \caption{$(a)$ The average vertical particle velocity during
    start-up of the two-phase simulation. 
    $(b)$ The time evolution of the friction-velocity-based Reynolds
    number around the start-up time. Particles are added at
    $t=t_0$. \dashed~case A; \solid~case B;
    \dotted~single-phase flow.} 
  \label{fig-uc-retau-init}
\end{figure}

Particles are introduced into fully-developed turbulent single-phase
flow at time $t_0$. The initial particle positions form a regular
array covering the computational domain, perturbed randomly 
with an amplitude 
of one particle diameter. Their initial velocities are
matched with the fluid velocities found at $t_0$ at the locations of
each particle's center; the initial angular particle velocities are
set to zero. During the first time step of the two-phase computation, the
coupling algorithm solidifies the fluid occupying the volume of each
particle and buoyancy forces set in. 
Figure~\ref{fig-uc-retau-init}$(a)$ shows how the 
average over all particles of the vertical particle velocity  
drops from a value close to the bulk fluid velocity to almost zero
over an initial interval of approximately 3 (1) bulk time units in
case A (B). 
Let us mention that the long-time averaged mean upward drift of the
particles amounts to $0.029u_b$ and $0.058u_b$, respectively. 
The data corresponding to the initial transient will not be considered
in the following and, therefore, an initial interval of $17$ bulk time
units has been discarded. The length of the observation interval for
the two cases is indicated in
table~\ref{tab-particles-channelp-params-num}.  

From figure~\ref{fig-uc-retau-init}$(b)$ it can be seen that the
friction velocity increases immediately after the addition of
particles, leading to a sharp rise of $Re_\tau$ which then oscillates
around a new mean value of approximately $225$ in both cases. The
amplitude and frequency of the oscillations is larger in case A due to
the smaller box size.
This higher friction velocity in the particulate cases reflects the
modification of the mean fluid velocity profile as discussed below
(cf.\ figure~\ref{fig-results-um}).  
%
From here on, it is understood that quantities given in wall units
(indicated by the usual superscript `$+$') are normalized with the
actual friction velocity, i.e.\ the increased values are used in the
particulate flow cases. 
\section{Eulerian statistics}
\label{sec-results-euler-stats}
\subsection{Computation and convergence of statistical quantities}
\label{sec-results-stats-def}
Let us introduce the notation for the different averaging operations
employed in the following. $\langle\xi\rangle_x$ refers to the streamwise
average of a quantity $\xi$, $\langle\xi\rangle_{xz}$ is the average
over wall-parallel planes and $\langle\xi\rangle_{t}$ is the temporal
average (over the observation interval given in
table~\ref{tab-particles-channelp-params-num}); the average over time
and wall-parallel planes is denoted by 
$\langle\xi\rangle=\langle\langle\xi\rangle_{xz}\rangle_{t}$;
finally, 
$\langle\xi\rangle_p$ refers to the instantaneous ensemble average
over all particles.  

The Eulerian space averages of quantities related to the dispersed
phase are carried out for discrete bins in the wall-normal
direction, defined as
$I_{yj}=[j-1,j)\cdot2h/N_{bin}$ $\forall\,1\leq j\leq N_{bin}$. 
In practice a value of $N_{bin}=80$ was chosen (except where otherwise 
stated), such that the width of each bin corresponds to one particle
radius. 

The time- and plane-averaged number density is then computed for each
bin by the following formula:  
\begin{equation}\label{equ-def-particle-bin-average-n}
  \langle n_s\rangle(y_j)=
  \frac{1}{n_{obs}L_xL_z|I_{yj}|}
  \sum_{l=1}^{n_{obs}}
    \mathop{\sum_{i=1}}_{y_{c}^{(i)}(t_l)\in I_{yj}}^{N_p}
    1
  \,,
\end{equation}
where $\mathbf{x}_c^{(i)}=(x_c^{(i)},y_c^{(i)},x_c^{(i)})$ are the
center coordinates of the $i$th particle, 
$y_j$ is the coordinate of the center of the bin $I_{yj}$, 
$n_{obs}$ the number of available records during the observation
interval and $|I_{yj}|$ is the width of $j$th wall-normal interval. 
The solid volume fraction simply becomes:
\begin{equation}\label{equ-def-particle-bin-average-phi}
  \langle \phi_s\rangle(y_j)=V_c\,\langle n_s\rangle(y_j)
  \,,
\end{equation}
denoting the particle volume by $V_c=D^3\pi/6$. 

The time- and plane-wise average of a quantity $\xi_c$ 
is computed as follows:  
\begin{equation}\label{equ-def-particle-bin-average}
  \langle\xi_c\rangle(y_j)=
  \frac{1}{\langle n_s\rangle(y_j)n_{obs}L_xL_z|I_{yj}|}
  \sum_{l=1}^{n_{obs}}
    \mathop{\sum_{i=1}}_{y_{c}^{(i)}(t_l)\in I_{yj}}^{N_p}
    \xi_c^{(i)}(t_l)
  \,.
\end{equation}
Here $\xi_c$ can stand for any component of the linear or angular
particle velocity vectors, which are denoted by $\mathbf{u}_c^{(i)}$
and $\boldsymbol{\omega}_c^{(i)}$, respectively. 

In order to analyze the motion of the dispersed phase, we have stored
the full set of particle-related quantities after every 10 (5)
time steps for case A (B); this corresponds to approximately
$1.5\cdot10^7$ ($5.2\cdot10^7$) samples. 
Statistical quantities were then computed in a post-processing stage.

\begin{figure}
  \centering
  \begin{minipage}{3.5ex}
    $\displaystyle\frac{\langle\phi_s\rangle}{\Phi_s}$
  \end{minipage}
  \begin{minipage}{.5\linewidth}
    \centerline{$(a)$}
    \includegraphics[width=\linewidth]{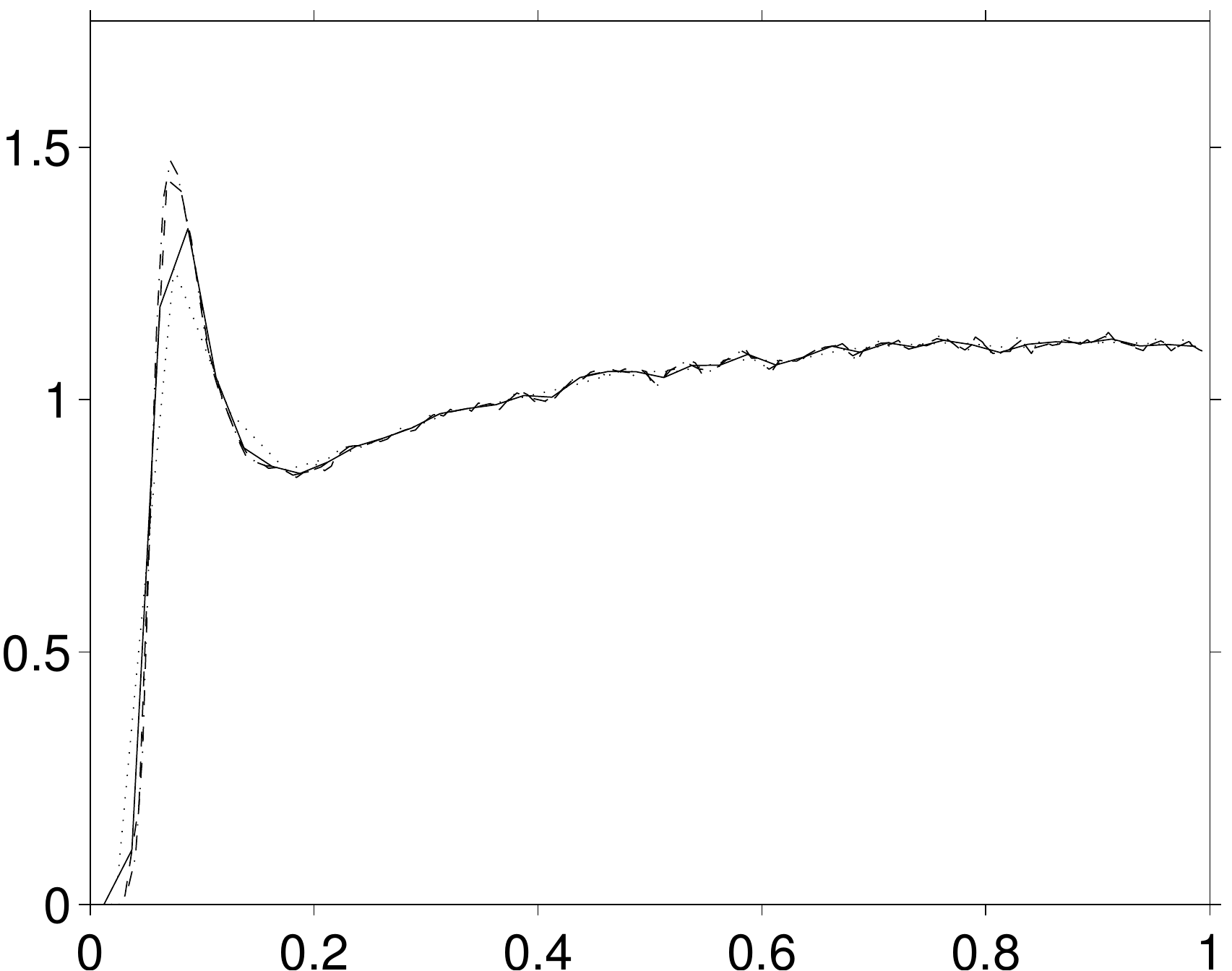}
    \centerline{$y/h$}
  \end{minipage}
  \\[1ex]
  \begin{minipage}{3.5ex}
    $\displaystyle\frac{\langle\phi_s\rangle}{\Phi_s}$
  \end{minipage}
  \begin{minipage}{.5\linewidth}
    \centerline{$(b)$}
    \includegraphics[width=\linewidth]{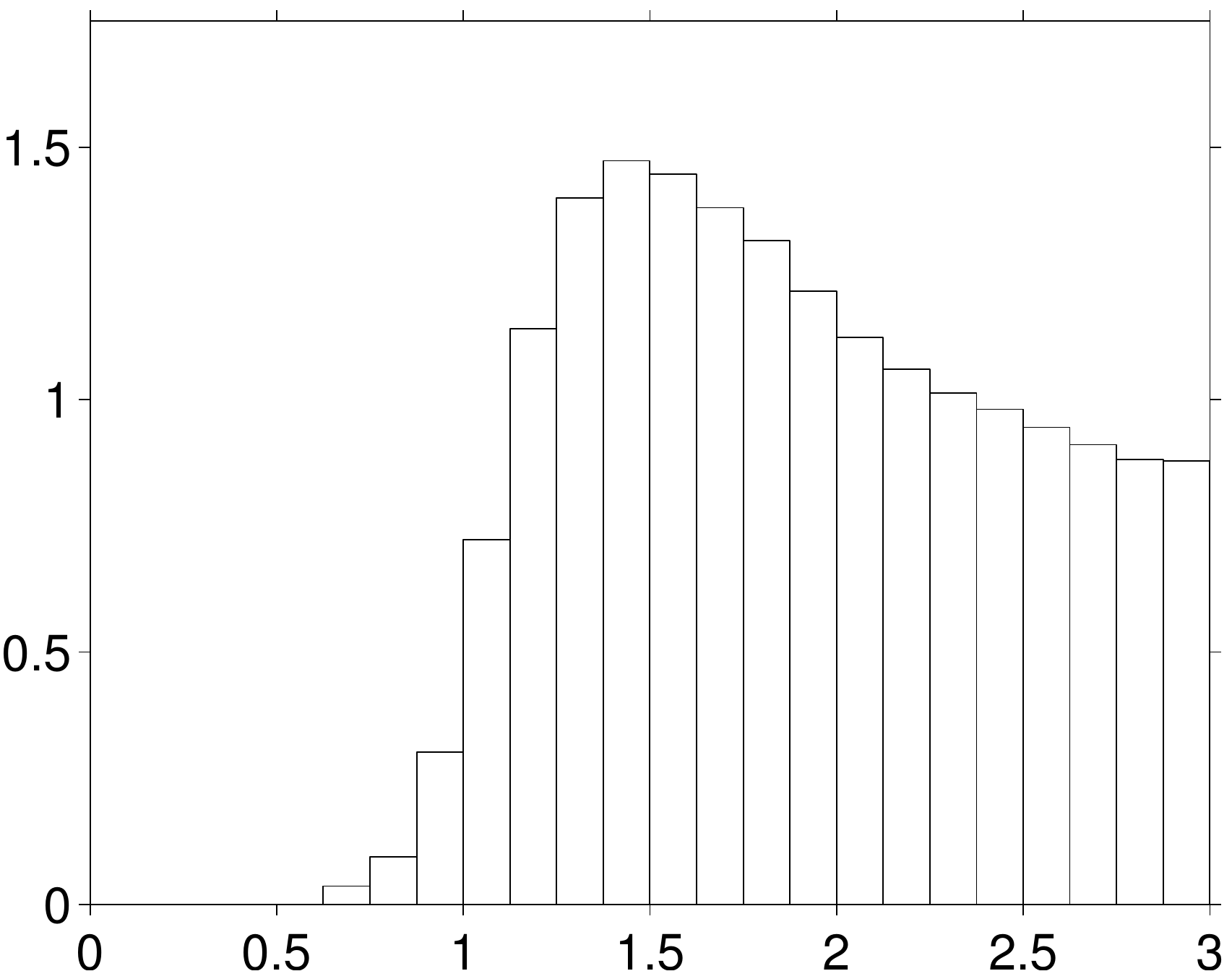}
    \centerline{$y/D$}
  \end{minipage}
  \caption{The effect of discrete binning upon the data for the mean
    solid volume fraction $\langle\phi_s\rangle$ in case B. 
    $(a)$ Profiles accumulated for different numbers of bins:
    \dotted~$N_{bin}=40$; \solid~$N_{bin}=80$;
    \dashed~$N_{bin}=160$; \chndot~$N_{bin}=320$.
    $(b)$ The data for $N_{bin}=320$ in a close-up near the
    wall; the abscissa is normalized with the particle diameter.
  }
  \label{fig-results-phi-bin}
\end{figure}

In the vicinity of sharp gradients, discrete binning of the data for
the dispersed phase tends to smooth out the profiles.  This effect can
be seen in figure~\ref{fig-results-phi-bin}$(a)$, where the solid volume
fraction of case B is shown for different numbers of wall-normal
bins varying from $N_{bin}=40$ to 320. 
The plot confirms the adequacy of the overall number of
discrete samples, because the profile of $\langle\phi_s\rangle$ does
not become significantly more noisy when increasing the number of bins.
Since the particles have a non-negligible size with respect to the
channel width ($D/h=1/20$), the apparent drop from the maximum to a 
value of zero at the wall could be an artifact of the
binning. Figure~\ref{fig-results-phi-bin}$(b)$ demonstrates that this is
not the case. Using fine bins with a width of $D/8$ reveals first that
there are no particles within $y_c\leq D/2$, which is the zone
`forbidden' by the geometric constraint of the wall. 
Secondly it shows that the solid
volume fraction smoothly increases for wall-distances larger than this
lower limit. 
%
Furthermore, let us point out that the mean wall-normal particle
velocity is negligible ($\max|\langle v_c\rangle|/u_b\leq
\{10^{-4},7\cdot10^{-4}\}$ in cases A, B), indicating that the
particle distribution has indeed achieved a statistically steady
state.  

\begin{figure}
  \raisebox{-5ex}{
    \rotatebox{90}{$x$-momentum balance}}
  \begin{minipage}{.47\linewidth}
    \centerline{$(a)$}
    {\includegraphics[width=\linewidth]
      {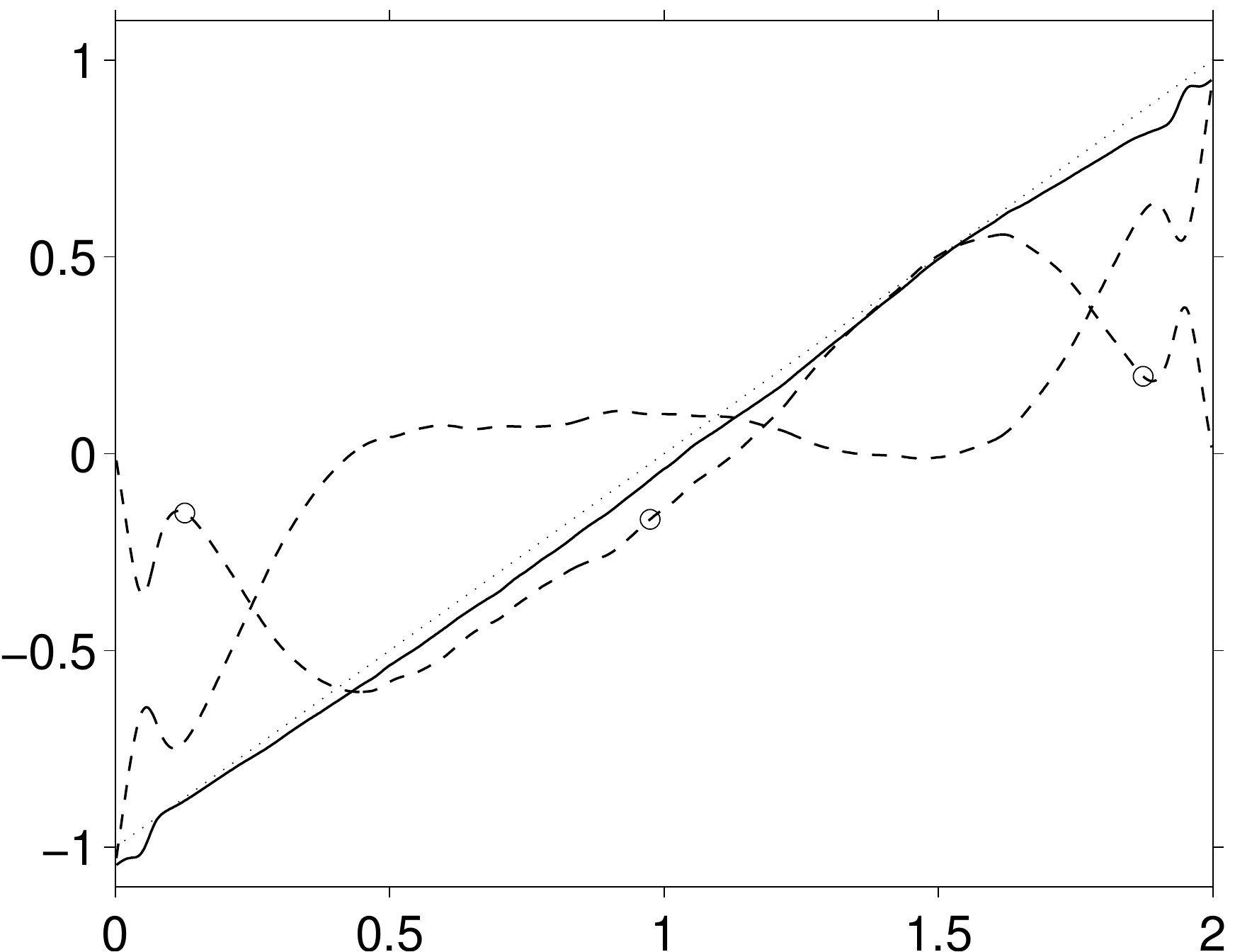}}
    \centerline{$y/h$}
  \end{minipage}
  \hfill
  \begin{minipage}{.47\linewidth}
    \centerline{$(b)$}
  {\includegraphics[width=\linewidth]
    {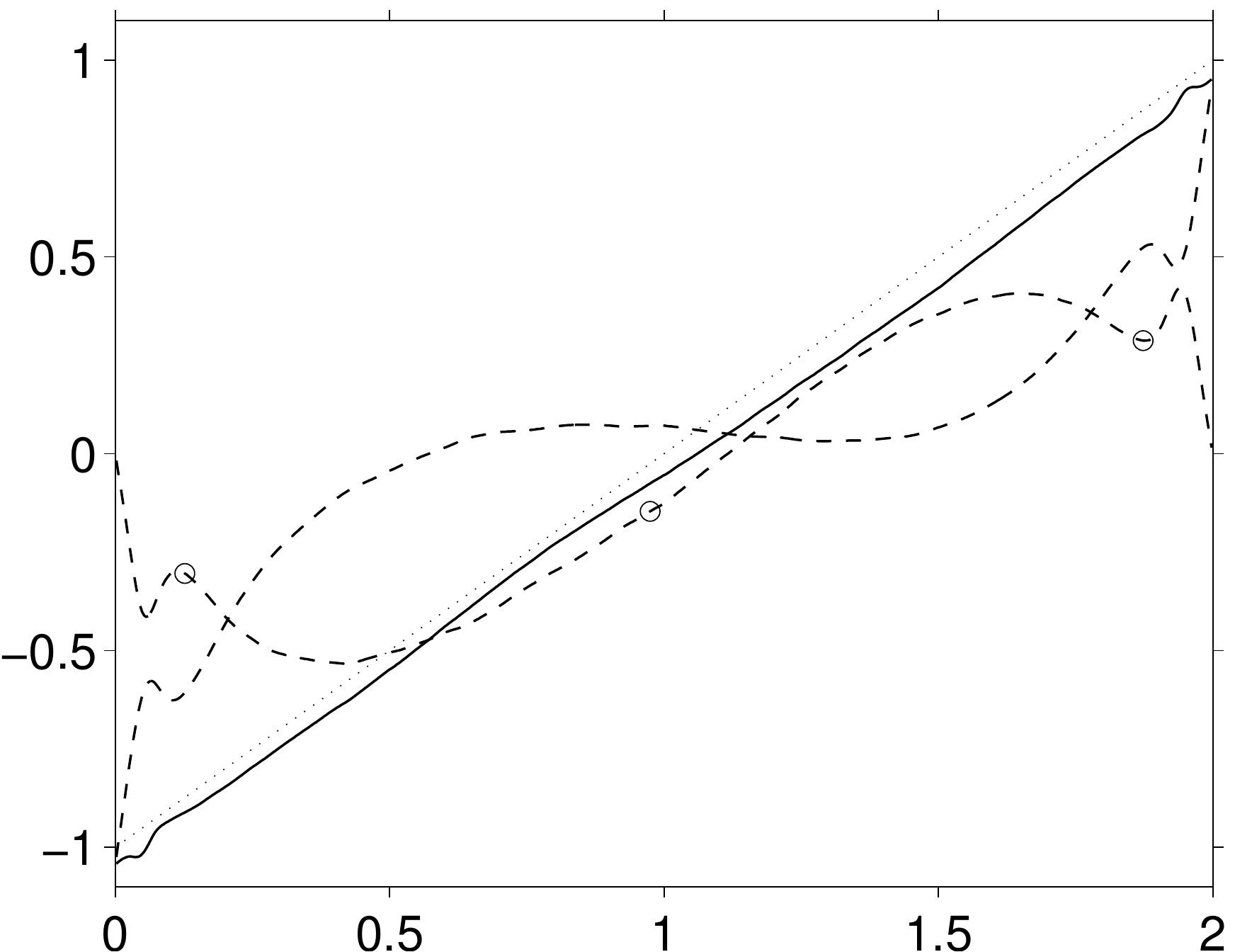}}
      \centerline{$y/h$}
  \end{minipage}
  \\[1ex]
  \raisebox{-5ex}{
    \rotatebox{90}{$x$-momentum balance}}
  \begin{minipage}{.47\linewidth}
    \centerline{$(c)$}
  {\includegraphics[width=\linewidth]
    {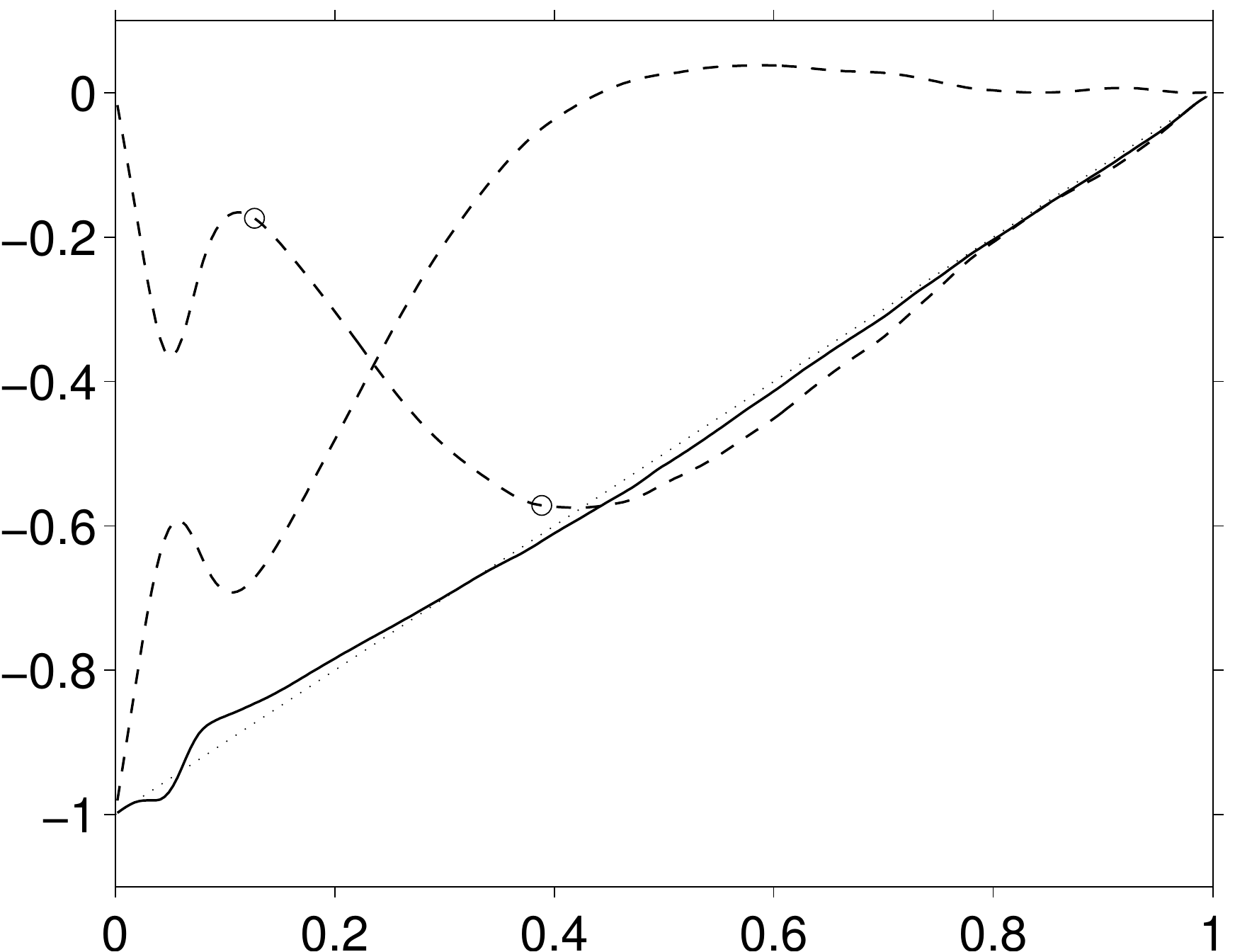}}
      \centerline{$y/h$}
  \end{minipage}
  \hfill
  \begin{minipage}{.47\linewidth}
    \centerline{$(d)$}
  {\includegraphics[width=\linewidth]
    {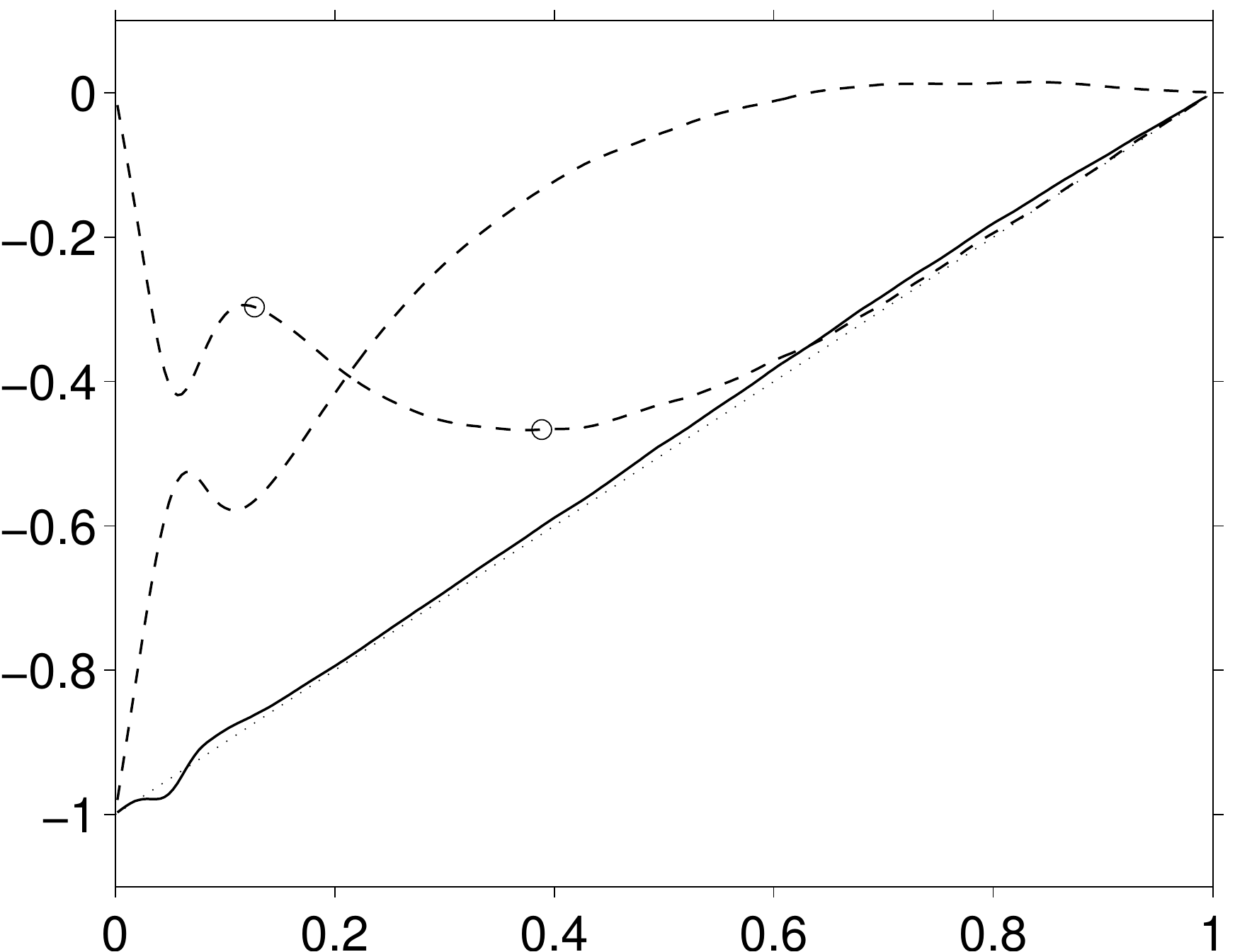}}
      \centerline{$y/h$}
  \end{minipage}
  \caption{The integral of the averaged streamwise momentum 
    equation (\ref{equ-particles-channelp-total-shear}), evaluated
    from the statistical data accumulated over the observation
    interval in cases A ($a$ and $c$) and B ($b$,$d$). $(a)$, $(b)$
    show the full channel width;  
    in $(c)$,$(d)$ the data is averaged over both channel half-widths,
    invoking odd symmetry. Statistics were accumulated over the
    composite flow field.  
    {The lines correspond to: 
      \solid~$\tau_{tot}+I_\phi$; 
      \dashed~$\tau_{tot}$; 
      \dashedcirc~$I_\phi$; 
      \dotted~$y/h-1$.}
  }
  \label{fig-results-shear-stress}
\end{figure}

Concerning the statistics of the carrier phase, we have accumulated
the usual low-order one-point moments (mean velocity, stress tensor)
during the simulations. 
For this purpose, averaging was performed over the entire composite 
flow field, i.e.\ containing the regions of the immersed solid
particles (the average velocity obtained by this procedure is denoted
by $\langle u\rangle$ etc).  
This decision was taken in favor of computational
efficiency, and its consequences are discussed below.
On the other hand, we have stored a number of full instantaneous flow
fields and corresponding particle data for visualization purposes and
{\it a posteriori} computation of additional quantities, such as
two-point correlation functions.  
From these fields we have estimated the difference between averaging
first and second moments over the composite field versus taking into
account the actual fluid nodes only; the fluid-only averaged velocity
is denoted by $\langle u\rangle_f$. 
For more details the reader is referred to
appendix~\ref{app-fluidonly}. 
This test has lead us to the conclusion that 
the employed method overestimates the streamwise velocity fluctuations (by
up to 6.7\%) and slightly underestimates the mean fluid velocity (by
less than 0.7\%); the wall-normal fluctuation and the Reynolds stress
are overestimated by approximately 2\%, the spanwise component by
0.7\%. 
{These differences affect the data presented in some of the
figures discussed in the present section as well as the following
section \S~\ref{sec-results-euler-stats-mom} (i.e.\ the figures
\ref{fig-results-shear-stress}, 
\ref{fig-results-um}, 
\ref{fig-results-uv}, 
\ref{fig-results-up}$b$, 
\ref{fig-results-uu},
\ref{fig-results-tke},
\ref{fig-results-uup}$b$).
As shown in appendix~\ref{app-fluidonly}, the over-prediction of the
streamwise velocity fluctuation amplitude can be quite accurately
represented by a linear function of the solid volume fraction,
allowing for a correction of the run-time statistics. This correction
has been included in figures~\ref{fig-results-uu},
\ref{fig-results-tke} and
\ref{fig-results-uup}$(b)$.}

The usual check of the statistical convergence of low-order moments in
single-phase plane channel flow includes a verification of the
linearity of the total shear stress profile. In the multi-phase case,
however, the streamwise momentum balance includes an additional
contribution. The equation, as derived in appendix~\ref{app-mombal},
reads: 
\begin{equation}\label{equ-particles-channelp-total-shear-int}
  \frac{\mbox{d}\langle{u^{\prime}v^{\prime}}\rangle^+}{\mbox{d}\bar{y}}
  -\frac{1}{Re_\tau}\,\frac{\mbox{d}^2\langle u\rangle^+}{\mbox{d}\bar{y}^2}
  +\left(\frac{\rho_p}{\rho_f}-1\right)
  \frac{g_xh}{u_\tau^2}
  \left({\Phi}_s-\langle\phi_s\rangle\right)
  =1\,,
\end{equation}
where $\bar{y}=y/h$, and $(u^\prime,v^\prime,w^\prime)$ are the components
of the fluctuations with respect to the mean velocity $(\langle
u\rangle,0,0)$. 
It should be noted that this equation is written for the composite
flow field (including the fluid located at the positions of the solid
particles), and all averaging operators extend over the entire
computational domain $\Omega$. 
It can be seen from
(\ref{equ-particles-channelp-total-shear-int}) that whenever the particle
distribution is not uniform (i.e.\ when $\Phi_s-\langle\phi_s\rangle$
is different from zero), the total shear stress does not vary
linearly with the wall-distance. 
In order to test whether the accumulated data has indeed achieved
statistical equilibrium, we have compared the terms of the integral
of equation (\ref{equ-particles-channelp-total-shear-int}), viz. 
\begin{equation}\label{equ-particles-channelp-total-shear}
  \underbrace{\langle{u^{\prime}v^{\prime}}\rangle^+
    -\frac{1}{Re_\tau}\,\frac{\mbox{d}\langle
      u\rangle^+}{\mbox{d}\bar{y}}}_{\displaystyle\tau_{tot}} 
  +\underbrace{\left(\frac{\rho_p}{\rho_f}-1\right)
    \frac{g_xh}{u_\tau^2}
    \int\left({\Phi}_s-\langle\phi_s\rangle\right)\mbox{d}\bar{y}}_{
    \displaystyle I_\phi}
  =\bar{y}-1\,.
\end{equation}
In practice, the integral of the solid volume fraction was evaluated  
numerically (after piecewise polynomial interpolation of the
statistical data, using cubic spline
functions). Figure~\ref{fig-results-shear-stress}$(a$-$b)$ shows the different
contributions to the balance
(\ref{equ-particles-channelp-total-shear}). 
It can be seen that the profile of $\tau_{tot}$ deviates quite
significantly from the usual straight line, the particle contribution
$I_\phi$ being significant at all wall-distances. However, the sum of
the stress terms and the particle contribution $\tau_{tot}+I_\phi$
does indeed follow the line $\bar{y}-1$ quite closely. It is further
observed that the data in figure~\ref{fig-results-shear-stress}$(a$-$b)$
does not verify the odd symmetry with respect to the center
at $\bar{y}=1$ as much as would be expected from the length of the
respective observation intervals and the size of the computational
boxes. We will return to this lack of symmetry in
\S~\ref{sec-results-fluid-structures} below.
Additional averaging over both channel halves clearly improves
the match (cf.\ figure~\ref{fig-results-shear-stress}$c$-$d$). Due to
the larger sample size (both for the carrier phase as well as the
dispersed phase), the statistical data for case B is found to be
closer to statistical equilibrium than case A. 
Therefore, in the following we will mainly discuss results from case
B.  
It should be noted that in both cases, a certain deviation from
linearity is observed in a limited region near the wall
($\bar{y}\le0.1$). 
This discrepancy can be attributed to the effect of
discrete binning of the solid volume fraction, as discussed above 
(cf.\ figure~\ref{fig-results-phi-bin}$a$).
Finally, it can be concluded that the statistical data---when averaged
over both channel halves---has approximately achieved a steady state
consistent with equation (\ref{equ-particles-channelp-total-shear}).  

\subsection{One-point moments}
\label{sec-results-euler-stats-mom}
\begin{figure}
  \centering
  \begin{minipage}{5.ex}
    $\displaystyle\frac{\langle u\rangle}{u_b}$
  \end{minipage}
  \begin{minipage}{.5\linewidth}
    \centerline{$(a)$}
    \includegraphics[width=\linewidth]{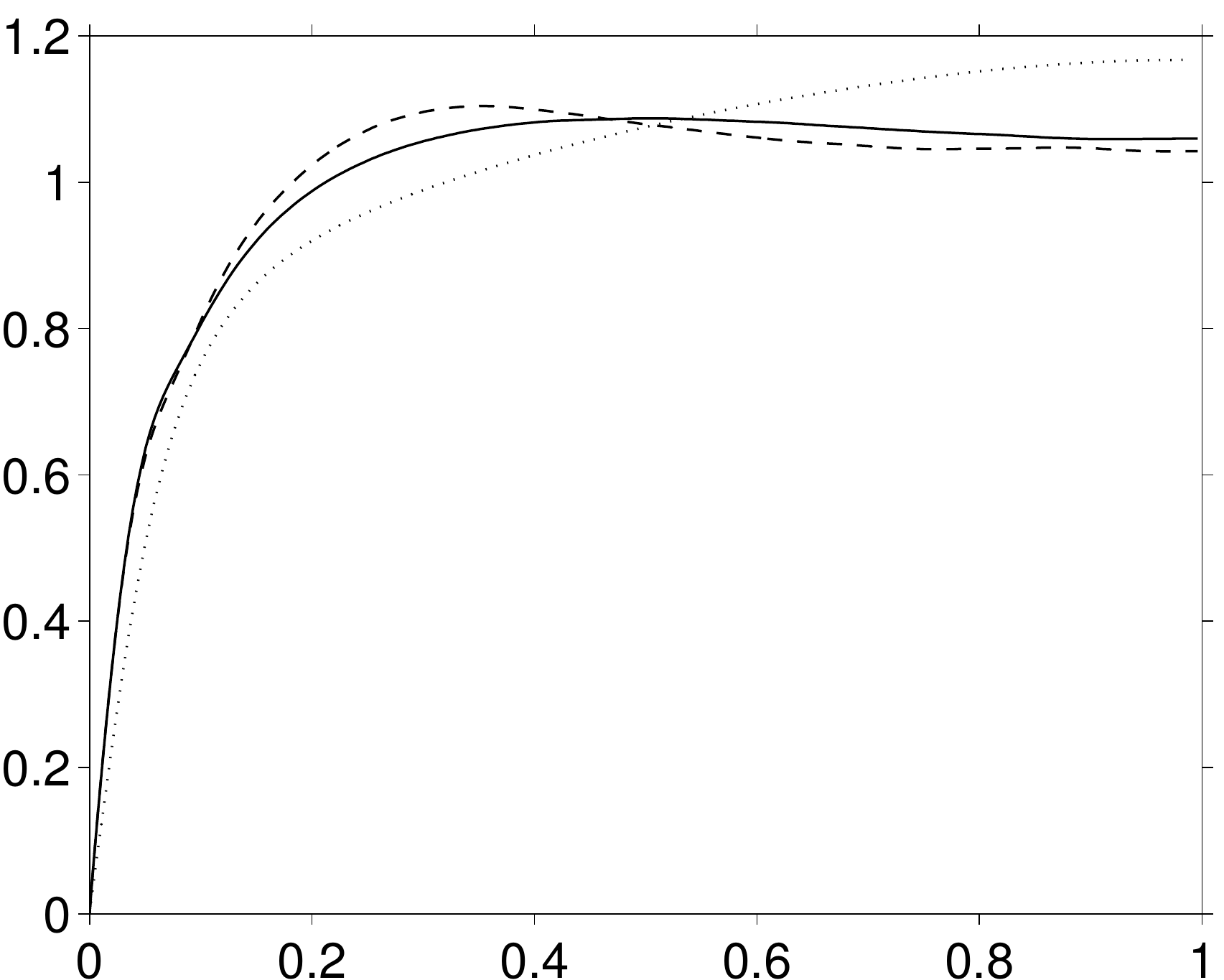}
    \centerline{$y/h$}
  \end{minipage}
  \\[1ex]
  \begin{minipage}{5.ex}
    $\langle u\rangle^+$
  \end{minipage}
  \begin{minipage}{.5\linewidth}
    \centerline{$(b)$}
    {\includegraphics[width=\linewidth]{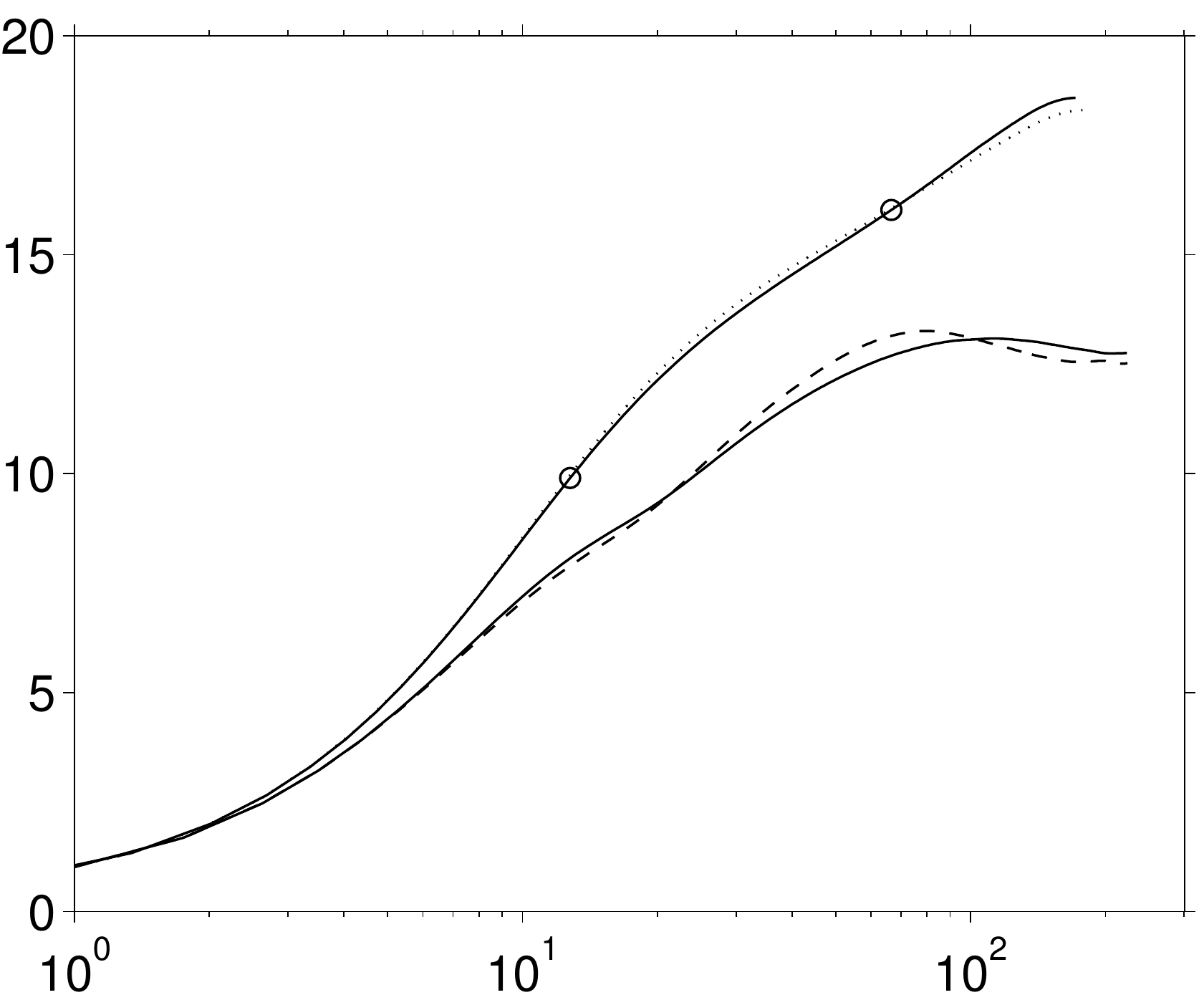}}
    \centerline{$y^+$}
  \end{minipage}
  \caption{Wall-normal profiles of the mean fluid velocity, $(a)$ in
    outer units, $(b)$ in wall units: 
    \dashed~case A; \solid~case B; \dotted~single phase flow
    \citep{kim:87}. Statistics were accumulated over the
    composite flow field.
    {In $(b)$ \solidcirc indicates the result of the
      single-phase computation with our code using the domain size of
      case A.} 
  }
  \label{fig-results-um}
\end{figure}

The mean fluid velocity profiles (computed by averaging over the
composite flow field) normalized by the bulk velocity are shown in
figure~\ref{fig-results-um}$(a)$. 
We observe a clear difference with respect to the single-phase case,
characterized by a tendency of the particulate flow to form a concave
profile with higher gradients at the walls and a flat section near the
center of the channel. 
The concave shape is slightly more pronounced in the high-inertia case
A.
On the other hand, when normalized in wall units (cf.\
figure~\ref{fig-results-um}$b$), the particulate flow cases exhibit a
considerably lower mean velocity than the single-phase counterpart
across the whole channel.  
As expressed by the streamwise momentum balance equation
(\ref{equ-particles-channelp-total-shear-int}) the mean velocity
of the composite field is determined by the Reynolds
shear stress and the particle distribution.
Before turning to the discussion of these two quantities, let us
mention that 
concave mean velocity profiles have previously been observed in
experiments on dilute particulate flow at various parameter points: in
vertical pipes \citep{tsuji:84,hadinoto:05} and---to a lesser extent---in
vertical channels \citep{sato:00}.
However, neither the distribution of the solid volume fraction nor the 
Reynolds shear stress profiles are available from these experimental
data-sets. 

\begin{figure}
  \centering
  \begin{minipage}{3.5ex}
    $\displaystyle\frac{\langle\phi_s\rangle}{\Phi_s}$
  \end{minipage}
  \begin{minipage}{.5\linewidth}
    \includegraphics[width=\linewidth]{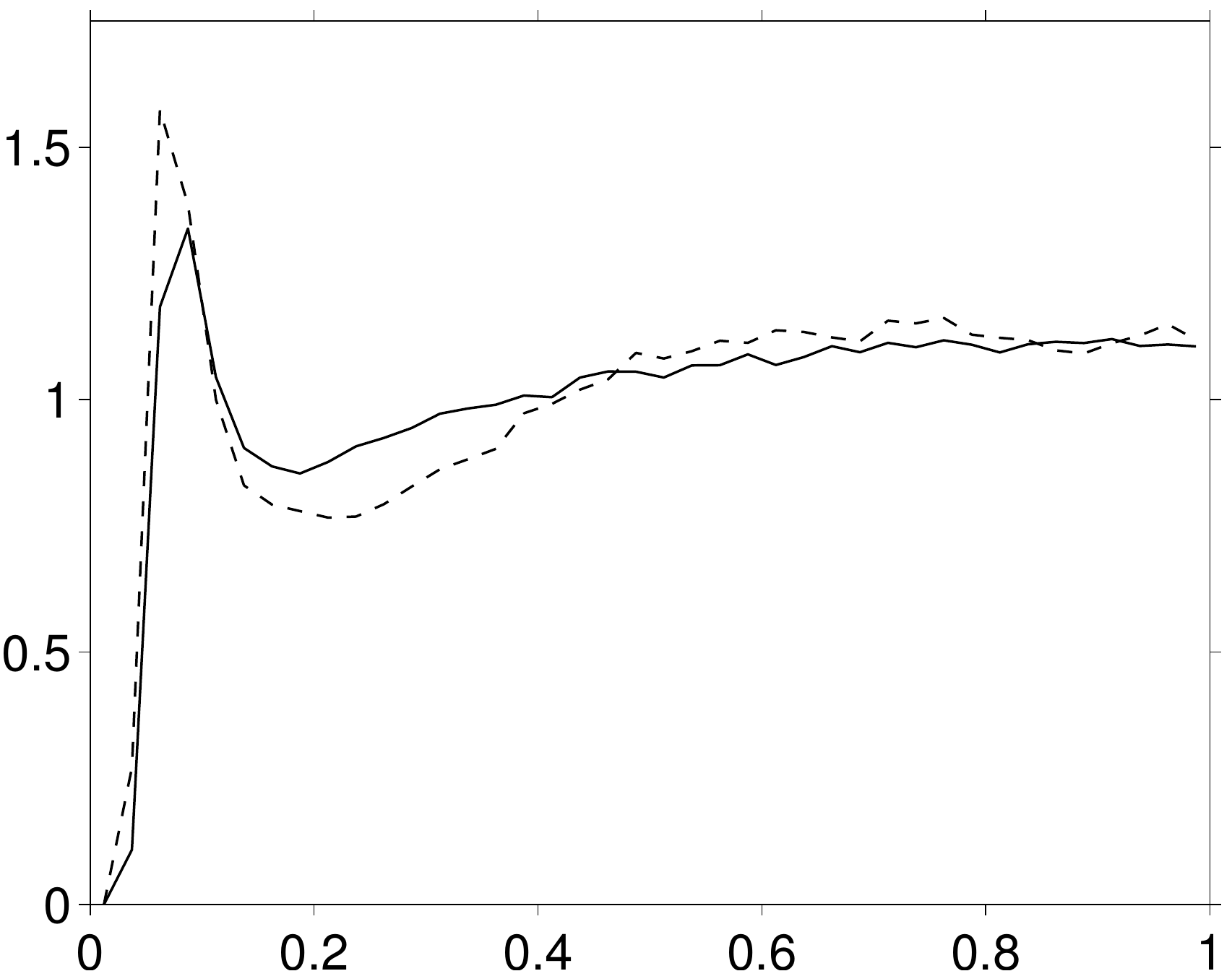}
      \centerline{$y/h$}
  \end{minipage}
  \caption{The wall-normal profile of the mean solid volume
    fraction for case A (\dashed) and case B (\solid), using
    $N_{bin}=80$.} 
  \label{fig-results-phi}
\end{figure}

The profiles of the mean solid volume fraction are shown in
figure~\ref{fig-results-phi}. 
In both cases we observe a sharp peak at a
wall distance of approximately $16$ wall units ($\bar{y}=0.07$), a
slightly below-average value around $\bar{y}=0.2$ and finally a mild
local maximum at the center of the channel. 
%
A sharp rise of the solid volume fraction near the wall was also
reported by \cite*{suzuki:00} in their experiment on downward channel
flow containing particles with a diameter of approximately $4$ wall
units. 

\begin{figure}
  \centering
  \begin{minipage}{11.ex}
    $\displaystyle
    -\frac{\mbox{d}\langle v_c^\prime v_c^\prime\rangle^+}{\mbox{d}\bar{y}}$
  \end{minipage}
  \begin{minipage}{.5\linewidth}
    \includegraphics[width=\linewidth]{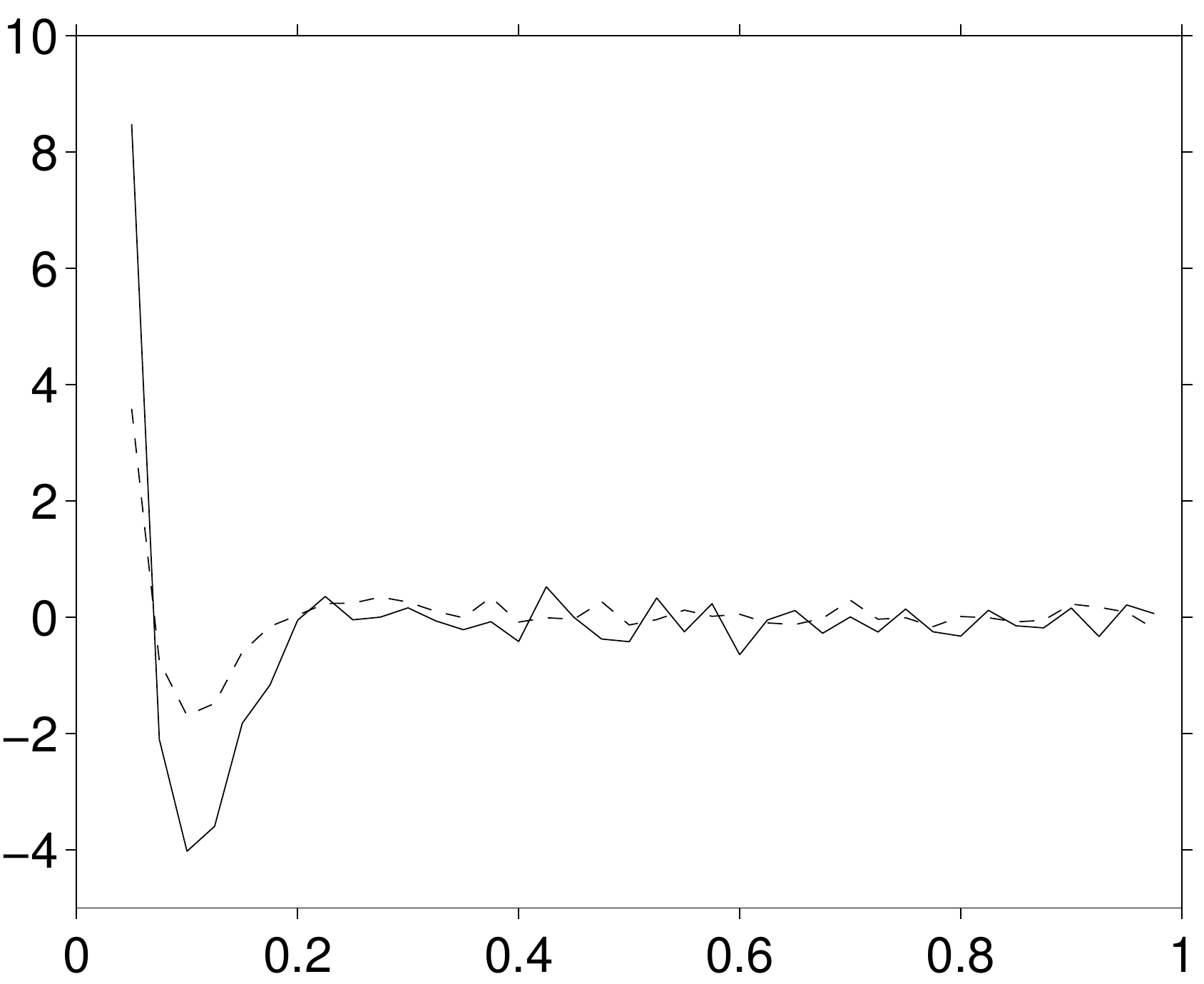}
    \centerline{$y/h$}
  \end{minipage}
  \caption{Wall-normal profiles of the derivative of the wall-normal
    particle velocity correlation; \dashed~case A; \solid~case B.
  }   
  \label{fig-results-dvv}
\end{figure}

A number of mechanisms which influence the wall-normal distribution of
particles have been discussed in the literature. 
One of the principal contributions 
has been termed {\it turbophoresis} \citep{caporaloni:75,reeks:83}, 
referring to an average migration of particles in the direction
opposite to gradients in the turbulence intensity. 
For an individual particle this means that it is less probable to
receive the necessary momentum driving it from a region of low
turbulence intensity towards a high-intensity region than vice versa; 
turbophoresis is then a statistical consequence.
%
An expression corresponding to a turbophoretic force appears formally
when considering a Eulerian formulation for the dispersed phase. In
this context the wall-normal particle momentum equation contains the
following term:   
$-\partial\langle v_c^\prime v_c^\prime\rangle/\partial y$
\citep{young:97}.  
The profile of this quantity is shown in figure~\ref{fig-results-dvv}
(the correlation $\langle v_c^\prime v_c^\prime\rangle$ itself is
shown in figure~\ref{fig-results-uup} and will be discussed below).  
It can be seen that the turbophoresis term is positive for
wall-distances below $y_c^+\approx16$ (i.e.\ driving particles towards
the center) and then negative (directed wall-ward) up to approximately
$\bar{y}=0.2$; for larger wall-distances its values are close to
zero. 
The point where the turbophoretic force changes sign corresponds to
the near-wall maximum of the mean solid volume fraction
$\langle\phi_s\rangle$ 
(cf.\ figure~\ref{fig-results-phi}), 
and the interval over which the force has significant values matches
the region where the corresponding bump in $\langle\phi_s\rangle$ is
observed. 
This close correspondence suggests that turbophoresis is indeed
important for the mean wall-normal particle distribution in the
presently investigated cases.   
%
Other effects---such as the mean shear-induced lift force and
the turbulent diffusion---are expected to contribute as well. 
However, the additional mechanisms are not directly quantifiable from
the available data, and, therefore, it is not possible to further 
investigate their relative importance in establishing the equilibrium
distribution.  

\begin{figure}
  \centering
  \begin{minipage}{18.ex}
    $\langle u^\prime v^\prime\rangle^+$
  \end{minipage}
  \begin{minipage}{.5\linewidth}
    \centerline{$(a)$}
    \includegraphics[width=\linewidth]{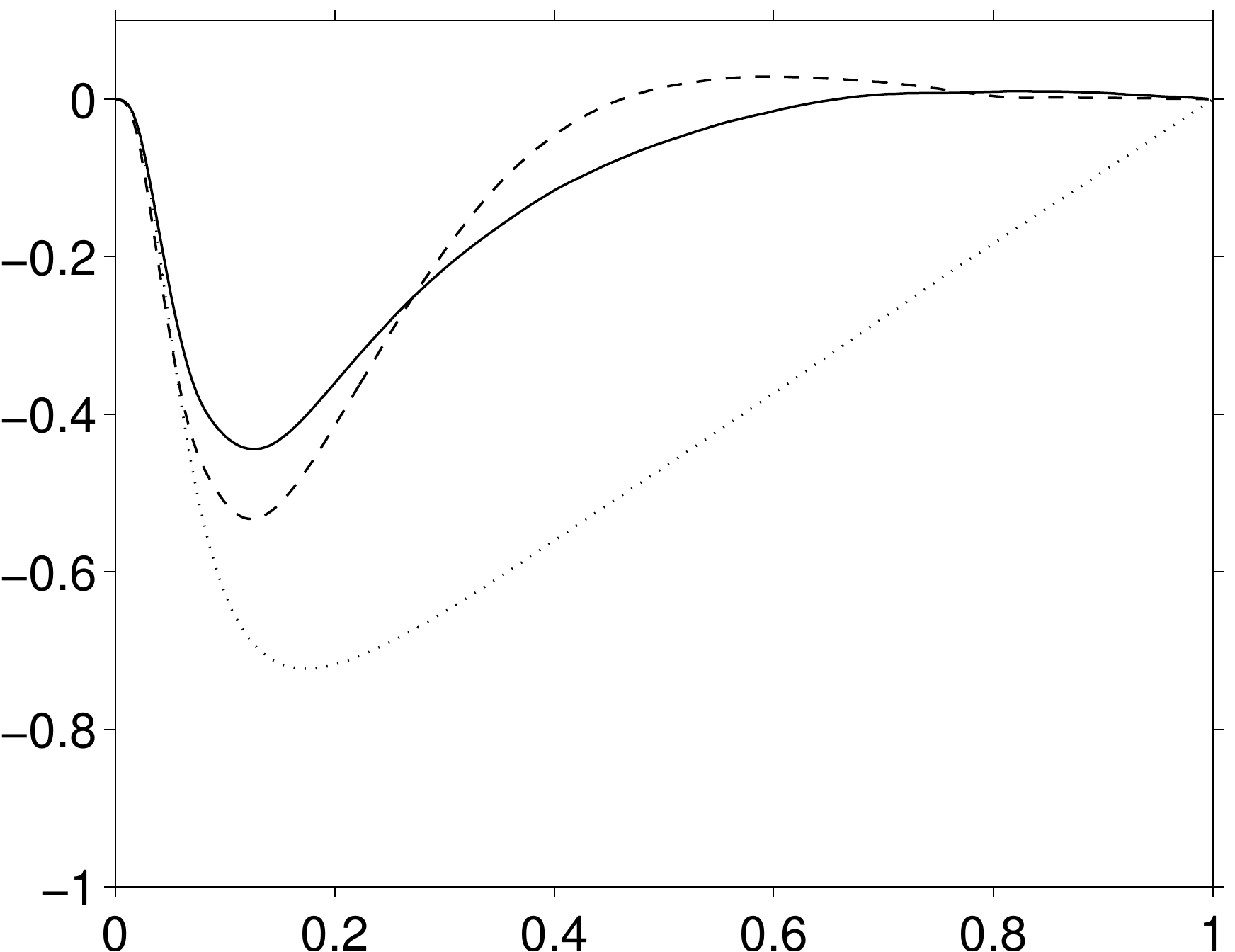}
    \centerline{$y/h$}
  \end{minipage}
  \\[1ex]
  \begin{minipage}{18.ex}
    $\langle u^\prime v^\prime\rangle^++I_\phi$
  \end{minipage}
  \begin{minipage}{.5\linewidth}
    \centerline{$(b)$}
    \includegraphics[width=\linewidth]{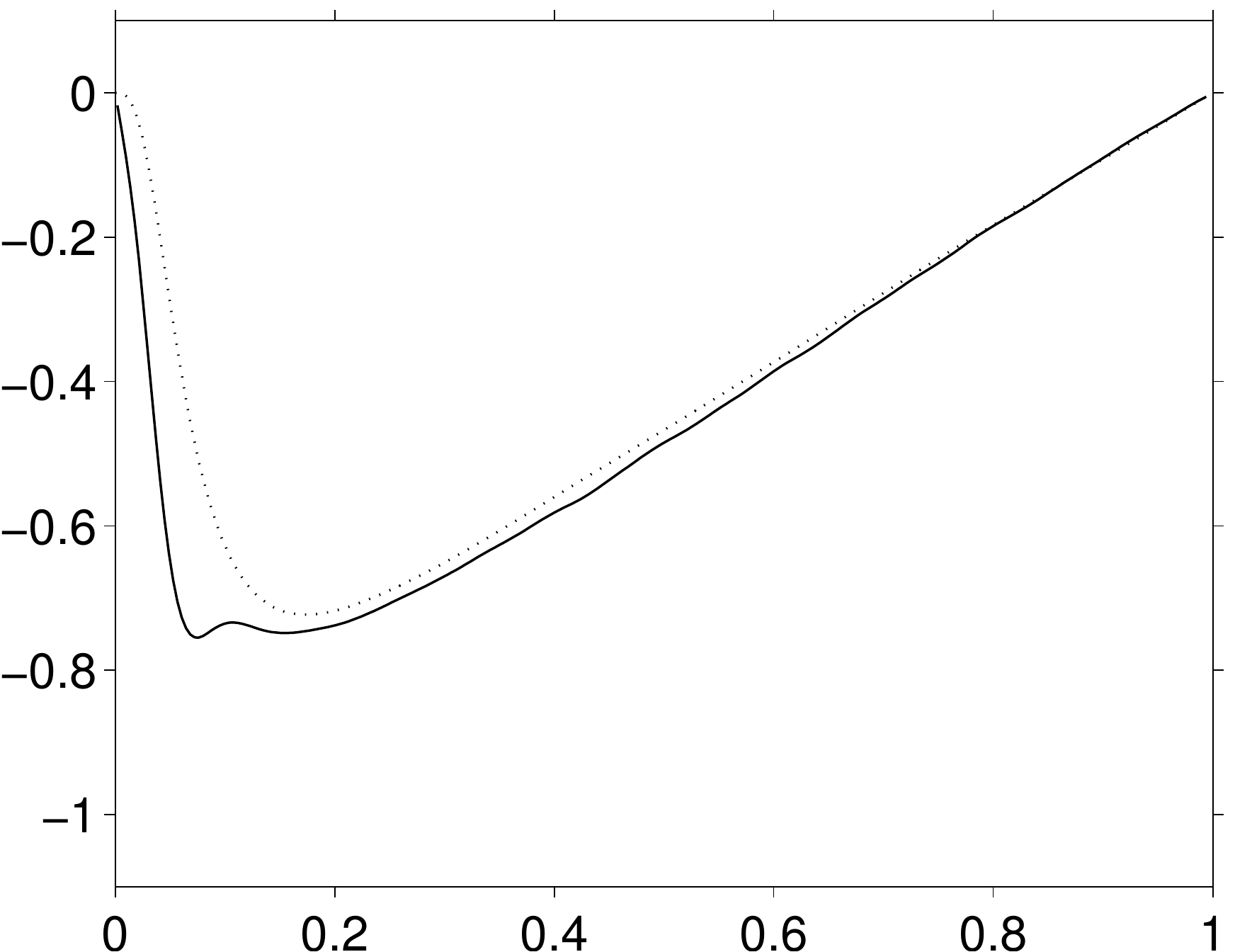}
    \centerline{$y/h$}
  \end{minipage}
  \caption{$(a)$ Wall-normal profiles of the Reynolds shear stress; 
    line styles as in figure~\ref{fig-results-um}.  
    In $(b)$ The sum of the Reynolds shear stress and the particle
    contribution to equation (\ref{equ-particles-channelp-total-shear}),
    $\langle{u^{\prime}v^{\prime}}\rangle^++I_\phi$, for case B
    (\solid) is compared to the Reynolds shear stress of the
    single-phase flow (\dotted). Statistics were accumulated over the
    composite flow field.
  }
  \label{fig-results-uv}
\end{figure}

The Reynolds stress normalized by the friction velocity
is shown in figure~\ref{fig-results-uv}$(a)$. The profiles of both
simulated cases are of a much smaller amplitude than 
the single-phase reference data across most of the channel
width. When normalized by the bulk velocity, the reduction is limited
to wall-distances $\bar{y}\gtrapprox0.2$. 
Notably, near the center of the channel a considerable region with
vanishing Reynolds stress is observed. This means that the correlation 
between the streamwise and the wall-normal fluid velocity fluctuations
is significantly reduced in the present cases. 
However, 
the sum of the Reynolds shear stress and the contribution from the
solid volume fraction to the integrated momentum balance
(\ref{equ-particles-channelp-total-shear}), i.e.\ 
$\langle u^\prime v^\prime\rangle^++I_\phi$,  
(shown in figure~\ref{fig-results-uv}$b$) 
is of larger absolute value than the Reynolds shear stress
of the single-phase flow when normalized in wall units. 
From equation (\ref{equ-particles-channelp-total-shear}) it follows
then  that the wall-normal gradient of the mean streamwise velocity
(in wall units) is smaller in the particulate case, leading to the
reduced value of $\langle u\rangle^+$ observed in
figure~\ref{fig-results-um}$(b)$.  
Furthermore, we note from figure~\ref{fig-results-uv}$(b)$ that the
largest difference between the sum
$\langle{u^{\prime}v^{\prime}}\rangle^++I_\phi$ and the Reynolds shear
stress of the single-phase case is obtained around $\bar{y}=0.075$,
where the profile of the former quantity exhibits a visible bump. This
is evidently the signature of the near-wall peak of the solid volume
fraction 
(cf.\ figure~\ref{fig-results-phi}) which is found at the same
location.  
Consequently, the accumulation of particles in the buffer region has
the effect of bringing more momentum towards the wall.
This in turn causes the flattening of the mean velocity profile and
generates a shear-free region near the center of the channel. 

\begin{figure}
  \centering
  \begin{minipage}{11ex}
    $\displaystyle\frac{\langle u_c\rangle}{u_b}$
  \end{minipage}
  \begin{minipage}{.5\linewidth}
    \centerline{$(a)$}
    \includegraphics[width=\linewidth]{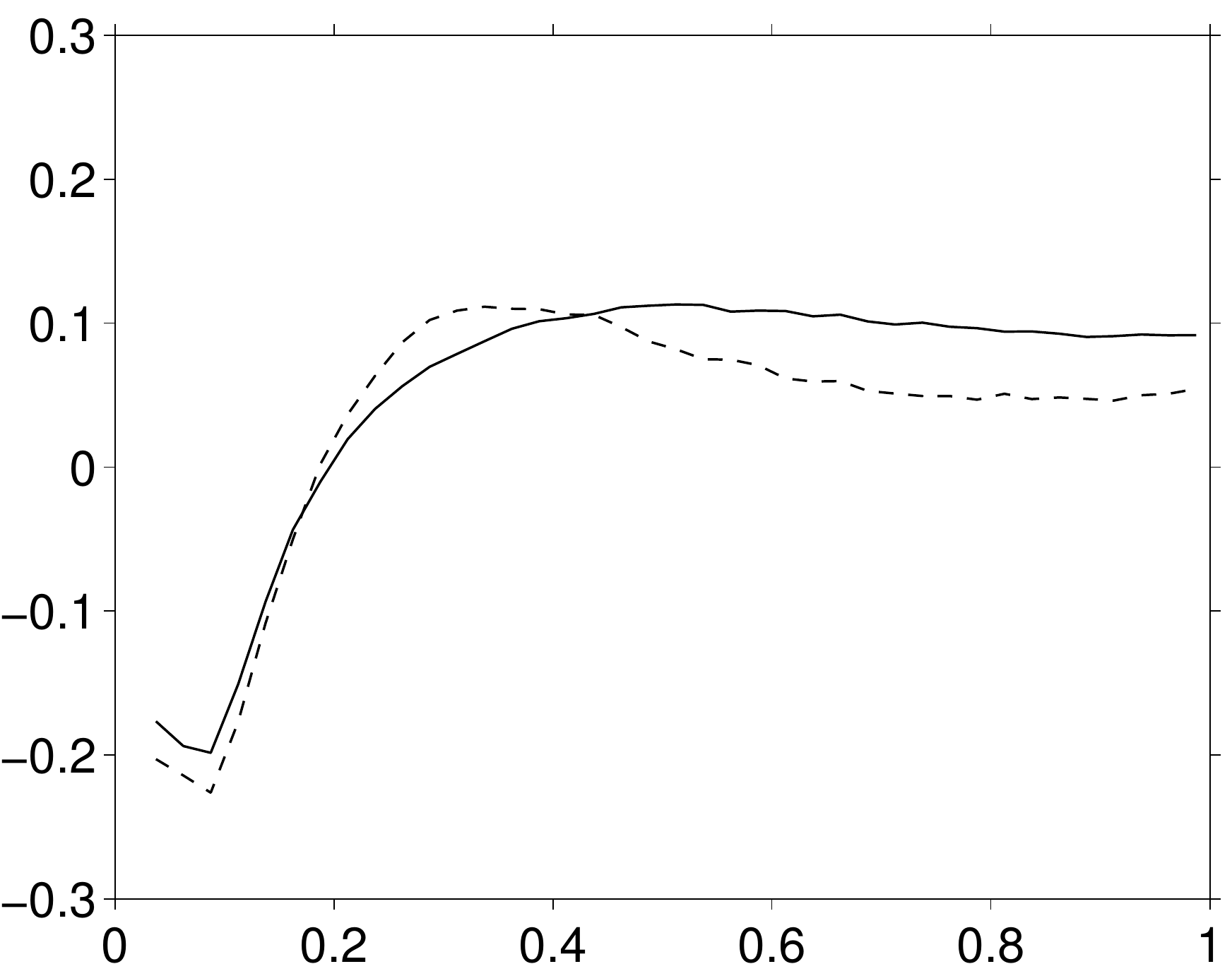}
      \centerline{$y/h$}
  \end{minipage}
  \\[1ex]
  \begin{minipage}{11ex}
    $\displaystyle\frac{\langle u_c\rangle-\langle u\rangle}{u_b}$
  \end{minipage}
  \begin{minipage}{.5\linewidth}
    \centerline{$(b)$}
    \includegraphics[width=\linewidth]{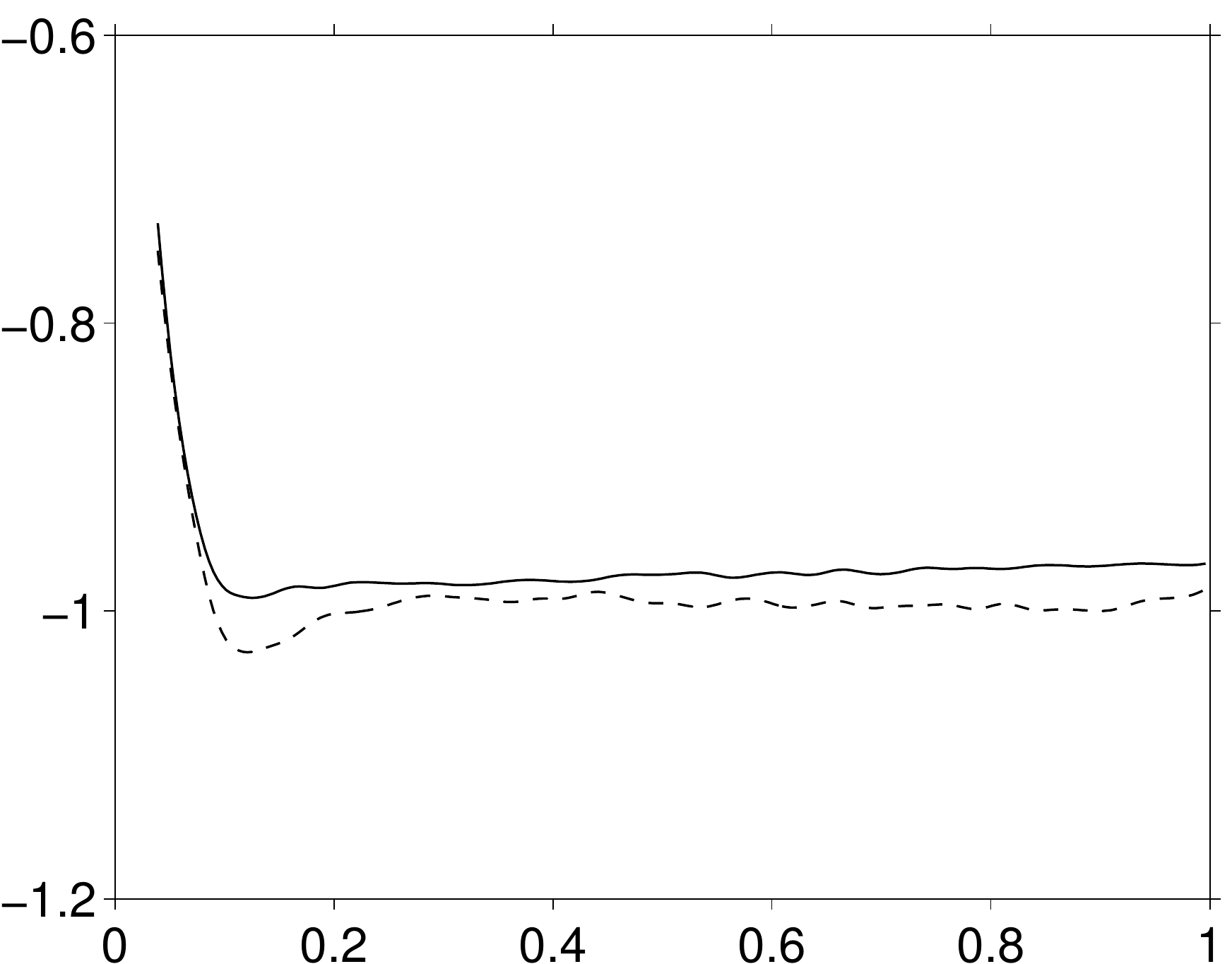}
      \centerline{$y/h$}
  \end{minipage}
  \caption{Wall-normal profiles of $(a)$ the mean particle velocity,
    and $(b)$ 
    {the difference between the mean velocities of the two phases:}
  \dashed~case A; \solid~case B. Statistics for $\langle u\rangle$
  were accumulated over the composite flow field.} 
  \label{fig-results-up}
\end{figure}

Figure~\ref{fig-results-up}$(a)$ shows the profiles of the mean
(streamwise) particle velocity, which is positive in the center of the
channel and changes sign at approximately $\bar{y}=0.2$. In both cases
the shape of the profile is very similar to the corresponding mean
fluid 
velocity profile (cf.\ figure~\ref{fig-results-um}$a$), but with a
negative shift. 
Close to the wall, the mean particle velocity profile differs in shape
from the fluid velocity counterpart, 
as will be discussed below. 
{In figure~\ref{fig-results-up}$(b)$ the difference between the mean
  velocities of the two phases, 
  $\langle u_c\rangle-\langle u\rangle$, is plotted; it is found to be
  approximately constant and nearly equal to $-u_b$. 
  For wall-distances $\bar{y}\leq0.1$,
  however, the magnitude of the relative mean velocity substantially
  decreases. This result indicates that the coefficient of the drag
  force acting on average upon the particles in this region is higher
  than at large wall-distances. 
  %
  Similar behavior is reported by \cite*{zeng:05} who have 
  investigated the drag force on a single sphere
  translating in a channel with quiescent fluid. These authors have
  detected a significant increase of the drag coefficient for wall
  distances below $y/D=2$ (i.e.\ for $\bar{y}\leq0.1$ in the present
  context), at similar values for the particle Reynolds number. 
  A drag coefficient based upon the relative mean velocity of our
  simulations can be computed by 
  formulating the equilibrium between buoyancy and drag forces,
  $C_{D,present}=(\rho_p/\rho_f-1)|{g}_x|\frac{4}{3}D/(\langle
  u_c\rangle-\langle u\rangle)^2$, and substituting the data of 
  figure~\ref{fig-results-up}$(b)$. 
  The ratio between this relative mean DNS drag coefficient 
  and the standard drag coefficient of an isolated sphere given by the
  Schiller-Naumann formula \citep*{clift:78} is
  $C_{D,present}/C_{D,SN}=1.46$ at a wall distance of $y/D=0.9375$
  (where the particle Reynolds number based upon the relative mean
  velocity is $107.2$).   
  As a comparison, the value obtained from the correlation formula given by
  \cite{zeng:05} is $C_{D,ZBF}/C_{D,SN}=1.44$. 
  The match is indeed remarkable, indicating that results obtained for
  the wall-effect on individual particles at ambient flow conditions are
  indeed relevant to turbulent particulate flow in the dilute regime.}

The r.m.s.\ fluctuations of fluid 
velocities, normalized by the friction velocity, are shown in
figure~\ref{fig-results-uu}. 
{This figure also includes profiles of the streamwise
  component from which the overestimation due to the computation
  of statistics over the composite flow field has been subtracted
  (using the fit of appendix~\ref{app-fluidonly}).}
The streamwise component in both particulate flow cases is strongly
enhanced as compared to the single-phase flow, except for a small
interval very close to the wall. 
%
The wall-normal component is slightly increased for $\bar{y}<0.1$ and
slightly decreased for wall-distances above that value;  
the spanwise component is decreased almost across the whole channel
width.  
When normalized by the bulk velocity, all three profiles exceed 
the single-phase reference data, except for the spanwise
component for intermediate wall-distances ($0.1\leq\bar{y}\leq0.6$). 
%
{Since the streamwise component is dominant, the turbulent kinetic
energy is significantly enhanced with respect to the single-phase
flow, as shown in figure~\ref{fig-results-tke}.  When subtracting the
overestimation due to the composite flow statistics, the 
ratio reaches approximately 4.6 (5.3) on the centerline in case A (B);
when normalized in bulk units, the ratio is 7.9 (9.0).}
%
The principal reason for the observed turbulence enhancement is the
generation of velocity fluctuations in the wakes trailing the
particles. Wakes are indeed found to be the most prominent flow
structures, as can be concluded from flow visualisations (cf.\ 
figure~\ref{fig-results-5-fluid-snapshot} below). 
The wake structures significantly increase the
turbulence intensity, but they do not contribute to the
generation of the mean Reynolds shear stress, since correlations
between $u^\prime$ and $v^\prime$ internal to each wake cancel out
upon averaging over an ensemble of mostly uncorrelated wakes (the
random-like distribution of particles will indeed be confirmed in
\S~\ref{sec-results-particle-structure}). 
In addition to the wakes, new large-scale velocity structures
representing streamwise velocity perturbations appear in the
particulate flow. These flow structures, which will be discussed in
\S~\ref{sec-results-fluid-structures}, further contribute to the
enhancement of turbulence intensity.
In case A somewhat lower values for the streamwise turbulence
intensity are obtained when compared to case B. This difference might
be a consequence of the smaller domain size in the former case, where
the large particle-induced structures are strongly constrained. 
%
A significant increase in streamwise velocity fluctuation levels has 
also been detected by \cite{suzuki:00}, although their solid volume
fraction was one order of magnitude lower. The authors already
attributed this observation to the presence of the wakes.   

{The r.m.s.\ velocity fluctuations of the dispersed phase are shown in
  figure~\ref{fig-results-uup}$(a)$; for convenience, a comparison of
  the r.m.s.\ velocity fluctuations of both phases in case B (i.e.\
  combining data of figures~\ref{fig-results-uu} and
  \ref{fig-results-uup}$a$) is shown in figure~\ref{fig-results-uup}$(b)$.}  
%
{A similarity between the corresponding components of
  the two phases can be observed on these graphs, except for the
  immediate vicinity of the wall.} 
The streamwise particle r.m.s.\ velocities also exhibit a mild peak in
the buffer layer, and the shape of the wall-normal and
spanwise components is similar as for the fluid data. 
This similarity shows that the present particles are able to respond to
the fluid velocity fluctuations. 
The kinetic energy of the particle velocity fluctuations (not shown)
is lower than the fluid turbulent kinetic energy across the whole
channel. This behavior is expected due to the particle inertia. 
In the high-inertia case A, the intensity of the fluctuations of all
particle velocity components is smaller than the fluid data. 
In case B, the streamwise component is also considerably smaller than
the corresponding fluid values. The level of the remaining two
components, however, slightly exceeds the fluid counterparts across
most of the channel. 
%
In addition, it can be seen in figure~\ref{fig-results-uup} that the
particle velocity fluctuation intensities flatten when approaching the
wall and increase again for small wall-distances
$\bar{y}\lessapprox0.06$ (corresponding to $14$ wall units). 
What looks like a `kink' of the curves is not an artifact of the data
analysis, since a refinement of the width of the bins used in the
computation of the Eulerian statistics has confirmed the observed
trend. It should be remarked that the particles are not subject to the 
no-slip condition at the wall, and, therefore, the asymptotic 
behavior of the particle velocity is expected to differ from the one
of the fluid velocity \citep[cf.][]{kulick:94}.  
However, it is difficult to imagine how the intensity of the
fluctuation components can rise near the wall, where the flow is
increasingly quiescent. 
Also, the observed behavior does not seem to be due to the numerical
treatment of particle-wall collisions, since those only concern the
wall-normal component of the particle motion, whereas the profiles of
all components exhibit a similar upward trend. 
It must be concluded that the reasons for the increased near-wall
variance of particle motion remain unclear at the time being,
requiring further research. 

The profiles of the spanwise component of the mean angular particle
velocity $\langle \omega_{c,z}\rangle$ are shown in 
figure~\ref{fig-results-uprot}$(a)$. In both cases there is a minimum
at a wall-distance of approximately $25$ wall units, and the profile
tends towards zero at the centerline and at the wall. 
In figure~\ref{fig-results-uprot}$(b)$ 
{the DNS data are compared} 
to the mean shear profile.  
In a linear shear flow the steady rotation of a cylinder is described
by the formula $-A\,\mbox{d}\langle u\rangle/\mbox{d}y$,
where $A$ is positive and decreases with the shear
Reynolds number \citep[cf.][]{ding:00,zettner:01}. 
Constrained simulations of a single neutrally-buoyant and spherical
particle placed in laminar pipe flow have likewise yielded a
proportionality between the steady-state angular particle velocity and
the wall-normal fluid velocity gradient over most of the pipe radius
\citep[][figure 3]{yang:05b}. 
It can be seen In figure~\ref{fig-results-uprot}$(b)$ that our data
for case B 
\revision{is}
{are} 
reasonably well represented when choosing $A=0.1$,
except for the region $y^+<25$ (similarly for case A, which is not
shown). 
%
The discrepancy in the near-wall region can probably be attributed to
a modification of the flow field around particles in the immediate
vicinity of the wall induced by the no-slip condition. 
A similar tendency was observed for a cylinder in linear shear
flow when the confinement ratio was lowered \citep{ding:00}.

\subsection{Probability density functions and quadrant analysis}
\label{sec-results-euler-stats-pdfs}
Probability density functions (pdfs) of the fluid velocity components
were computed from 12 instantaneous flow fields of case B, taking into
account only values at grid nodes inside the fluid domain. The sample
size was doubled by making use of the symmetry with respect to the
center of the channel.
In the following we will discuss the normalized pdfs in the buffer
region ($y^+=20$) and at the center of the channel, in comparison to
the  single-phase data of \cite*{moser:99} at $Re_\tau=180$. The
values of the skewness at these wall-distances are given in
table~\ref{tab-results-pdf-skewness-fluid}. It can be seen that the
skewness of the spanwise velocity component computed from our data set
is close to zero 
{as a consequence of the spanwise periodicity, indicating
that sufficient sample has been obtained.}
In figures~\ref{fig-results-pdf-fluid-1} and
\ref{fig-results-pdf-fluid-2}
we observe that the pdf for the streamwise component is strongly
non-symmetric with respect to the mean value. While the branch for 
positive fluctuations matches those of the single-phase data closely,
there exists a higher probability of finding strongly negative
fluctuations in the present particulate case. This tendency is
confirmed for both wall-distances and it reflects the negative
perturbations induced by the streamwise momentum deficit in the wakes
trailing the particles. 
The comparison with the single-phase data shows that the influence of
the particles upon the shape of the pdfs for the wall-normal and
spanwise velocity components is relatively small, manifesting itself
through a small increase of the non-Gaussian tails. 
The skewness of the wall-normal component in the buffer layer is
reduced from its value in single-phase flow.

The corresponding pdfs of the particle velocity are shown
in figure~\ref{fig-results-pdf-particles-12} for different
wall-distances (in the buffer region, the logarithmic layer and
in the center of the channel); the skewness values are given in
table~\ref{tab-results-pdf-skewness-particles}. 
The curves for the streamwise component are found to largely resemble
Gaussian functions. However, they exhibit discernible non-Gaussian
tails on the negative side which grow with the wall-distance. 
This feature indicates that the
particles respond to some degree to the negative fluid velocity
fluctuations which they experience in the wakes of other particles. 
On the other hand, the pdfs for the wall-normal component are
characterized by much more pronounced non-Gaussian tails. 
Additionally, in the buffer region one can observe a strong skewness
of $v^\prime_c$ towards negative values, i.e.\ high-velocity 
motion (with intensity above four standard deviations) directed
towards the wall is occurring more frequently than away from
it. Since the pdf of wall-normal fluid velocity at the same
wall-distance was not found to exhibit a comparable skewness (cf.\
figure~\ref{fig-results-pdf-fluid-1}$b$, comparison between 
tables~\ref{tab-results-pdf-skewness-fluid} and
\ref{tab-results-pdf-skewness-particles}), the present result suggests 
that the particles respond selectively to certain coherent structures in
the buffer layer. Data accumulated for the wall-normal particle
velocities in case A (not shown) features a less pronounced negative
skewness. 
Unfortunately, no experimental data for the skewness of the
wall-normal particle velocity pdf in channel flow could be obtained. 
Therefore, in order to clarify this point, more in-depth investigations
of the interaction between buffer-layer structures and particles with
different 
values of the Stokes number 
should be undertaken. In particular,
conditional averaging of wall-normal particle motion based upon the
type of the surrounding flow structures \citep[similar to the
point-particle study of][]{marchioli:02}
should provide valuable insight. 
Finally, we observe in figure~\ref{fig-results-pdf-particles-12}$(c)$
that the pdf of the spanwise component of the particle velocity 
(having a skewness near zero by virtue of symmetry) takes a shape very
similar to the one found for the wall-normal component outside the
buffer layer. 
%

In order to analyze the different contributions to the Reynolds
shear stress, we have computed joint probability density functions of
the fluid velocity fluctuations $u_f^\prime$ and $v^\prime_f$. The
data for wall-distances of $y^+=20$ and $74$ is shown in
figures~\ref{fig-results-quaduv-fluid}$(a,b)$; an exhaustive
discussion of the corresponding single-phase case can be found in
\cite{kim:87}.  
Quadrants are defined in the usual manner, such that e.g.\ the second
quadrant corresponds to an ejection of low-speed fluid (a positive
value of $v_f^\prime$ coupled with negative $u_f^\prime$). 
The joint pdfs for the present case B are found to take an elliptical
shape, elongated in a direction which is slightly inclined with respect
to the axis $v^\prime_f=0$. Therefore, the contributions from the
second and the fourth quadrants dominate as in single-phase flow. 
It can be seen in both figures~\ref{fig-results-quaduv-fluid}$(a,b)$
that the most extreme negative fluctuations of $u^\prime_f$ are not
significantly correlated with fluctuations $v^\prime_f$ of any sign,
therefore not contributing to the mean Reynolds shear stress. In the
logarithmic region the joint pdf tends towards a less anisotropic shape.

The fractional contributions from each quadrant to the Reynolds shear
stress are given in
figure~\ref{fig-results-quaduv-fluid-frac-contrib}. In the buffer
layer ($y^+=20$) and in the logarithmic region ($y^+=74$) the total 
contribution from the second quadrant is the strongest, with the
fourth quadrant closely following. The main difference with respect to
the single phase data \citep[as discussed in][]{kim:87} is the
relative importance of the quadrants one and three (i.e.\ those which
correspond to positive Reynolds shear stress). In the particulate
case, the fractional contributions of all four quadrants are
significantly higher than in the case without particles, most probably
due to fluctuations generated in the wake of particles. As
discussed above, the wakes are expected to have the effect of adding
additional contributions of similar magnitude to each quadrant,
therefore not contributing to an increase in the average Reynolds
shear stress. 
The plots in figure~\ref{fig-results-quaduv-fluid-frac-contrib} also
show the fractional contributions from each quadrant when applying a
threshold to the magnitude of the correlation $u^\prime_f
v^\prime_f$. At both wall-distances the contributions from the second
and third quadrants become completely dominant for thresholds above
five times the product of the r.m.s.\ intensities of $u^\prime_f$ and
$v^\prime_f$. This reflects the large value of the negative skewness
of the streamwise velocity pdf, i.e.\ the importance of the wake
structures. 

The joint pdfs for the streamwise and wall-normal components of the
particle velocity fluctuations at two corresponding wall-distances
($y^+=17$ and $73$) are shown in
figure~\ref{fig-results-quaduv-part}. The overall 
shape of the contourlines is found to be similar to the fluid
counterparts in figure~\ref{fig-results-quaduv-fluid}, yet events
with highly negative values of the streamwise component are
absent here. 
The fractional contributions from each quadrant to the correlation
$u_c^\prime v_c^\prime$ (figure omitted) agree with the respective
contributions of the fluid velocity to the Reynolds shear stress. 
However, a faster decay of all quadrants' contribution with increasing
threshold is observed for the particle velocity correlation
$u_c^\prime v_c^\prime$, in accordance with the near-Gaussian shape of
the pdf for the streamwise component (cf.\
figure~\ref{fig-results-pdf-particles-12}$a$).   
%

\section{Lagrangian statistics}
\label{sec-results-lag}
In order to get further insight into the particle response to fluid
motion, we consider statistics which describe the evolution of
the particle velocity along individual particle trajectories. 
The Lagrangian auto-correlation of the particle velocity components
as a function of the temporal separation $\tau$ can be defined as
follows \citep*[similar to][]{ahmed:01}: 
\begin{equation}\label{equ-particles-def-lag-corr}
  {R}_{Lp,\alpha}(\tau)=
  \frac{\langle\langle u_{c,\alpha}^{\prime}(t)\cdot 
    u_{c,\alpha}^{\prime}(t+\tau)\,\rangle_p\rangle_t}
  {\langle\langle(u_{c,\alpha}^\prime(t))^2\rangle_p\rangle_t}
  \,,
\end{equation}
where
$u_{c,\alpha}^{\prime(i)}(t)=u_{c,\alpha}^{(i)}(t)-
\langle u_c\rangle(y_j|x_{c,2}^{(i)}(t)\in I_j)\,\delta_{\alpha 1}$ 
is the instantaneous velocity of the $i$th particle in the $\alpha$
direction, from which the mean streamwise particle velocity 
at the corresponding bin $y_j$ has been subtracted.
By way of this definition, the correlation of the fluctuating particle
motion can be studied while the contribution from the mean particle
drift is eliminated. 
%

Our present results are shown in figure~\ref{fig-results-lag-corr}. 
It can be observed that all Lagrangian two-time correlation functions
of case B exhibit a damped oscillation with a period of approximately
$8h/u_b$ superposed upon the typical decaying curves. For case A this
feature is less pronounced, but it can still be observed with a period
of approximately $4h/u_b$, in particular in the wall-normal and the
spanwise components. 
This phenomenon is believed to be related to the finite streamwise
length of the computational domain. Considering that 
{the difference between the mean velocities of the two phases} 
is approximately equal to $u_b$ across
most of the channel width (cf.\ figure~\ref{fig-results-up}$b$),
particles will encounter the same material fluid elements upon
average after a relative turnover period of
$T_{rel}=L_x/u_b$. Substituting the domain lengths of cases A and B
($L_x=4h$ and $8h$) yields a value for $T_{rel}$ which indeed matches the
observed period separating the successive peaks of the Lagrangian
two-time velocity correlation functions. 
However, these relative maxima of the correlation functions can only be
caused by a significant interaction of particles with structures which
are temporally coherent over time scales comparable to the turnover
period $T_{rel}$.
\cite*{fede:07} have considered the problem of statistical bias caused
by self-correlation of the fluid velocity along the path of
heavy point-particles in homogeneous-isotropic turbulence (by means of 
one-way coupled simulations). They showed that choosing the parameters
such that $T_{rel}\geq4\tau_E$, where $\tau_E$ is the Eulerian
integral time scale of the fluid velocity field, is sufficient in
order to avoid bias of the Lagrangian correlation of the fluid
velocity `seen' by the particles. 
With the purpose of applying this criterion to the present case, the
Eulerian integral time scale can be estimated from the Eulerian
integral length scale and invoking Taylor's hypothesis. The streamwise
integral length scale of the streamwise velocity component is defined
as: 
\begin{equation}\label{equ-results-eul-integral-length}
  L_{f,u}=\int_0^{{L_x}/2}{R_{uu}}(x)\,\mbox{d}x
  \,,
\end{equation}
where $R_{uu}$ is the two-point auto-correlation function of
fluid velocity. In single-phase channel flow, the data of
\cite{kim:87} yields $L_{f,u}=0.87h$ at the centerline ($L_{f,u}=1.2h$
at $y^+=21$). Supposing a convection velocity equal to the local mean
velocity \citep{kim:93c}, 
this leads to an estimate of $\tau_E=0.74h/u_b$ at the center and
$1.5h/u_b$ in the buffer layer. Therefore, based upon the single-phase
data the streamwise domain size $L_x$ in case B verifies the condition
of \cite{fede:07}, while the domain in case A seems to be slightly too
short in order to guarantee unbiased Lagrangian statistics. 
%
%
However, this a priori estimation does not seem to be able to explain
that the bias effect is stronger in the case with the longer domain,
where several repeated peaks up to the fifth one can be observed. 
Since a repeated positive correlation over four relative turnover
periods requires that particles interact with fluid features of
extreme longevity, the data shown in figure~\ref{fig-results-lag-corr}
suggests that there exist some large-scale coherent structures in
the particulate flow which are not found in single-phase turbulence. 
In \S~\ref{sec-results-fluid-structures} below we will indeed identify
flow structures correlated over the whole box-length. 

Let us return to the discussion of the Lagrangian particle velocity
correlations themselves.
In the absence of reference data for channel flow, we will refer to
the results of \cite{ahmed:01} obtained by DNS of homogeneous shear
flow with point-particles. Their most relevant conclusions with
respect to the present work are: 
(a) particle inertia tends to increase the Lagrangian velocity
correlations; 
(b) mean relative velocity (caused by gravity) leads to a decrease of 
the correlations due to the so-called `crossing trajectories effect'
\citep{yudine:59};
(c) the anisotropic structure of the shear flow is reflected in the
particle velocity correlations.

Concerning the results shown in figure~\ref{fig-results-lag-corr}, we
remark that for all velocity components the decay is faster in case B
than in the high-inertia case A, in accordance with the results of
\cite{ahmed:01}. A rough match of the curves for both cases can be
obtained by rescaling the temporal separations in case A with a factor
of $1.5$. 
%
%
%

We further observe from figure~\ref{fig-results-lag-corr} that the
streamwise (vertical) particle velocity component exhibits a much 
longer decorrelation time than the two cross-stream (horizontal)
components, with a first zero-crossing occurring after as much as 
$85$ ($45$) bulk time units in case A (B). The anisotropy is also
evidenced by the integral time scales computed from the Lagrangian
correlation functions, viz.
\begin{equation}\label{equ-results-lag-integral-time}
  T_{Lp,\alpha}=\int_0^\infty{R}_{Lp,\alpha}(\tau)\,\mbox{d}\tau
  \,,
\end{equation}
which are shown in table~\ref{tab-results-lag-integral-scales}. 
%
%
It is well-known that the time scales of fluid structures in channel
flow are strongly anisotropic. \cite*{choi:04} have computed
Lagrangian velocity correlations of fluid elements in the absence of
particles. Their data at the center of the channel suggests a ratio of
decorrelation time scales of roughly $1:0.4:0.5$ for the three fluid
velocity components $u_f:v_f:w_f$.
By comparison, the corresponding correlations for the particle
velocity in the present cases exhibit a much higher anisotropy of the
streamwise component. 
This difference is believed to be related to the modified flow
structure in the particulate case, where we have already observed a
strongly enhanced intensity of the streamwise velocity fluctuations
(cf.\ figure~\ref{fig-results-uu}).  
Unfortunately, no Lagrangian correlation data for fluid elements is
available for the present cases. However, we will provide evidence of
increased Eulerian two-point fluid velocity correlations in
\S~\ref{sec-results-fluid-structures} below. 

Concerning the crossing trajectories effect, the usual measure for
gauging its importance is the ratio between 
{the relative mean velocity} 
and the r.m.s.\ fluid velocity fluctuations in the 
mean slip direction, 
viz.\ $|\langle u_c\rangle-\langle u_f\rangle|/\sqrt{\langle
  u_f^\prime u_f^\prime\rangle}$ \citep{csanady:63}. 
From figures~\ref{fig-results-up}$(b)$ and \ref{fig-results-uu}$(a)$
we deduce that the ratio for both cases ranges between 4 and 5 across
most of the channel width. The relative motion can therefore be judged
as sufficiently strong for the particles to experience the
consequences of their continuous change of the surrounding fluid
environment, which is expected to contribute to a reduction of the
temporal auto-correlation of particle velocities. 
However, in order to isolate the impact of the crossing trajectories
effect upon the correlation functions, results from cases with
different drift velocities and otherwise identical parameters would be
necessary.  

The definition of the Lagrangian velocity correlation discussed so far
does not distinguish the particles
according to their location along the inhomogeneous direction. This
distinction can be accomplished by modifying relation
(\ref{equ-particles-def-lag-corr}) such that the sums are performed
only over those particles which are located within a wall-normal
interval $I_j$ at the initial time $t$. In doing so, one obtains
correlation functions which turn out to have a similar shape to the
globally averaged functions shown in
figure~\ref{fig-results-lag-corr}, but where the decay depends upon
the particle's initial wall-distance.  
Figure~\ref{fig-results-lag-corr-integral-scale-y} shows the
dependency between the integral time scale and the initial particle
coordinate $y$ in case B. The Lagrangian integral time scales
associated with the streamwise and wall-normal particle velocity can
be seen to increase progressively with the wall-distance; the spanwise
component exhibits a similar behavior up to $\bar{y}\approx0.8$, then
it decreases again towards the center of the channel. 
The trend of faster decorrelation near the wall is consistent with the
fact that the length scales of the dominant turbulent structures in
channel flow increase with the wall-distance. 
\cite{choi:04} have shown for single-phase flow that this spatial
inhomogeneity leads to a faster decay of the velocity correlations
of fluid elements when they are released from smaller wall-distances. 
It is believed that a similar reasoning applies to our solid
particles, with smaller fluid scales near the wall leading to a
faster decorrelation of the particle velocity in that region.  

\section{Spatial structure of the dispersed phase}
\label{sec-results-particle-structure}
In this section we are concerned with the possible formation of
instantaneous particle agglomerations in the flow. As discussed in the
introduction, at least two mechanisms for clustering appear to be
available: interaction with turbulent flow structures (i.e.\
`preferential concentration') and the particle-wake effect.
In order to determine whether agglomerations play a significant role, 
an analysis of inter-particle distances and local number densities is
required. For this purpose we have applied four different diagnostic
methods to our data: 
the average distance to the nearest neighbor, 
cluster detection, 
the probability of local particle concentration and  
the particle-pair correlation function. 

Let us define the average distance to the nearest particle as follows:
\begin{equation}\label{equ-particles-channelp-res-dmin}
  {d}_{min}=\frac{1}{N_p}\sum_{i=1}^{N_p}\min_{\stackrel{j=1}{j\neq
      i}}^{N_p}(d_{i,j})\,,
\end{equation}
where $d_{i,j}=|\mathbf{x}_c^{(i)}-\mathbf{x}_c^{(j)}|$ is the
distance between the centers of particles $i$ and $j$. 
The time evolution of the quantity 
$d_{min}$, normalized by its value for a homogeneous distribution with
the same solid volume fraction, $d_{min}^{hom}=(||\Omega||/N_p)^{1/3}$,
has been plotted in figure~\ref{fig-results-dist}. 
Here the ratio $d_{min}/d_{min}^{hom}$ has a
lower bound of $0.2$ (corresponding to each particle being in contact
with at least one other particle) and an upper bound of unity
(homogeneous distribution). 
Actual values recorded during our simulations vary mildly between
$0.5$ and $0.6$, which is in fact very close to the value of a random
distribution (numerically determined as $0.5621$). 
It is interesting to note that direct simulations of pure
sedimentation in triply periodic domains also yield values for the
mean minimum inter-particle distance which (when normalized with
$d_{min}^{hom}$) fluctuate around $0.55$ \citep{kajishima:04}.  

An alternative way of characterizing the spatial structure of the
dispersed phase is by searching directly for the presence of 
particle clusters. Here we define a cluster 
as a set of particles of which each member is within a distance 
$l_c$ of at least one other member \citep{wylie:00}. 
Since cluster detection is a computationally expensive task, it is
necessary to resort to a fast numerical algorithm. 
We have implemented the technique suggested by \cite{melheim:05}
which was applied to 400 (50) instantaneous particle distributions of
case A (B), from which the probability of the occurrence of clusters
with different numbers of members $n_c$ was determined for each case.
Figure~\ref{fig-results-clusters}$(a,b)$ shows our results for two
different cut-off lengths $l_c=2.5D$ and $4D$. It can be observed that
the pdf for the occurrence of clusters is a rapidly decaying function
of the number of its members, and that the curve flattens as expected
when increasing the cut-off length. Both DNS cases yield fairly
similar results.
The plots also include the corresponding pdf for a random particle
distribution with the same solid volume fraction (numerically
determined from 100 realizations), which is found to collapse with the
DNS data. 

Since particles can interact over long distances via their wakes, the
cluster definition was then modified in order to take the non-isotropic
structure of the wakes into account. Instead of using a spherical
cut-off radius, we define clusters through an ellipsoidal definition. 
More specifically, two particles $i$ and $j$ are considered as members
of a cluster if their positions obey
\begin{equation}\label{equ-def-ellipsoid}
  \sum_{\alpha=1}^3\frac{\left(x_{c,\alpha}^{(i)}-x_{c,\alpha}^{(j)}\right)^2}
  {l_{c,\alpha}^2}
  \leq1\,,
\end{equation}
for a given cut-off axis vector with components $l_{c,\alpha}$. 
Figure~\ref{fig-results-clusters}$(c)$ shows the pdf of the occurrence
of clusters using a streamwise-elongated elliptical definition with a
$3:1:1$ axis ratio.
It is found that the probability for large clusters indeed increases
with respect to a spherical definition with the radius corresponding
to the minor axis length. 
However, the random particle distribution again exhibits the same pdf
as the DNS data to within statistical uncertainty. 

Several authors have used the statistics of local particle
concentration as a measure of the tendency for preferential
concentration \citep*{squires:91,wang:93,fessler:94,aliseda:02}. 
The concentration is determined by subdividing the flow domain into 
smaller boxes (with linear dimensions $H_x,H_y,H_z$),
counting the number of particles in each box, $n_{p,box}$, and then
dividing by the box volume, i.e.\ $C=n_{p,box}/(H_xH_yH_z)$. 
This process is repeated for each field, and finally statistics
are computed over the homogeneous directions and the number of snapshots. 
In the present case, the width of all boxes in the wall-normal
direction is taken as one particle diameter ($H_y=D$), therefore
allowing for a distinction of the results with respect to the
wall-distance; the box-size in the two wall-parallel directions was
varied between $H_x=H_z=5D$ and $40D$. 
Figure~\ref{fig-results-local-concentr}$(a)$ shows the pdf of the local
particle concentration in case B at a wall-distance of $y^+=28$ for a
cross-stream (horizontal) box-dimension of $H_x=H_z=20D$. The
concentration $C$ is normalized by the average concentration at the
same wall-distance, $C_0$. 
It can be seen that the DNS data agrees very well with a
Poisson distribution which characterizes the probability of a random
process \citep{squires:91}. The difference between the pdf from the
DNS data and the Poisson distribution can be quantified by summing the
square of the difference between both probabilities \citep{wang:93}, viz. 
\begin{equation}\label{equ-results-disp-diff-pdf}
  D_2=\sum_{n=1}^{N_p}\left(P(n)-P_{Poisson}(n)\right)^2\,,
\end{equation}
where $P(n)$ is the probability of finding $n$ particles in a given
box. 
The quantity $D_2$ is shown in
figure~\ref{fig-results-local-concentr}$(b)$ as a function of the 
linear box-dimension and for different wall-distances. We observe that
the difference always remains below $10^{-3}$.
Again, this definition can be extended to the non-isotropic case by
defining elongated boxes (i.e. $H_x>H_z$). The results (which are not
shown here) lead to the same conclusion: the local concentration
closely follows a Poisson distribution.

As an additional measure of the spatial structure of the dispersed phase, we
have determined the pairwise particle distribution function. This
quantity corresponds to the probability of finding a second particle
at a certain distance from a given particle. In order to investigate
the two homogeneous directions separately, we have defined a directional (as
opposed to radial) distribution function for streamwise and spanwise
separations. 
In both directions, the curves are found to be
essentially constants (the figures are omitted), indicating that all 
possible inter-particle distances are roughly of equal probability,
irrespective of the wall-distance. 
%

Finally we can conclude 
from the different diagnostic techniques employed here 
that the instantaneous spatial distribution of
particles in the present cases is not significantly different from the
distribution generated by a random process. This point will be further
discussed in \S~\ref{sec-conclusion}.

\section{Coherent flow structures}
\label{sec-results-fluid-structures}
We have come across indications of a significantly modified turbulence
structure with respect to single-phase channel flow at several points
in the previous discussion. Here we attempt to identify the coherent
flow structures which are responsible for the observed differences in
the various statistical quantities. 

Figure~\ref{fig-results-5-fluid-snapshot} shows 
isosurfaces of the fluctuations of the streamwise velocity component
for an instantaneous flow field in case B. The isovalues
($u^\prime=\pm3.6u_\tau$) are chosen such that in the corresponding
single-phase flow almost exclusively buffer-layer streaks are
visualized. In the present case, however, the picture is fundamentally 
different.  
First, we recognize in figure~\ref{fig-results-5-fluid-snapshot}$(b,d)$
the wakes trailing individual particles as streamwise elongated surfaces
with negative fluctuation values (due to fluid upflow and negative
particle buoyancy). 
These wakes can be distinguished more clearly in
figure~\ref{fig-results-5-fluid-snapshot-zoom} where a zoom 
into the graphs of  
figure~\ref{fig-results-5-fluid-snapshot}$(b,d)$ is provided.
At the present particle Reynolds number, particle
wakes have a characteristic streamwise extension of less than one
channel half width and a cross-stream extension comparable to the
particle diameter. 
In addition, we observe in figure~\ref{fig-results-5-fluid-snapshot}
very large coherent structures with streamwise dimensions of the order
of the length of the current domain ($8h$) and cross-stream scales
comparable to the channel half-width. Very large structures with both 
signs of the isovalue for $u^\prime$ are found, the
negative-valued structures being more difficult to discern due to the
cluttering of the image by the large number of particle wakes. 
Visualizations of successive snapshots reveal that these flow
structures evolve on a relatively large time scale. An example of such
a temporal sequence is shown in
figure~\ref{fig-results-5-fluid-sequence}, where we can identify
structures over an interval of approximately $45$ bulk time units
while they slowly revolve around each other in the cross-stream
plane. 
This temporal coherence of the large-scale structures is at the origin
of the slow convergence of the Eulerian statistics in the present
case. The lack of symmetry of the streamwise momentum balance with
respect to the midplane 
\revision{(observed in
\S~\ref{sec-results-stats-def},
figure~\ref{fig-results-shear-stress}$a,b$)}
{(observed in \S~\ref{sec-results-stats-def},
figure~\ref{fig-results-shear-stress}$a,b$  as well as
figure~\ref{fig-app-fluid-only}$a$ of appendix~\ref{app-fluidonly})} 
is due to the marginal sampling of these large scales over the present
observation interval.  
Furthermore, the presence of flow structures with a lifetime of tens
of bulk time units explains the repeated peaks of the Lagrangian
particle velocity auto-correlation function discussed in
\S~\ref{sec-results-lag} (cf.\ figure~\ref{fig-results-lag-corr}). 
Finally, it should be remarked that both particle wakes and the very
large structures contribute to the strong increase in the streamwise
r.m.s.\ fluid velocity, leading to the overall turbulence enhancement
(cf.\ figure~\ref{fig-results-uu}). 

In order to investigate whether the particles respond to the
large-scale perturbations of the streamwise fluid velocity, we have
visualized the instantaneous particle motion. The particles have been
classified according to their streamwise velocity fluctuation: those
with a high positive fluctuation value ($u_c^\prime\geq0.2u_b$),
highly negative fluctuation ($u_c^\prime\leq-0.2u_b$) and small
fluctuation ($-0.2u_b<u_c^\prime<0.2u_b$). The instantaneous locations
of the three classes of particles are shown in
figures~\ref{fig-results-5-particle-snapshot} and
\ref{fig-results-5-particle-snapshot-neutral} at the same time as
the flow field in figure~\ref{fig-results-5-fluid-snapshot}. 
The high-speed particles of both signs are clearly found in regions
with a streamwise elongated shape, whereas the particles with
streamwise velocity fluctuation amplitudes below the threshold are
more or less equally distributed everywhere else. Comparing the
particle distribution with the locations of the fluid structures
(figure~\ref{fig-results-5-particle-snapshot} versus
\ref{fig-results-5-fluid-snapshot}), a strong spatial correlation
between high-intensity velocity fluctuations of the two phases is
indeed observed. It has been confirmed by visualizing time-series that
the spatial distribution of particles with intense streamwise velocity
fluctuations exhibits the same high temporal coherence as the fluid
counterpart. 
In particular, animations show that the large-scale organization
of the flow starts to be visible approximately $12$ bulk time
units after introducing the particles into the flow. Subsequently, it
remains a prominent feature over the entire observation
interval.  
%

In order to further confirm the statistical significance of the
observed instantaneous flow structures we have computed 
Eulerian two-point correlations of fluid velocity (taking into account
only points in the actual fluid domain) from $12$ instantaneous flow
fields.  The data for case B is shown in 
figure~\ref{fig-results-5-autocorrelations-dx} (for streamwise
separations) and figure~\ref{fig-results-5-autocorrelations-dz}
(spanwise separations). 
For all components and in both spatial directions the decay at
small separations is found to be enhanced in the present case as
compared to the single-phase reference data. 
This result reflects the presence of smaller scales in the vicinity
of the particles' surfaces as well as in their wakes.
For larger separations the correlation functions of the particulate
case and the single-phase flow cross over.
The most pronounced features are the finite correlation values of the
streamwise velocity component for the largest separations. These
correlation values are positive (negative) for large streamwise
(spanwise) separations. 
This result is fully consistent with the presence of the observed
large-scale structures. 
As a further consequence of the modified flow structure, all 
streamwise correlation amplitudes increase with the wall-distance
contrary to what is found in single phase flow 
(figure~\ref{fig-results-5-autocorrelations-dx}). 
It is also remarkable here that for intermediate spanwise separations in
the range of $50\leq r_z^+\leq200$ the correlation functions for the
streamwise and spanwise velocity components do not exhibit the
characteristic negative minimum which is directly linked to the
average spacing of low- and high-speed buffer-layer streaks
\citep{kim:87}. We still observe a weaker local minimum of the
wall-normal component at the smallest wall-distance
(figure~\ref{fig-results-5-autocorrelations-dz}$b$). 
The evidence from the fluid velocity correlation functions shows
that the structures which we have observed instantaneously have a
large significance for the flow statistics, masking in part even the
signatures of the usual near-wall turbulence structures. 

The equivalent to the Eulerian two-point velocity correlation for the
dispersed phase is the so-called particle pair velocity correlation
\cite*[cf.][]{reeks:06}.
It was confirmed that the spatial decay of the particle pair velocity
correlations (figures omitted) exhibits the same characteristic
features as the counterparts for the fluid velocity. 
Thereby we have statistically substantiated 
the correspondence between particle motion and large-scale fluid
structures observed in instantaneous flow fields.

\section{Summary and discussion}
\label{sec-conclusion}
We have conducted a DNS study of turbulent particulate
channel flow in a doubly-periodic domain, 
considering finite-size rigid particles with numerically resolved
phase interfaces. 
The simulations were performed in the dilute regime,
allowing for an approximate treatment of direct particle encounters. 
In our simulations we have considered spherically-shaped heavy particles
whose buoyancy was adjusted such that the terminal velocity matches the
bulk flow velocity. 
The fluid was driven upwards by a mean pressure gradient, thereby
fluidizing particles in the center of the channel while particles near
the walls moved downwards upon average. 
For the present bulk Reynolds number of $2700$ and
particle diameter of $1/20$ of the channel half-width (corresponding to 
approximately $11$ wall units), this leads to a terminal particle
Reynolds number of $136$. 
Two different density ratios were considered, differing by a factor of
$4.5$. The corresponding Stokes numbers of the two types of particles
were 
${\cal O}(10)$ in the near-wall region and ${\cal O}(1)$ in the outer
flow. 
The case with lower particle inertia was run in a computational domain
with a length of $8$ and $4$ channel half-widths in the two
wall-parallel directions, involving $4096$ particles and integration
times of ${\cal O}(100)$ bulk flow time units.  

We have presented statistical data for the Eulerian one-point
moments of the velocities of both phases, velocity probability density
functions, Lagrangian particle velocity correlations, various measures
for the spatial structure of the dispersed phase, flow visualizations
and Eulerian two-point correlations.
The main results can be summarized as follows. 
\begin{enumerate}
\item The presence of particles strongly affects the carrier flow. 
  On the one hand the mean flow profile tends towards a concave shape
  due to the interplay between the modified Reynolds shear stress
  profile and the mean force exerted by the particles upon the
  fluid. The latter contribution is directly linked to the
  non-homogeneity of the mean solid volume fraction which has the
  effect of bringing more streamwise momentum towards the wall. 
  On the other hand the intensity of turbulence is strongly enhanced
  with respect to the single-phase flow at the same bulk Reynolds
  number. This enhancement is mostly due to an increase in the
  streamwise velocity fluctuations, whereby the anisotropy of the
  Reynolds stress tensor is considerably modified. 
\item At the chosen parameter values there is no significant
  clustering of particles. 
  By `clustering' we understand deviations from the local mean solid 
  volume fraction beyond the extent that can be expected from a random
  process.  
\item The two dominating flow features are particle wakes and 
  very large jet-like structures.
  The wakes constitute regions with low streamwise velocity (due to
  mean upflow and negative particle buoyancy), causing a marked 
  negative skewness of the corresponding pdf. 
  The large-scale structures appear as regions with positive or
  negative streamwise velocity fluctuations extending over eight channel
  half widths (the entire domain length) in the streamwise direction
  and approximately one half width in the cross-stream directions. 
  As a consequence, the Eulerian two-point correlation of the
  streamwise fluid velocity exhibits finite values at large streamwise
  and spanwise separations.  
  Furthermore, it was observed that the particle motion is strongly
  affected by these large-scale flow structures, whose signatures were 
  found in Lagrangian particle velocity correlations and Eulerian
  particle pair velocity correlations. 
\item The differences between the two flow cases were relatively
  small. 
  The fact that the high-inertia case was computed in a smaller
  computational domain might account partly for the observed
  differences. 
  An exception are the Lagrangian particle velocity correlations which 
  clearly exhibit longer decorrelation times in the high-inertia
  case. 
\end{enumerate}
The fact that no particle agglomerations were found while 
the formation of new large-scale flow structures was detected merits
further discussion.  
With respect to particle agglomerations it must be concluded that the
two known mechanisms for their formation are not effective in the
present context. 
For the preferential accumulation effect to generate significant
segregation of particles by centrifugal ejection from vortical
structures, the particles need to be small compared to the dominant
(i.e.\ high vorticity) flow structures. 
The particles studied here are probably too large 
in order to accumulate outside of the high-vorticity regions. 
This is particularly true in the near-wall region, 
where the present particles' size is comparable to
the typical cross-stream dimensions of the streamwise vortices. 

Concerning the wake attraction effect, i.e.\ the formation of clusters
due to the decreased drag experienced by trailing particles, 
it can be inferred from our results that the particle Reynolds number
in the present study is probably too low for this mechanism to be
of importance. 
In the pure sedimentation case \citep{kajishima:04} 
particle clusters start to appear at a particle Reynolds number of
$200$; 
in the case with grid-generated turbulent background flow
\citep{nishino:04} trains of particles were found for particle
Reynolds numbers of $140$. 
It should be noted that in both of those studies the suspension was
even more dilute than here. 
In turbulent channel flow wake-induced clusters have to our
knowledge not been reported in the literature. Therefore, it would be 
interesting to perform future simulations with higher slip velocities
in order to establish whether the wake-attraction mechanism is able to
actuate in channel flow and to determine a critical Reynolds
number.  

Even in the absence of particle agglomerations (i.e.\ significant
perturbations of the solid volume fraction) the flow is found to
exhibit large-scale organization in the form of streak-like velocity
perturbations.  
It should be emphasized that this feature is clearly particle-induced
since no corresponding flow structures exist in single-phase channel
flow.  
Since these large-scale flow structures were found to have a
substantial effect on various statistical quantities (two-point
velocity correlations, Lagrangian particle velocity correlations,
r.m.s.\ fluid velocity fluctuations) the question about their origin
is of great interest. 
It is plausible to believe that a mechanism related to hydrodynamic
instability is responsible for the formation of the large-scale flow 
structures. In this scenario it can be imagined that the addition of
particles to the turbulent channel flow triggers an intrinsic
instability eventually leading to the observed velocity perturbations. 
However, at the present time there is no confirmation of the existence
of such an instability mechanism. 
%
In this context it should be remarked that \cite*{matas:03} found 
a decrease of the critical Reynolds number when adding
neutrally-buoyant particles above a certain size  
($D\geq d_{pipe}/65$) to laminar pipe flow. 
These authors speculated that the lowered transition threshold is due
to the particle-induced velocity fluctuations triggering subcritical
transition. 
The influence of the presence of dispersed particles upon the
stability characteristics of pipe or channel flow deserves further
investigation in the future. 

One further question which naturally arises in relation with the
emergence of flow scales of the order of the streamwise period of the
computational domain is whether the finite box size might play a
decisive role. In particular, the typical streamwise size of these
structures cannot be precisely determined if no decorrelation within
the fundamental period takes place. 
It should be mentioned that these largest scales are not clearly
observed in the case with the smaller domain (case A). On the other
hand, computing the flow in much larger domains than the one
used in case B is a challenging task, since those simulations 
are already on the limit of what can be afforded with the present
algorithm on the largest available hardware. 
Nevertheless, we have recently initiated a companion simulation with
twice the streamwise period of case B and otherwise identical
conditions, corresponding to a Eulerian grid with more than
$2.6\cdot10^{9}$ nodes. 
The results of this extended simulation---when available---will
hopefully shed light on the scaling of the large structures. 
\section*{Acknowledgments}
The author is grateful to A.\ Pinelli for helpful comments
contributing to the improvement of an earlier version of this
manuscript.  
This work was supported by the Spanish Ministry of Education and
Science under the `Ram\'on y Cajal' program (contract
DPI-2002-040550-C07-04) and through grant ENE2005-09190-C04-04/CON. 
The computer resources, technical expertise and assistance provided by
the Barcelona Supercomputing Center - Centro Nacional de
Supercomputaci\'on are thankfully acknowledged. 
\begin{appendix}
\section{The effect of evaluating Eulerian statistics over the
  composite flow field}
\label{app-fluidonly}
In order to assess the effect of computing the Eulerian one-point
statistics of the carrier phase over the composite flow field
(including solid and fluid nodes), we have applied this procedure to
{12 instantaneous flow fields of case B which were taken over an
  interval of approximately $55$ bulk time units.} 
For the purpose of comparison,
we have computed the corresponding quantities by averaging only
over nodes which are located in the fluid domain, using the same set
of flow fields. This `fluid-only' average shall be denoted by 
$\langle\cdot\rangle_f$. 
Both results are compared in figure~\ref{fig-app-fluid-only}. It can
be seen that a significant difference is only obtained for the
streamwise normal stress component, whereas the other stress
components as well the mean velocity are not noticeably affected. The
maximum normalized difference between the results of the two methods,
as defined by  
\begin{equation}\label{equ-app-fluidonly-def-diff}
  {\cal E}(\langle u\rangle)=\frac{max_y|
{\langle u\rangle_f}-{\langle u\rangle}|}
  {max_y{\langle u\rangle_f}}
  \,,
\end{equation}
(and similarly for the other quantities) is given in
table~\ref{tab-app-fluidonly-diff}. It can be seen that the difference
has a maximum of 6.7\% in the case of the streamwise velocity fluctuation,
while it is significantly smaller for all other quantities. In
particular, the difference for the mean velocity is below 0.7\%.
The larger difference for the quantities related to the streamwise
velocity components is due to the fact that the virtual fluid velocity
inside the heavy particles in upward flow represents principally a
low-velocity perturbation with respect to the streamwise velocity of the
surrounding fluid. Therefore, averaging over the composite flow field
yields an underestimation of the mean velocity and an overestimation
of the r.m.s.\ value of the streamwise velocity fluctuations. 
Figure~\ref{fig-app-fluid-only-diff} shows the difference between the
two methods of computing these two moments of the streamwise velocity,
plotted as a function of the wall-normal coordinate. It can be seen
that the difference is directly proportional to the local solid volume
fraction (computed from the corresponding 12 particle fields) with a
coefficient of ${\cal O}(1)$. 
{In particular, the least-squares fit of the data yields:
  \begin{subequations}\label{equ-app-fluid-fit-diff-phis}
    \begin{eqnarray}\label{equ-app-fluid-fit-diff-phis-um}
      (\langle u\rangle_f-\langle
      u\rangle)/u_b&\approx&-1.4682\,\langle\phi_s\rangle\\
      \label{equ-app-fluid-fit-diff-phis-uu}
      (\langle u^\prime u^\prime\rangle_f-\langle
      u^\prime
      u^\prime\rangle)/u_b^2&\approx&1.3634\,\langle\phi_s\rangle\,,
    \end{eqnarray}
  \end{subequations}
  with standard deviations of $\sigma_u=2.1966\cdot10^{-4}$ and 
  $\sigma_{uu}=2.7847\cdot10^{-4}$, respectively.}  

The result of the preceding paragraph is consistent with the
following estimation. Let us assume that particles maintain a velocity
equal to their terminal settling velocity with respect to the 
surrounding fluid velocity at all times and particle rotation can be
neglected, i.e.\ the streamwise component of the virtual fluid
velocity inside the solid regions can be expressed as
$u(\mathbf{x}\in{\cal S})\approx\langle u\rangle_f-|u_{c,\infty}|$; 
fluctuations of the solid volume fraction are also neglected.
Then it can be shown that for the mean velocity: $\langle u\rangle-\langle
u\rangle_f=-|u_{c,\infty}|\langle\phi_s\rangle$. Likewise, for the streamwise
component of the stress tensor: $\langle u^\prime u^\prime\rangle-
\langle u^\prime u^\prime\rangle_f=-\langle u^\prime
u^\prime\rangle_f\langle\phi_s\rangle
+|u_{c,\infty}|^2(\langle\phi_s\rangle-\langle\phi_s\rangle^2)$; 
since we have imposed $|u_{c,\infty}|=u_b$ (and the following two
relations hold: $\langle\phi_s\rangle\ll1$ and $\langle u^\prime
u^\prime\rangle_f\ll u_b$), the dominant term is
$|u_{c,\infty}|^2\langle\phi_s\rangle$. This suggests that for both
quantities the 
difference scales with the local solid volume fraction, as was indeed
observed in figure~\ref{fig-app-fluid-only-diff}.

From this comparison we can conclude that the statistics accumulated
at run-time of the simulation, while not distinguishing between fluid
and solid locations, overestimate the r.m.s.\ value of the streamwise
velocity fluctuations by up to 6.7\%, while affecting the other
fluctuating components much less and to an even lesser extent the
mean velocity. 
{The excellent fit of the present data with a linear
  function of the mean solid volume fraction enables us to correct the
  statistics of $\langle u^\prime u^\prime\rangle$ which were
  accumulated at run-time.} 
  \section{The mean streamwise momentum balance}
\label{app-mombal}
Since the driving pressure gradient is not fixed in our simulations,
we need to determine it first. For this purpose 
we consider the streamwise component of the momentum equation,
\begin{eqnarray}\label{equ-n-s-mom}
\partial_t\mathbf{u}+(\mathbf{u}\cdot\nabla)\mathbf{u}+\nabla
p&=&\nu\nabla^2\mathbf{u}+\mathbf{f}
\end{eqnarray}
where $\mathbf{u}$ is the vector of fluid velocities, $p$ the pressure
normalized with the fluid density ($\rho_f$), $\nu$ the kinematic
viscosity and $\mathbf{f}$ the volume force term which serves
to impose the rigid body motion upon the fluid at the location
of the particles. 
Integrating (\ref{equ-n-s-mom}) over the whole spatial domain $\Omega$
and a sufficiently long temporal interval $T$ to allow for a
statistically steady state where the velocity field only depends upon
the wall-normal coordinate, viz. 
\begin{equation}\label{equ-particles-channelp-x-mom-int-over-all-omega}
  \frac{\partial\langle p\rangle}{\partial x}2h
  +2u_\tau^2
  =\frac{1}{TL_xL_z}\int_{t_0}^{t_0+T}\int_\Omega f_x
  \,\mbox{d}\mathbf{x}\,\mbox{d}t 
  =\frac{1}{TL_xL_z}\int_{t_0}^{t_0+T}
  \sum_{i=1}^{N_p}\int_{x\in{\cal S}^{(i)}}
  f_x
  \,\mbox{d}\mathbf{x}\,\mbox{d}t 
  \,,
\end{equation}
where the last integral extends over the subvolume ${\cal
  S}^{(i)}$ occupied by the $i$th particle. 
The force integral 
can be eliminated by making use
of the Newton equation for the linear particle motion \citep[][equ.\
13]{uhlmann:04}, which states:
\begin{equation}\label{equ-particles-newton}
  V_c\left(\frac{\rho_p}{\rho_f}-1\right)\dot{\mathbf{u}}_c^{(m)}
  =-\int_{x\in{\cal S}^{(i)}}\mathbf{f}\,\mbox{d}\mathbf{x}
  +\left(\frac{\rho_p}{\rho_f}-1\right)V_c\mathbf{g}
  \,,
\end{equation}
where $\mathbf{u}_c^{(m)}$ is the vector of the $m$th particle's
translational velocity. 
Substituting the streamwise component of (\ref{equ-particles-newton})
into (\ref{equ-particles-channelp-x-mom-int-over-all-omega}) and
making use of the fact that the flow is statistically stationary
(i.e.\ the average particle acceleration vanishes) yields the
mean streamwise pressure gradient:
\begin{equation}\label{equ-particles-newton-dpdx}
  \frac{\partial\langle p\rangle}{\partial x}
  =
  -\frac{u_\tau^2}{h}+\Phi_s\left(\frac{\rho_p}{\rho_f}-1\right)g_x
  \,.
\end{equation}
In order to derive the $y$-dependent streamwise momentum balance, we
now integrate equation (\ref{equ-n-s-mom}) over wall-parallel planes
and time:
\begin{equation}\label{equ-particles-channelp-x-mom-int}
  \frac{\mbox{d}\langle{u^{\prime}v^{\prime}}\rangle}{\mbox{d}y}
  -\frac{u_\tau^2}{h}+\Phi_s\left(\frac{\rho_p}{\rho_f}-1\right)g_x
  =\nu\,\frac{\mbox{d}^2\langle u\rangle}{\mbox{d}y^2}+\langle f_x\rangle
  \,,
\end{equation}
where the Reynolds decomposition $\mathbf{u}=(\langle
u\rangle,0,0)+\mathbf{u}^\prime$ has been substituted. 
Equation (\ref{equ-particles-channelp-x-mom-int}) could be used for
evaluating the convergence of the statistics. However, the local
average force density $\langle f_x\rangle$ at the Eulerian grid nodes
is not directly available in our data-sets. 
Alternatively, the quantity $\langle f_x\rangle$ can be
expressed by averaging the equation for the linear particle
acceleration (\ref{equ-particles-newton}) over finite wall-normal
bins, similar to the operation defined in
(\ref{equ-def-particle-bin-average}).  
This procedure amounts to associating the sum of the forces applied
to each particle with the particle's center location, i.e.\ the force
density is computed as follows:
\begin{equation}\label{equ-particles-channelp-force-int-slab}
  \langle f_x\rangle(y_j)
  \approx
  \frac{1}{n_{obs}L_xL_z|I_{yj}|}
  \sum_{l=1}^{n_{obs}}
    \mathop{\sum_{i=1}}_{y_{c}^{(i)}(t_l)\in I_{yj}}^{N_p}
    \int_{x\in{\cal S}^{(i)}}
    f_x(t_l)\mbox{d}\mathbf{x}
    \,.
\end{equation}
The same operator applied to the equation of particle acceleration
(\ref{equ-particles-newton}), assuming statistical stationarity,
yields: 
\begin{equation}\label{equ-particles-channelp-force-int-partacc}
  \langle f_x\rangle(y_j)=\left(\frac{\rho_p}{\rho_f}-1\right)g_x\,
  \langle\phi_s\rangle(y_j)
  \,.
\end{equation}
Figure~\ref{fig-results-converge-newton} shows that the equality
(\ref{equ-particles-channelp-force-int-partacc}) 
is indeed verified across the entire channel by the data accumulated
during our simulations.  
%
Finally, we substitute
(\ref{equ-particles-channelp-force-int-partacc}) into
(\ref{equ-particles-channelp-x-mom-int}) to obtain the desired
streamwise momentum balance equation
(\ref{equ-particles-channelp-total-shear-int}) given in the main text
(cf.\ \S~\ref{sec-results-stats-def}).   

\end{appendix}

\clearpage
\begin{table}
  \centering
  \setlength{\tabcolsep}{5pt}
  \begin{tabular}{*{7}{c}}
    case&$\frac{\rho_p}{\rho_f}$&$|\mathbf{g}|h/u_b^2$&$St^+$&$St_{b}$&
    $N_p$&$\Omega$\\[1ex] 
    A&$10$&$1.625$&$67$&$3.75$&$512$&$4h\!\times\!2h\!\times\! h$\\
    B&$2.21$&$12.108$&$15$&$0.83$&$4096$&$8h\!\times\!2h\!\times\!4h$\\
  \end{tabular}
  \caption{Physical parameters used in the simulation of
    particulate flow in a vertical plane channel.
    In all cases the particle diameter is chosen as $D=h/20$, the
    global solid volume fraction is set to $\Phi_s=0.0042$.
  }    
  \label{tab-particles-channelp-params-phys}
\end{table}
\begin{table}
  \begin{center}
    \begin{tabular}{*{5}{c}}
      case&$N_x\times N_y\times N_z$&
      $\Delta tu_b/\Delta x$&$t_{obs}u_b/h$\\[1ex] 
      A&
      $1024\times513\times256$&$0.5761$&$700$\\
      B&
      $2048\times513\times1024$&$0.4148$&$105$\\ 
    \end{tabular}
  \end{center}
  \caption{Numerical parameters employed in the simulations. $N_i$ is the
    number of grid nodes in the $i$th direction, 
    $t_{obs}$ is the observation interval after
    discarding the initial transient.  
    The grid spacing in all cases is fixed at $\Delta x=h/256$,
    corresponding to $N_L=515$ Lagrangian force points per particle.
  }    
  \label{tab-particles-channelp-params-num}
\end{table}
\begin{table}
  \begin{center}
    \begin{tabular}{*{7}{r}}
      &\multicolumn{3}{c}{present results, case B}&
      \multicolumn{3}{c}{single-phase \citep{moser:99}}\\
      &$S(u_f^\prime)$&$S(v_f^\prime)$&$S(w_f^\prime)$&
      $S(u_f^\prime)$&$S(v_f^\prime)$&$S(w_f^\prime)$\\[1ex] 
      $y^+=20$&
      $-0.948$&$-0.135$&$0.019$&
      $-0.323$&$-0.204$&$-0.006$\\
      $\bar{y}=1$&
      $-1.289$&$0.029$&$0.031$&
      $-0.629$&$0.030$&$-0.046$
    \end{tabular}
  \end{center}
  \caption{Skewness of the pdfs of fluid velocity components at two
    different wall distances, compared to single-phase reference data
    from \cite{moser:99}.}     
  \label{tab-results-pdf-skewness-fluid}
\end{table}
\begin{table}
  \begin{center}
    \begin{tabular}{*{4}{r}}
      &$S(u_c^\prime)$&$S(v_c^\prime)$&$S(w_c^\prime)$\\[1ex] 
      $y^+=17$&
      $0.089$&$-1.221$&$0.009$\\
      $y^+=73$&
      $0.038$&$0.003$&$0.022$\\
      $y^+=220$&
      $0.025$&$0.014$&$0.032$
    \end{tabular}
  \end{center}
  \caption{Skewness of the pdfs of particle velocity components in
    case B at different wall distances.}     
  \label{tab-results-pdf-skewness-particles}
\end{table}
\begin{table}
  \begin{center}
    \begin{tabular}{*{4}{r}}
      case&$T_{Lp,1}u_b/h$&$T_{Lp,2}u_b/h$&$T_{Lp,3}u_b/h$\\[1ex]
      A&$13.58$&$0.86$&$2.30$\\
      B&$8.15$&$0.48$&$0.79$
    \end{tabular}
  \end{center}
  \caption{Integral time scales (in bulk units) computed from the
    Lagrangian particle velocity auto-correlations, as defined in
    equation (\ref{equ-results-lag-integral-time}).}      
  \label{tab-results-lag-integral-scales}
\end{table}
\begin{table}
  \begin{center}
    \begin{tabular}{*{5}{c}}
      ${\cal E}(\langle u\rangle)$&
      ${\cal E}(\sqrt{\langle u^\prime u^\prime \rangle})$&
      ${\cal E}(\sqrt{\langle v^\prime v^\prime \rangle})$&
      ${\cal E}(\sqrt{\langle w^\prime w^\prime \rangle})$&
      ${\cal E}(\langle u^\prime v^\prime \rangle)$
      \\[1ex] 
      $0.0069$&$0.0672$&$0.0213$&$0.0070$&$0.0200$
    \end{tabular}
  \end{center}
  \caption{The maximum relative difference between the computation of the
    Eulerian statistics by averaging over the composite flow field and
    averaging over the fluid nodes only, estimated from 12
    instantaneous flow fields. The definition of the 'error' is given
    in (\ref{equ-app-fluidonly-def-diff}).}    
  \label{tab-app-fluidonly-diff}
\end{table}

\clearpage
{\begin{figure}
    \centering
    \begin{minipage}{13.ex}
      $\sqrt{\langle u_\alpha^\prime u_\alpha^\prime\rangle}^+$
    \end{minipage}
    \begin{minipage}{.5\linewidth}
      \centerline{$(a)$}
      \includegraphics[width=\linewidth]{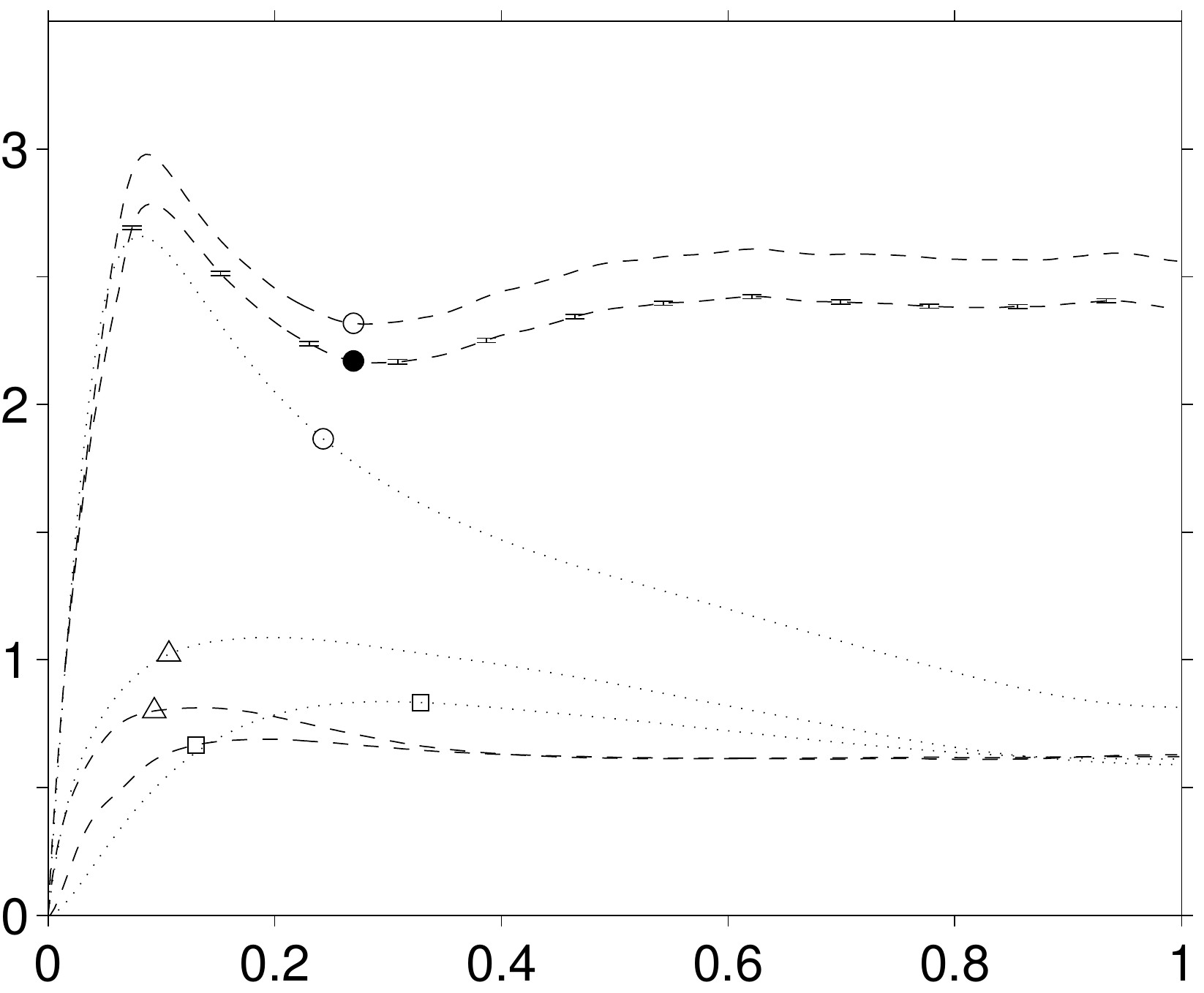}
      \centerline{$y/h$}
    \end{minipage}\\[1ex]
    \begin{minipage}{13.ex}
      $\sqrt{\langle u_\alpha^\prime u_\alpha^\prime\rangle}^+$
    \end{minipage}
    \begin{minipage}{.5\linewidth}
      \centerline{$(b)$}
      \includegraphics[width=\linewidth]{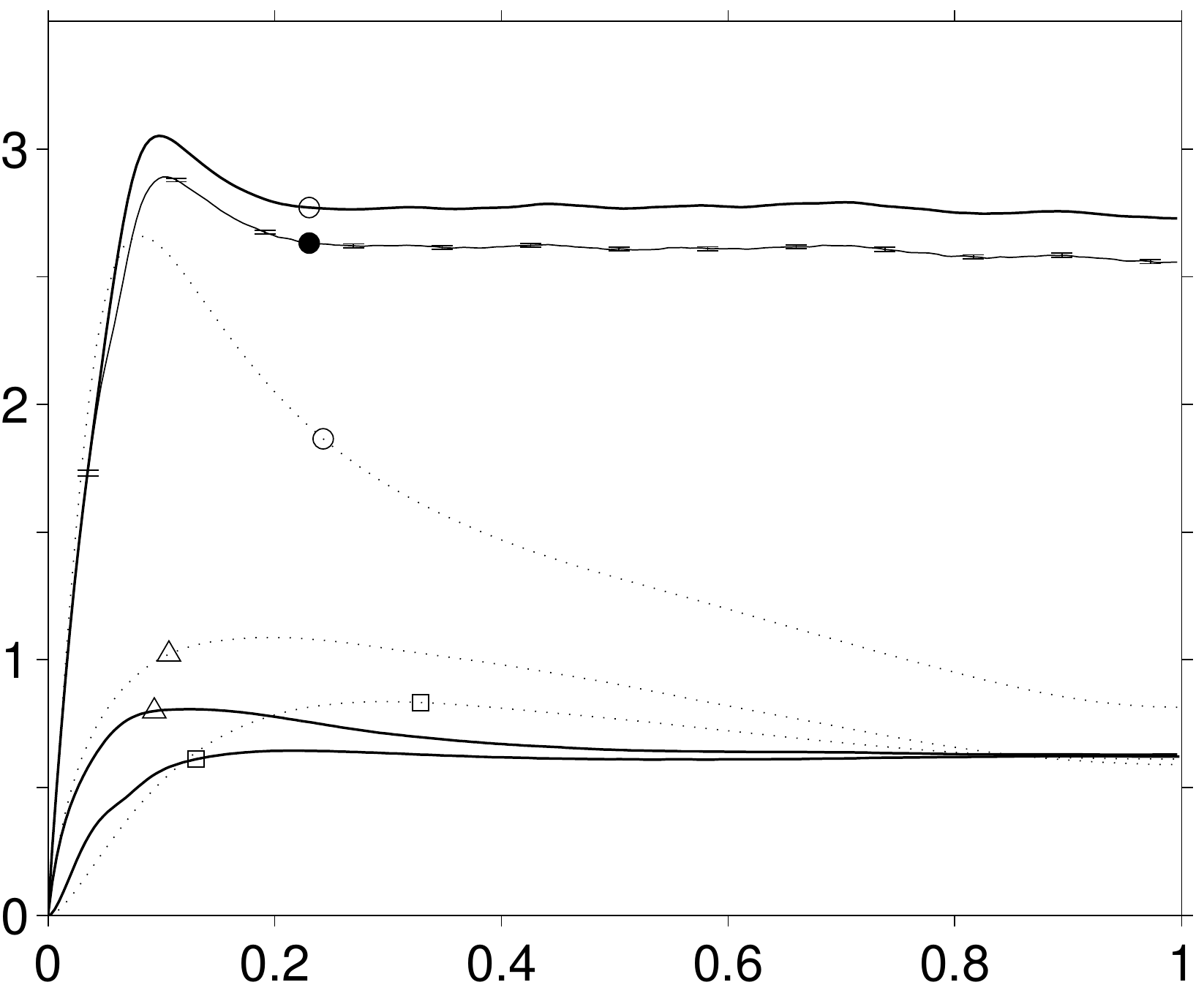}
      \centerline{$y/h$}
    \end{minipage}
    \caption{
      {Wall-normal profiles of the r.m.s.\ fluid velocity
        fluctuations; line styles as in figure~\ref{fig-results-um}. 
        $(a)$ Data from case A vs.\ single-phase flow; $(b)$ case B vs.\
        single-phase flow. 
        The coordinate directions are indicated by the symbols:
        $\circ$~$\alpha=1$; $\square$~$\alpha=2$;
        $\vartriangle$~$\alpha=3$.  
        The lines with the symbol ``$\bullet$'' indicate the profiles of the
        streamwise component, corrected for the overestimation due to the 
        accumulation of statistics over the composite flow field (cf.\
        appendix~\ref{app-fluidonly}); errorbars show the
        standard deviation of the fit.}
    }   
    \label{fig-results-uu}
  \end{figure}
}
\begin{figure}
  \centering
  \begin{minipage}{3.ex}
    $k^+$
  \end{minipage}
  \begin{minipage}{.5\linewidth}
    \includegraphics[width=\linewidth]{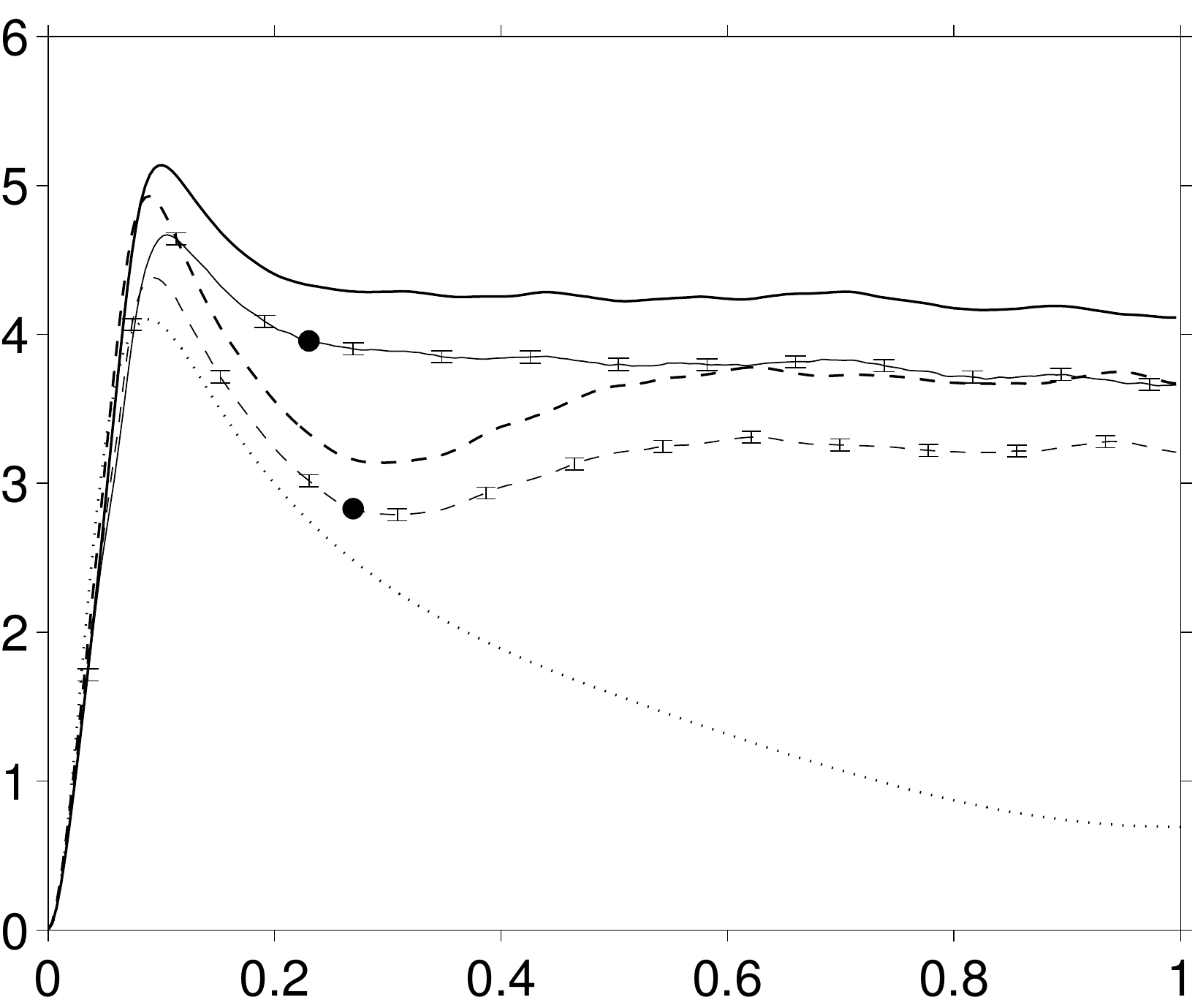}
    \centerline{$y/h$}
  \end{minipage}
  \caption{
    {Wall-normal profiles of the turbulent kinetic energy
      $k=\sum_{i=1}^3\langle u_i^\prime u_i^\prime\rangle/2$; 
      line styles as in figure~\ref{fig-results-um}.  
      The lines with the symbol ``$\bullet$'' indicate the profiles
      corrected for the overestimation of the streamwise stress
      component, due to the accumulation of statistics over the
      composite flow field (cf.\ appendix~\ref{app-fluidonly});
      errorbars show the standard deviation of the fit.} 
  }   
  \label{fig-results-tke}
\end{figure}
\begin{figure}
  \centering
  \begin{minipage}{13ex}
    $\sqrt{\langle u^\prime_{\alpha,c} u^\prime_{\alpha,c}\rangle}^+$
  \end{minipage}
  \begin{minipage}{.5\linewidth}
    \centerline{$(a)$}
    \includegraphics[width=\linewidth]{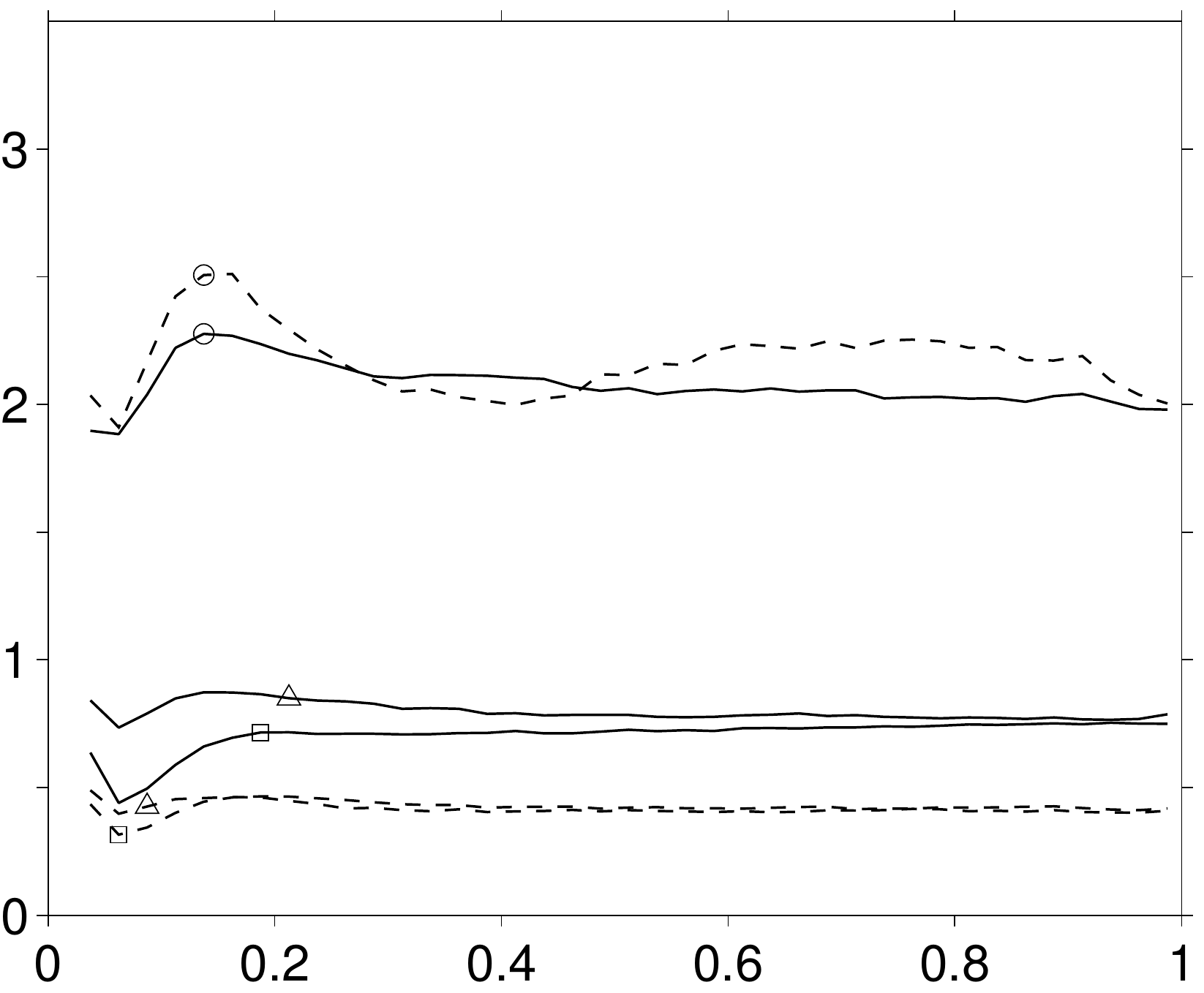}
    \centerline{$y/h$}
  \end{minipage}\\
  \begin{minipage}{13ex}
    $\sqrt{\langle u^\prime_{\alpha} u^\prime_{\alpha}\rangle}^+$\,,\\
    $\sqrt{\langle u^\prime_{\alpha,c} u^\prime_{\alpha,c}\rangle}^+$
  \end{minipage}
  \begin{minipage}{.5\linewidth}
    \centerline{$(b)$}
    \includegraphics[width=\linewidth]{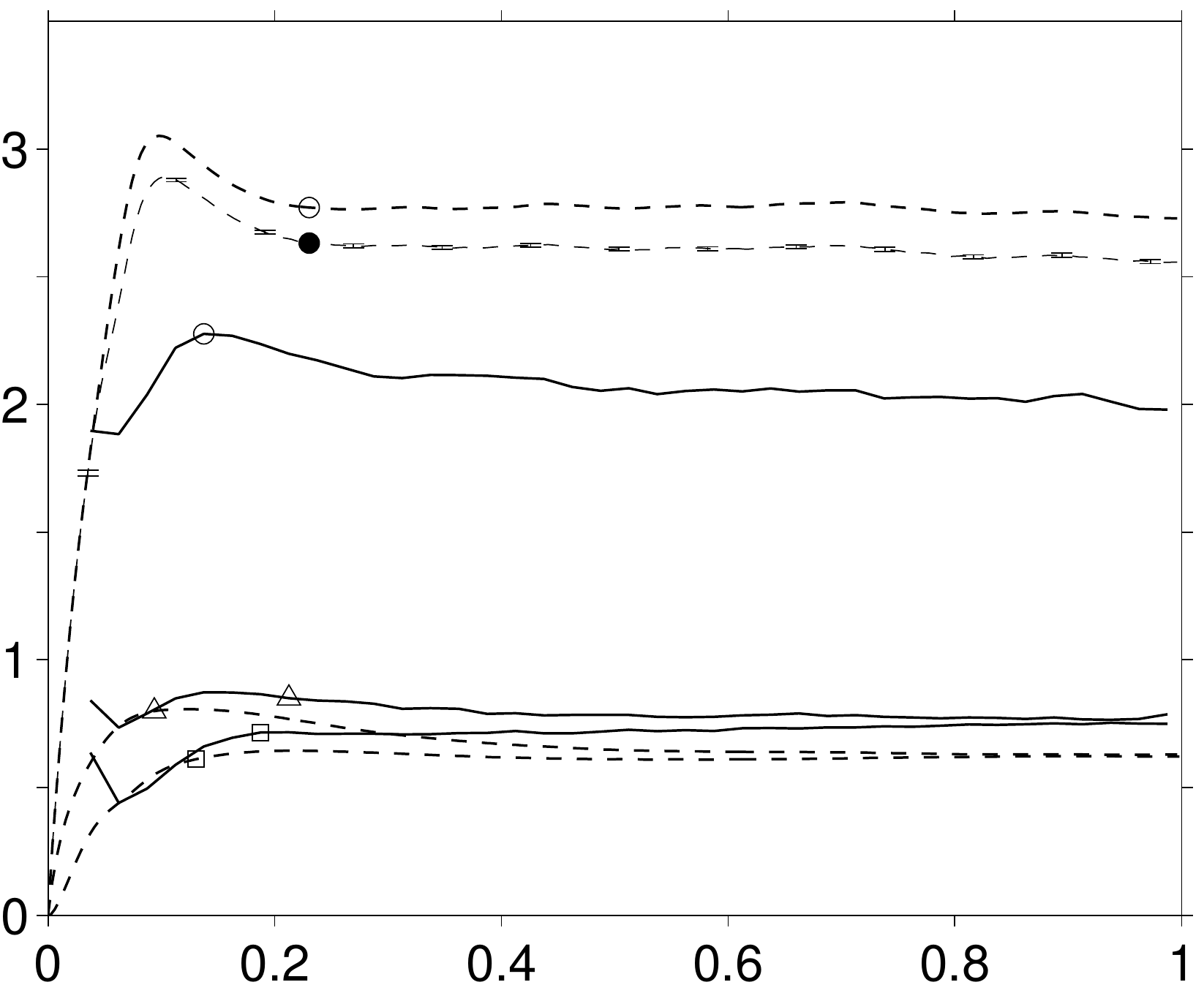}
    \centerline{$y/h$}
  \end{minipage}
  \caption{
    {Wall-normal profiles of the r.m.s.\ particle velocity
      fluctuations: $(a)$ case A vs.\ case B with line styles as in
      figure~\ref{fig-results-up}; 
      $(b)$ data of case B (solid lines) compared to the r.m.s.\ fluid
      velocity (dashed lines).
      The coordinate directions are indicated by the symbols: 
      $\circ$~$\alpha=1$; $\square$~$\alpha=2$;
      $\vartriangle$~$\alpha=3$.
      In $(b)$ the dashed line with the symbol ``$\bullet$'' indicates 
      the profile of the streamwise fluid velocity component,
      corrected for the overestimation due to the accumulation of
      statistics over the composite flow field (cf.\ 
      appendix~\ref{app-fluidonly}); errorbars show the
      standard deviation of the fit.} 
  }
  \label{fig-results-uup}
\end{figure}
\begin{figure}
  \centering
  \begin{minipage}{11ex}
    {$\displaystyle\frac{\langle \omega_{c,z}\rangle h}{u_b}$}
  \end{minipage}
  \begin{minipage}{.5\linewidth}
    \centerline{$(a)$}
    \includegraphics[width=\linewidth]{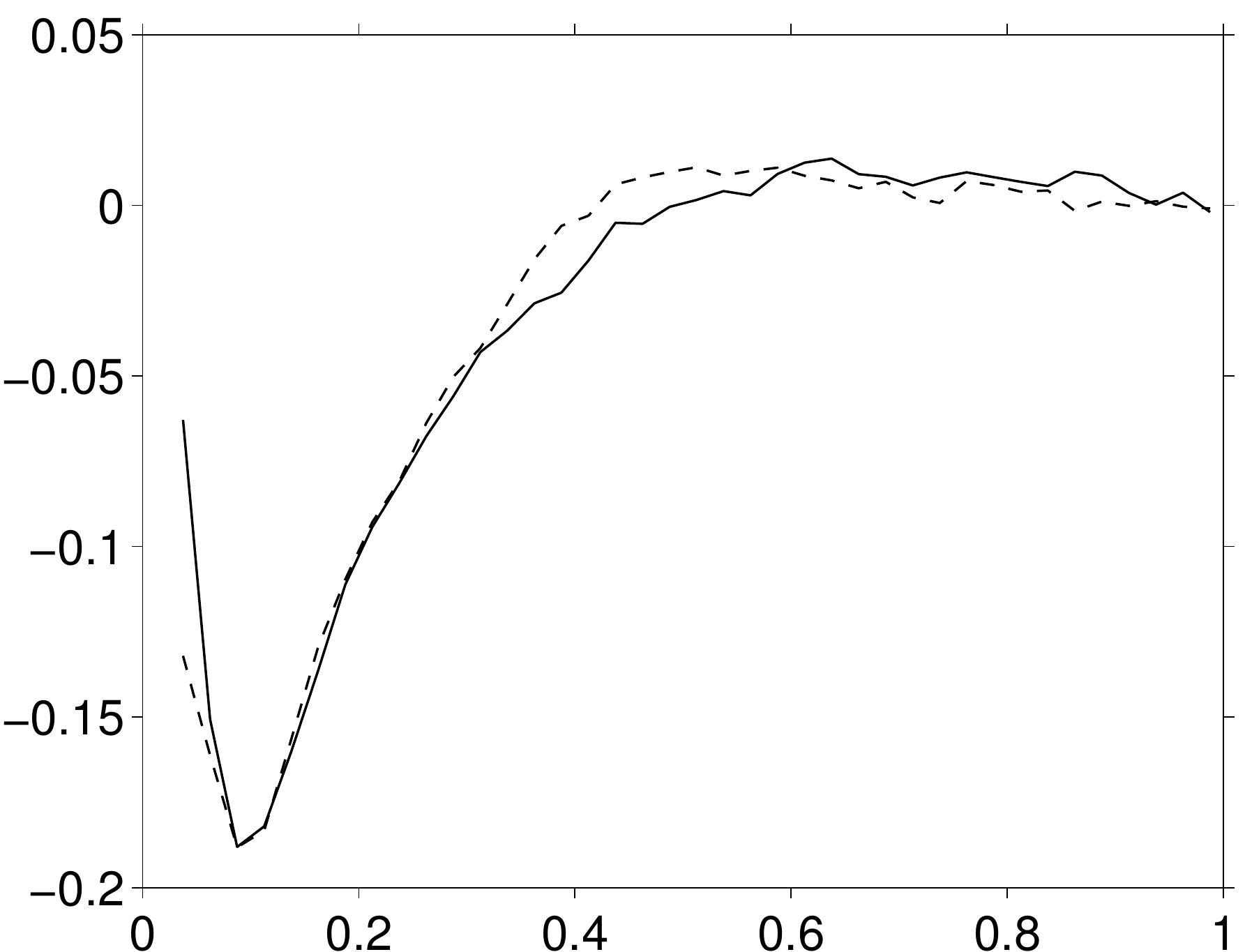}
    \centerline{$y/h$}
  \end{minipage}
  \\[1ex]
  \begin{minipage}{11ex}
    {$\displaystyle\frac{\langle \omega_{c,z}\rangle h}{u_b}$}
  \end{minipage}
  \begin{minipage}{.5\linewidth}
    \centerline{$(b)$}
    \includegraphics[width=\linewidth]{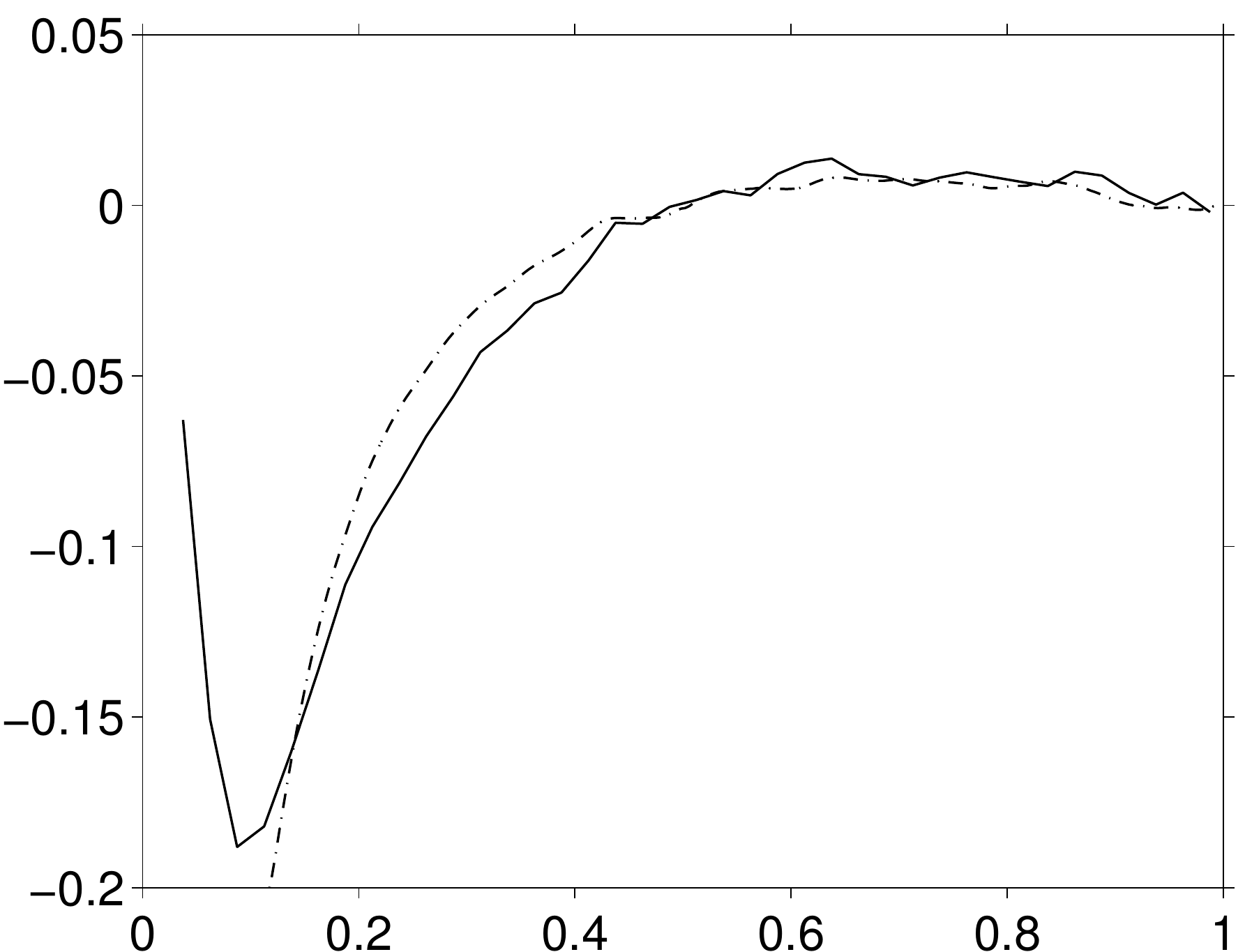}
    \centerline{$y/h$}
  \end{minipage}
  \caption{Wall-normal profiles of 
    the mean angular particle velocity; 
    line styles as in figure~\ref{fig-results-up}.
    In $(b)$ the data for case B is compared to the formula
    $-0.1\cdot\mbox{d}\langle u\rangle/\mbox{d}y$ (\chndot). }
  \label{fig-results-uprot}
\end{figure}
\begin{figure}
  \centering
  \rotatebox{90}{
    $\sigma\cdot pdf$}
  \begin{minipage}{.5\linewidth}
    \centerline{$(a)$}
    \includegraphics[width=\linewidth]
    {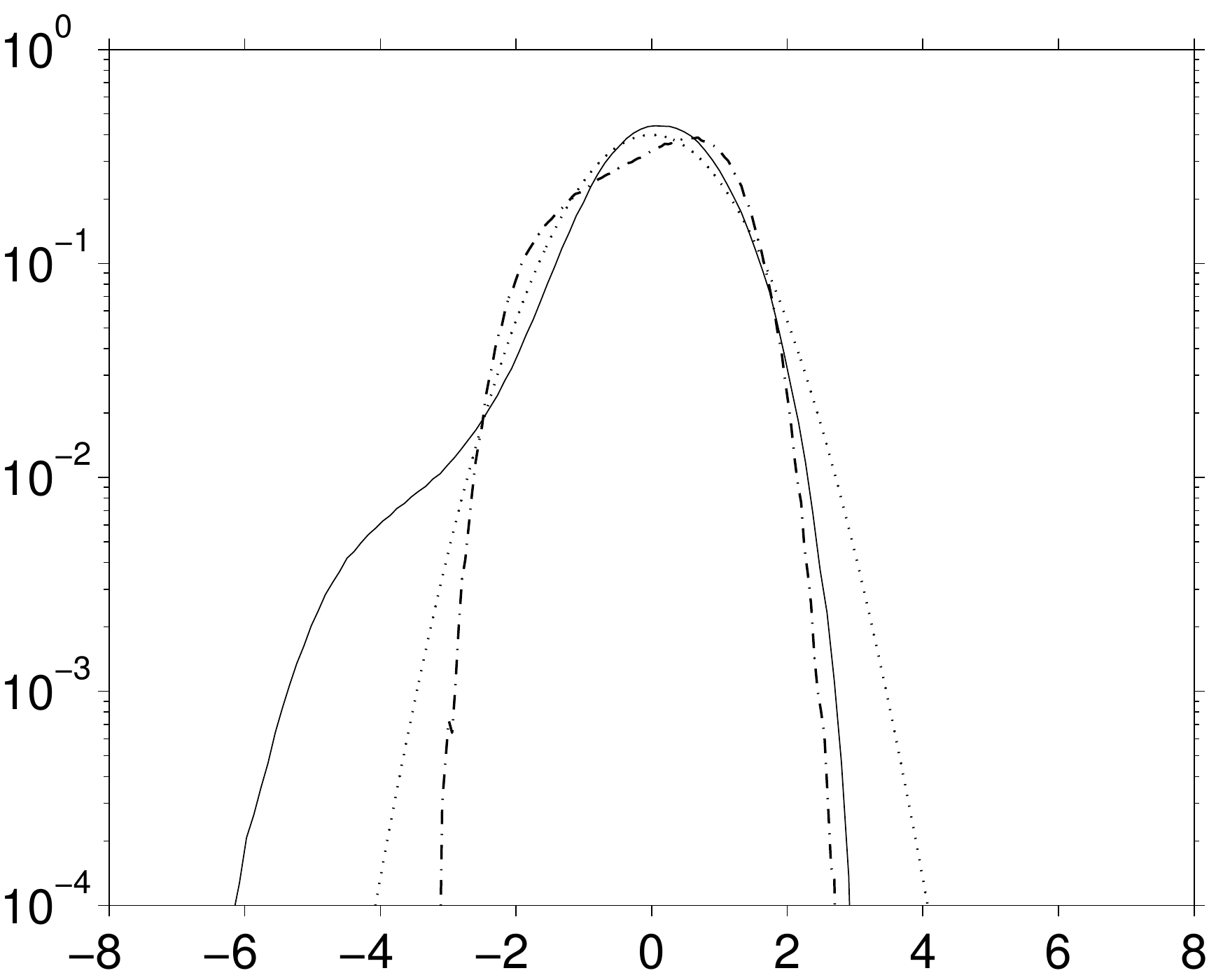}
    \centerline{$u_f^\prime/\sigma$}
  \end{minipage}
  \\[1ex]
  \rotatebox{90}{
    $\sigma\cdot pdf$}
  \begin{minipage}{.5\linewidth}
    \centerline{$(b)$}
    \includegraphics[width=\linewidth]
    {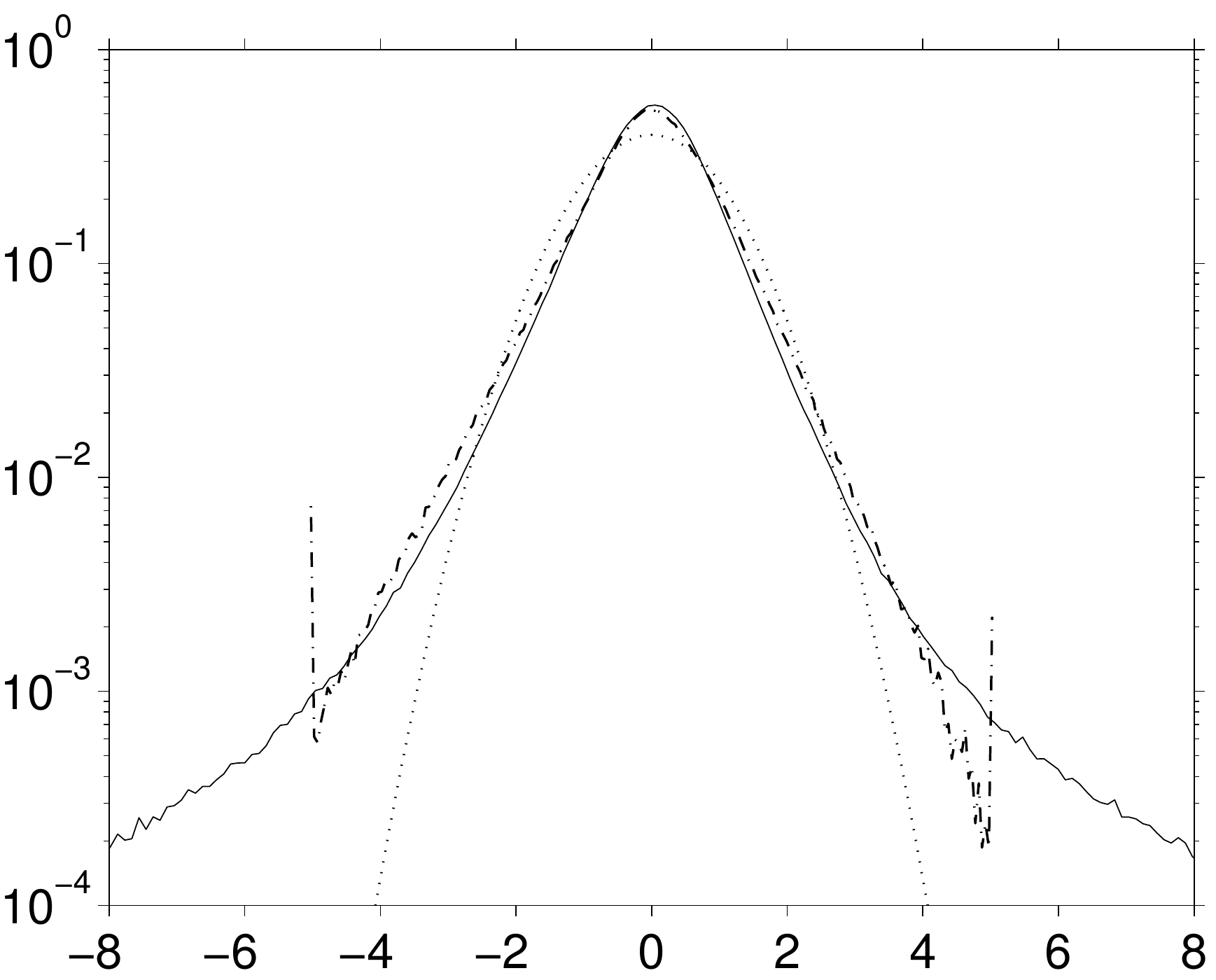}
    \centerline{$v_f^\prime/\sigma$}
  \end{minipage}
  \\[1ex]
  \rotatebox{90}{
    $\sigma\cdot pdf$}
  \begin{minipage}{.5\linewidth}
    \centerline{$(c)$}
    \includegraphics[width=\linewidth]
    {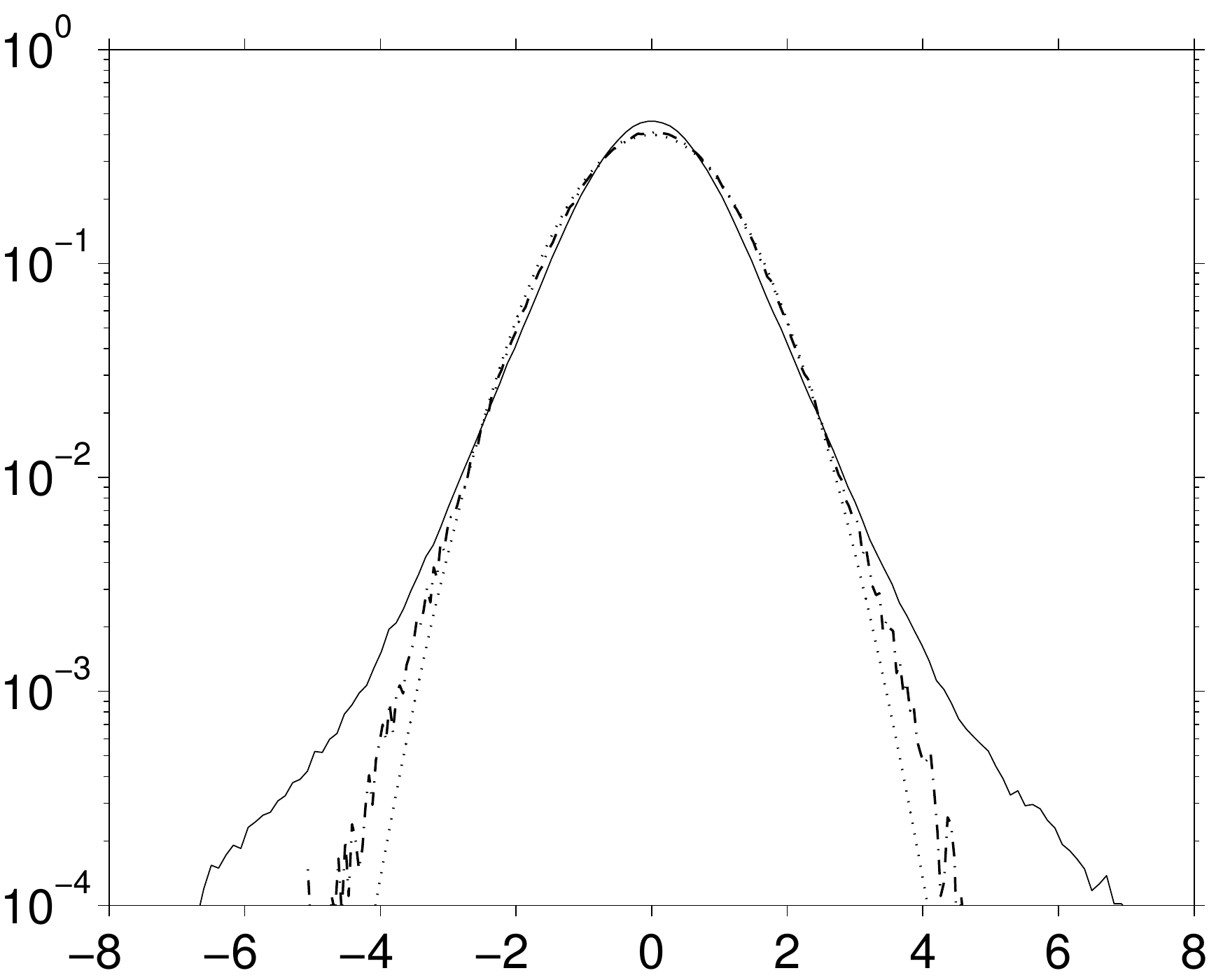}
    \centerline{$w_f^\prime/\sigma$}
  \end{minipage}
  \caption{Normalized probability density functions of fluid 
    velocity components, computed from 12 instantaneous flow fields of
    case B (taking into account only nodes inside the fluid domain) at
    $y^+=20$: $(a)$ streamwise, $(b)$ wall-normal, and $(c)$ spanwise;   
    \solid~present results (case B);
    \chndot~single-phase data from \cite{moser:99} at $Re_\tau=180$; 
    \dotted~Gaussian reference curve.
  }
  \label{fig-results-pdf-fluid-1}
\end{figure}
\begin{figure}
  \centering
  \rotatebox{90}{
    $\sigma\cdot pdf$}
  \begin{minipage}{.5\linewidth}
    \centerline{$(a)$}
    \includegraphics[width=\linewidth]
    {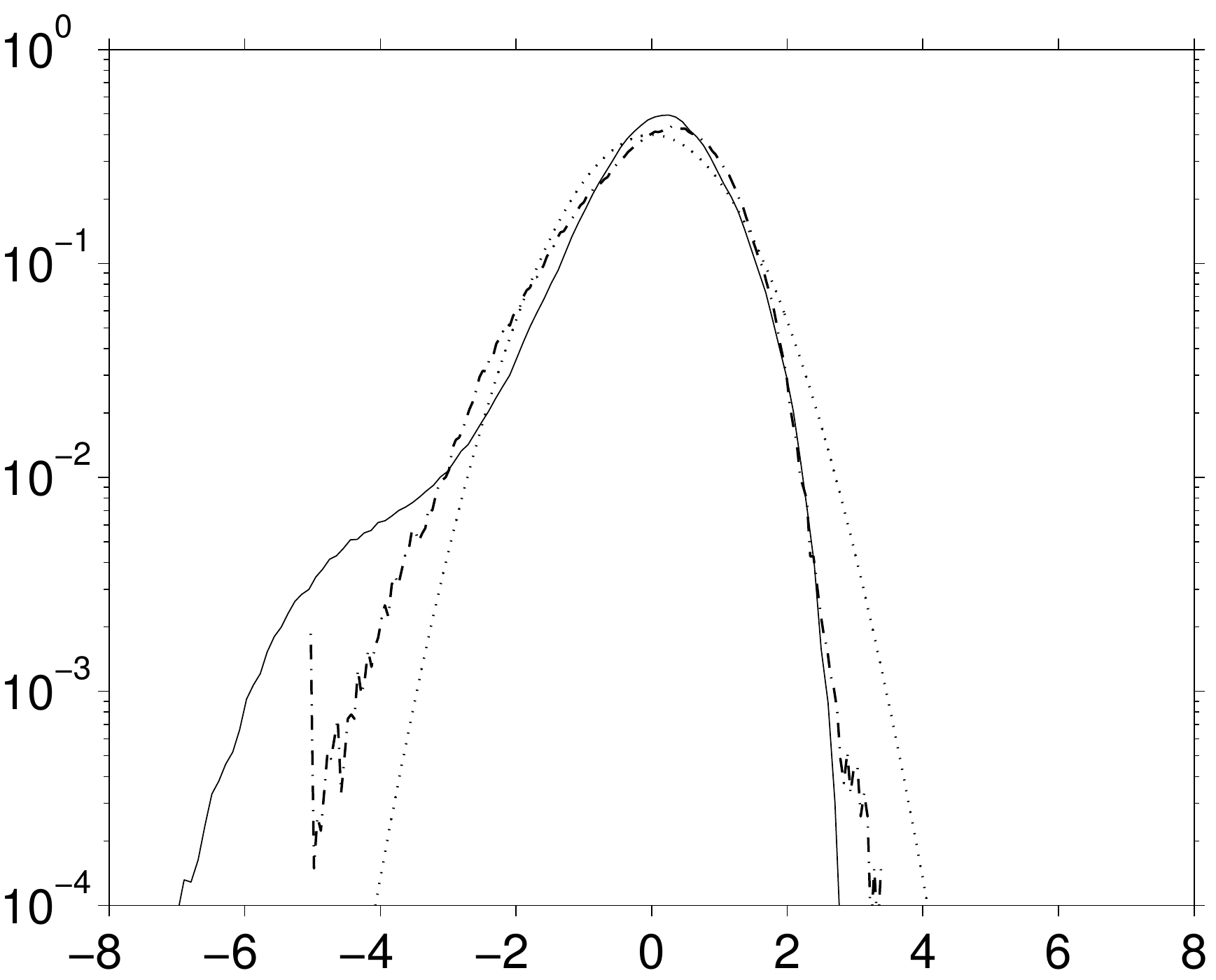}
    \centerline{$u_f^\prime/\sigma$}
  \end{minipage}
  \\[1ex]
  \rotatebox{90}{
    $\sigma\cdot pdf$}
  \begin{minipage}{.5\linewidth}
    \centerline{$(b)$}
    \includegraphics[width=\linewidth]
    {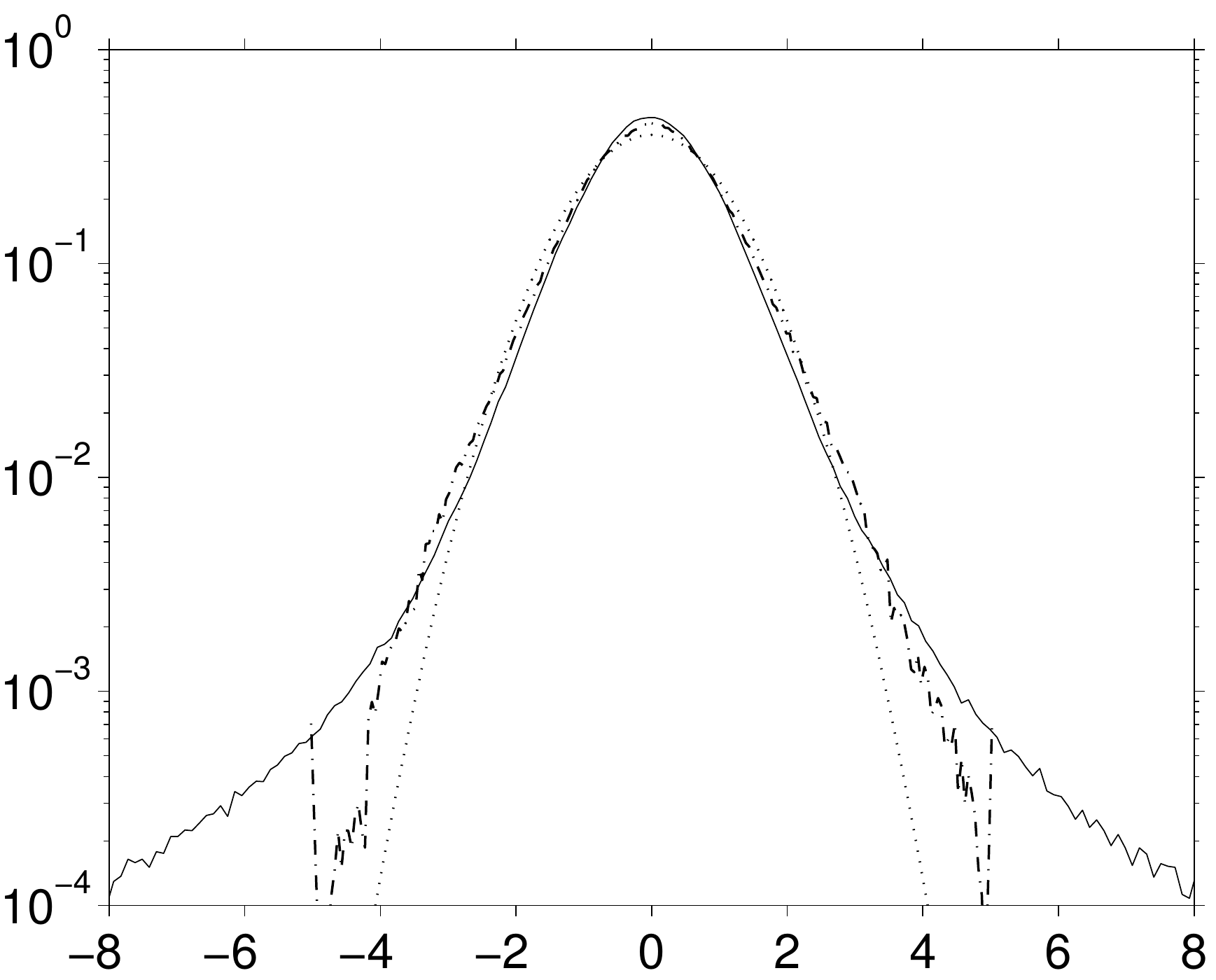}
    \centerline{$v_f^\prime/\sigma$}
  \end{minipage}
  \\[1ex]
  \rotatebox{90}{
    $\sigma\cdot pdf$}
  \begin{minipage}{.5\linewidth}
    \centerline{$(c)$}
    \includegraphics[width=\linewidth]
    {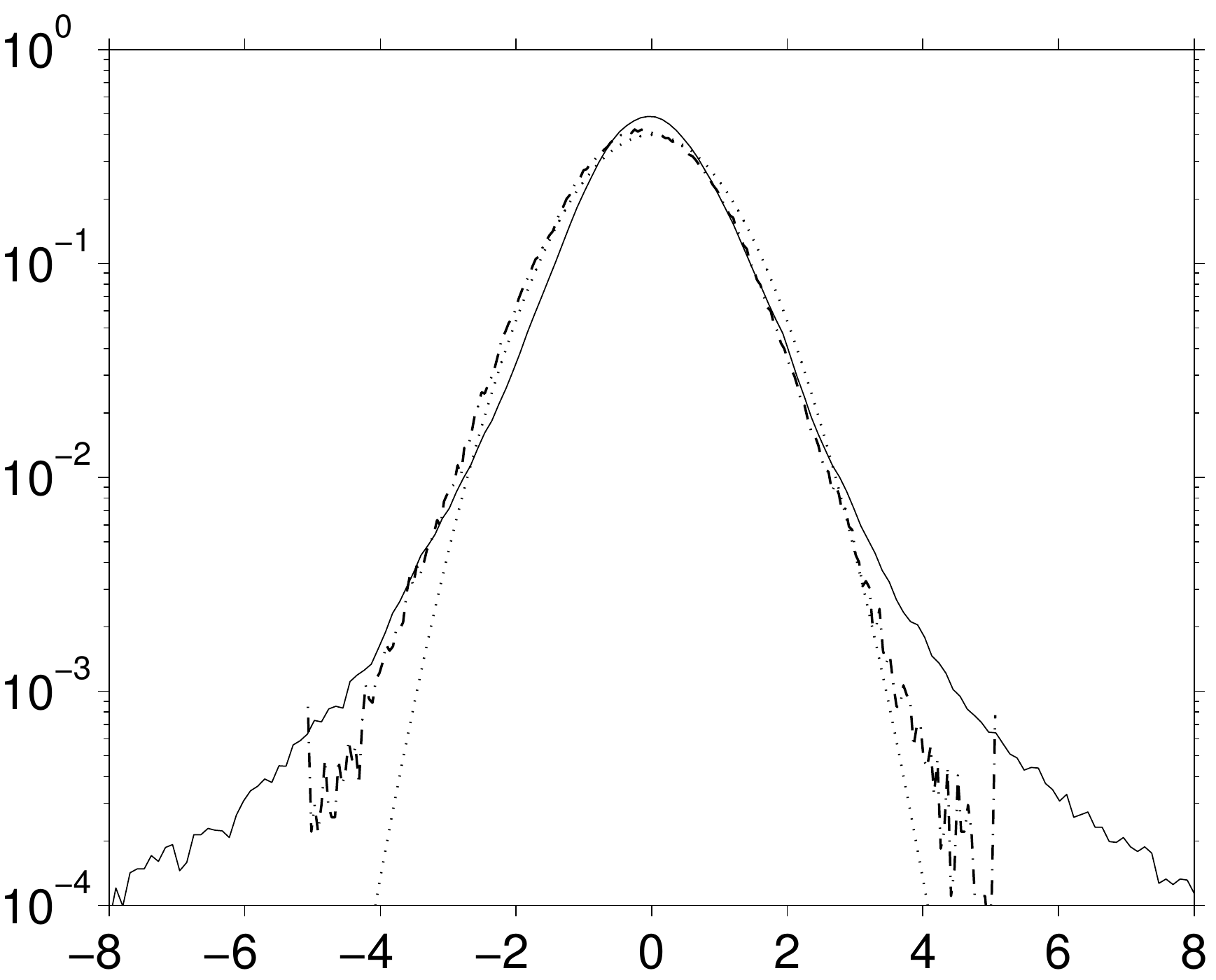}
    \centerline{$w_f^\prime/\sigma$}
  \end{minipage}
  \caption{As figure~\ref{fig-results-pdf-fluid-1}, but for the center
    of the channel ($y/h=1$).
  }
  \label{fig-results-pdf-fluid-2}
\end{figure}
\begin{figure}
  \centering
  \rotatebox{90}{
    $\sigma\cdot pdf$}
  \begin{minipage}{.5\linewidth}
    \centerline{$(a)$}
    \includegraphics[width=\linewidth]{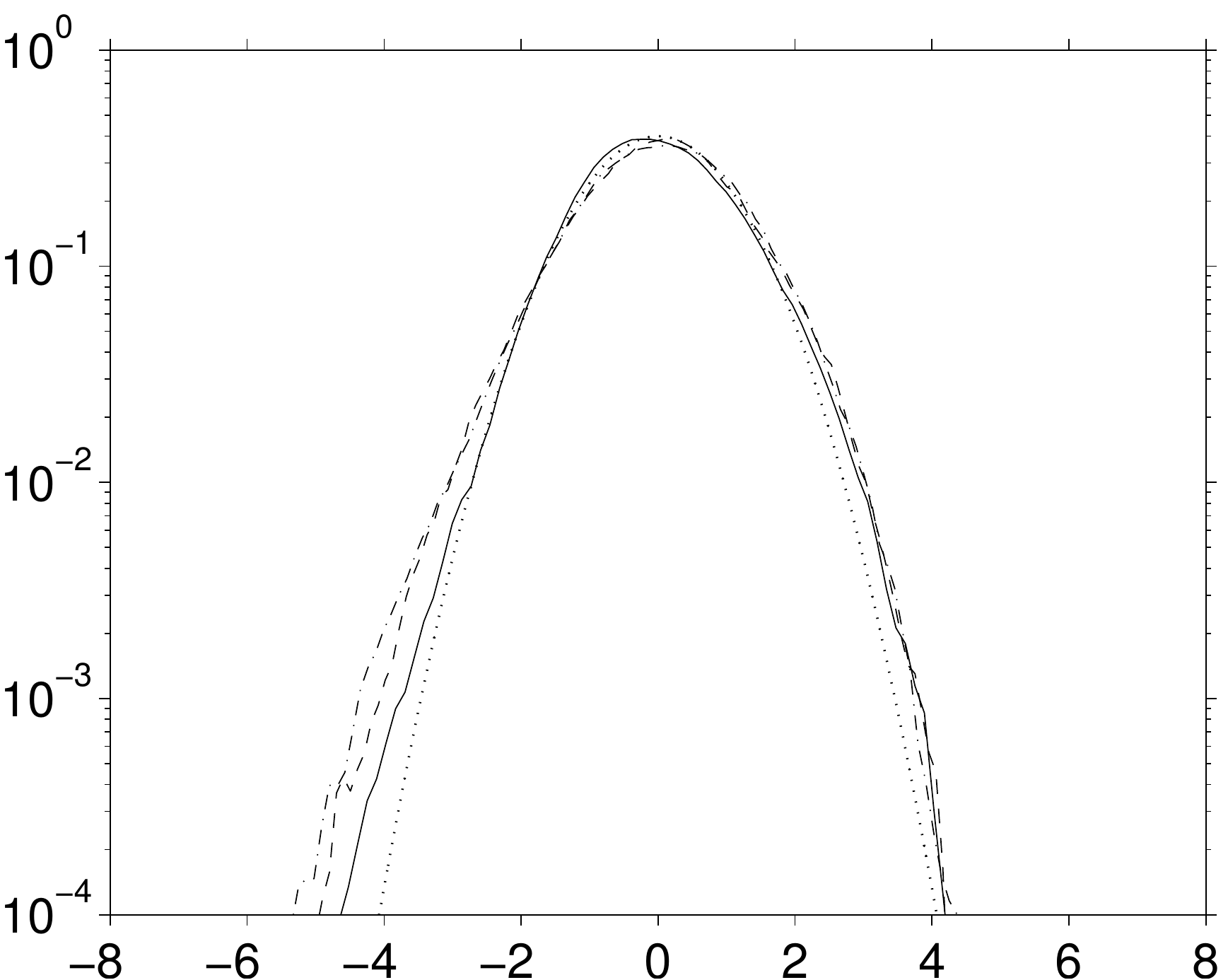}
    \centerline{$u_c^\prime/\sigma$}
  \end{minipage}
  \\[1ex]
  \rotatebox{90}{
    $\sigma\cdot pdf$}
  \begin{minipage}{.5\linewidth}
    \centerline{$(b)$}
    \includegraphics[width=\linewidth]{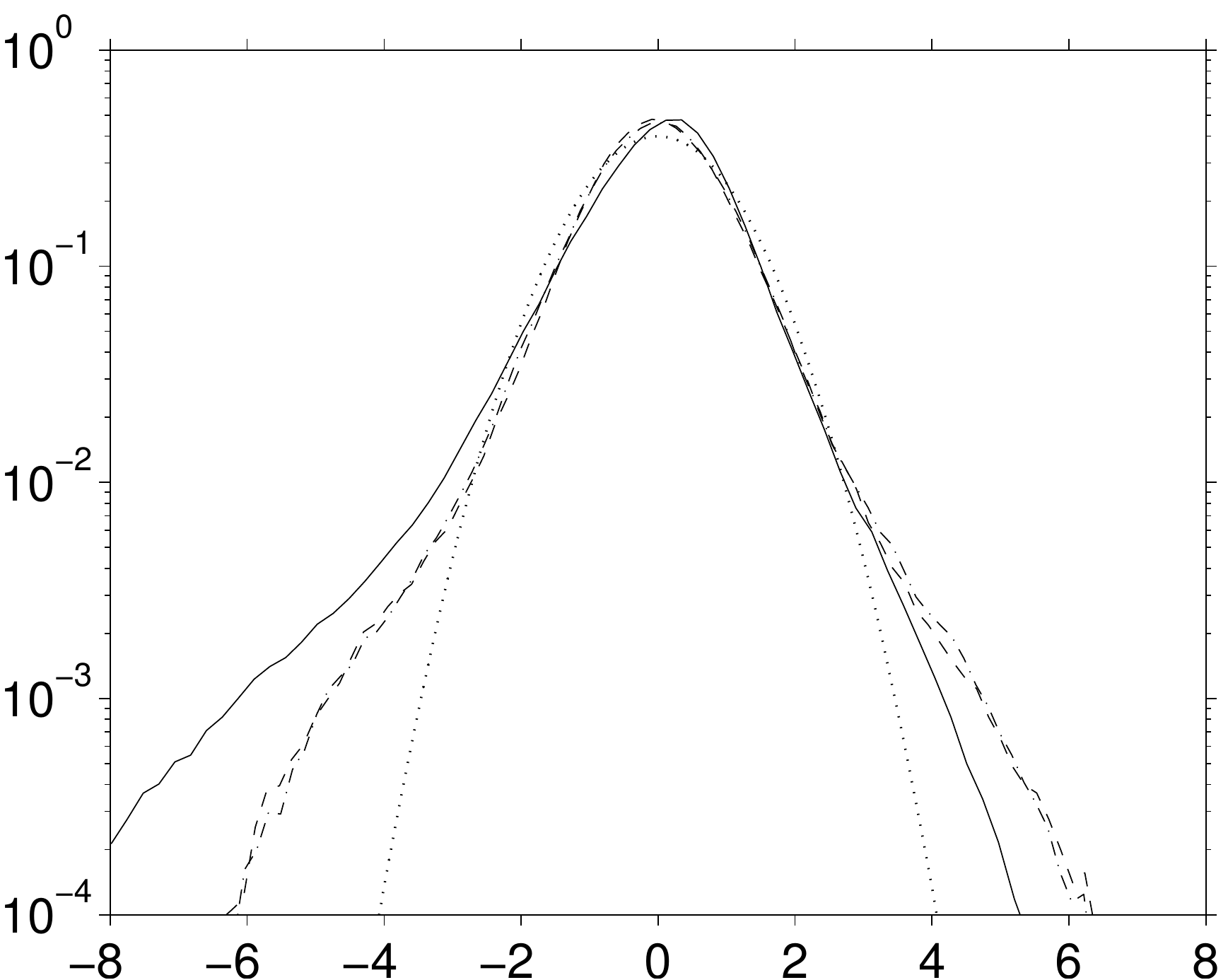}
    \centerline{$v_c^\prime/\sigma$}
  \end{minipage}
  \\[1ex]
  \rotatebox{90}{
    $\sigma\cdot pdf$}
  \begin{minipage}{.5\linewidth}
    \centerline{$(c)$}
    \includegraphics[width=\linewidth]{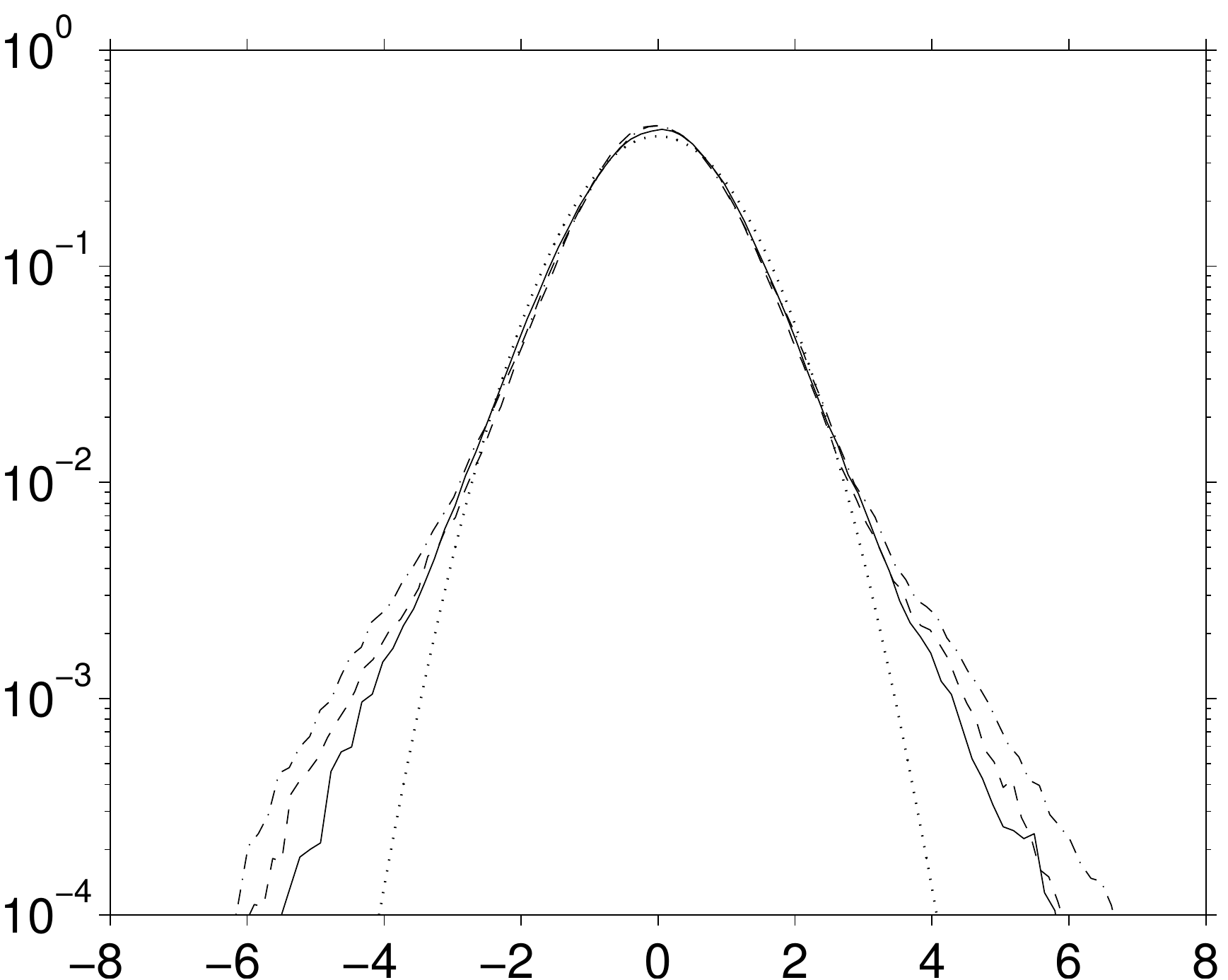}
    \centerline{$w_c^\prime/\sigma$}
  \end{minipage}
  \caption{Normalized probability density functions of particle
    velocity components in case B at different wall-distances:
    \solid~$y^+=17$; \dashed~$y^+=73$; \chndot~$y^+=220$;
    \dotted~Gaussian reference curve.
  }  
  \label{fig-results-pdf-particles-12}
\end{figure}
\clearpage
\begin{figure}
  \centering
  \begin{minipage}{5ex}
    $\displaystyle\frac{v^\prime_f}{u_b}$
  \end{minipage}
  \begin{minipage}{.5\linewidth}
    \centerline{$(a)$}
    \includegraphics[width=\linewidth]
    {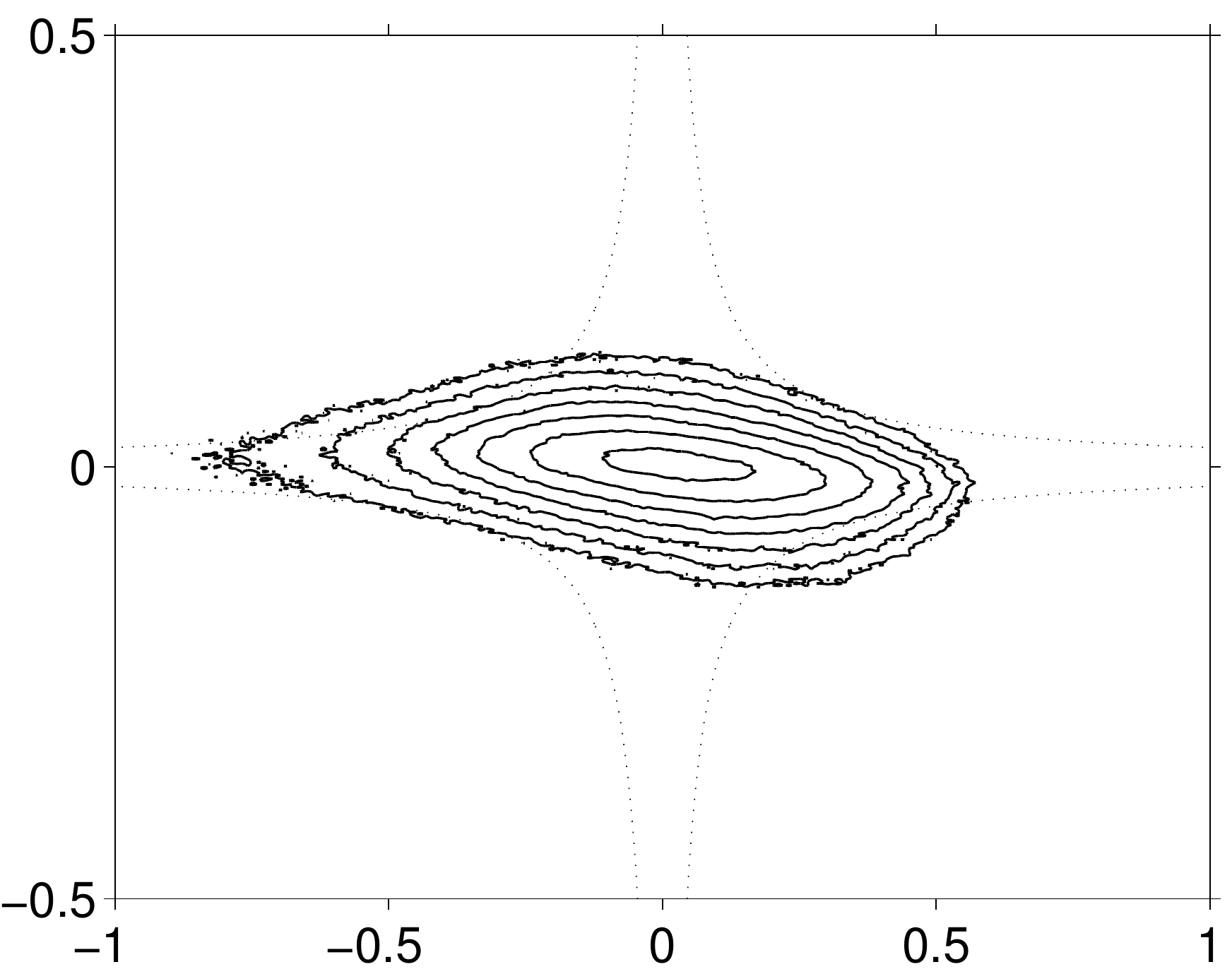}
    \centerline{$u^\prime_f/u_b$}
  \end{minipage}
  \\[1ex]
  \begin{minipage}{5ex}
    $\displaystyle\frac{v^\prime_f}{u_b}$
  \end{minipage}
  \begin{minipage}{.5\linewidth}
    \centerline{$(b)$}
    \includegraphics[width=\linewidth]
    {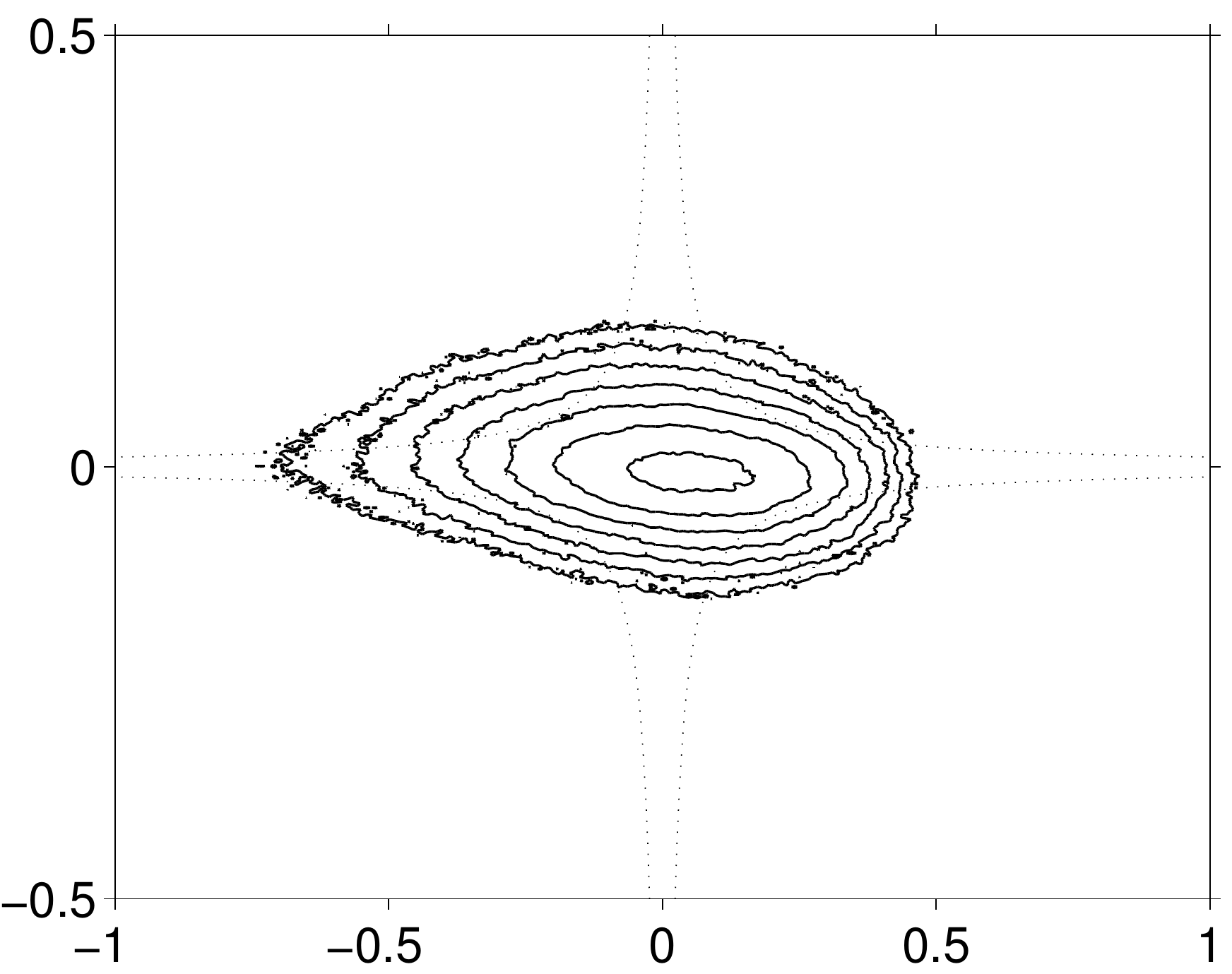}
    \centerline{$u^\prime_f/u_b$}
  \end{minipage}
  \caption{Isocontours of the joint probability density function of
    streamwise and wall-normal fluid velocity fluctuations in case B
    (evaluated from 12 instantaneous flow fields, taking into account
    only points in the fluid domain).
    The values for the contours vary between 0.0125 and 0.8 of the
    maximum probability, increasing by a factor of 2 between each level. 
    $(a)$ $y^+=20$; $(b)$ $y^+=74$.
    The dotted lines indicate $|u^\prime_f v^\prime_f|=-8\langle
    u^\prime_f v^\prime_f\rangle$. 
  }
  \label{fig-results-quaduv-fluid}
\end{figure}
\begin{figure}
  \centering
  \begin{minipage}{.5\linewidth}
    \centerline{$(a)$}
    \includegraphics[width=\linewidth]
    {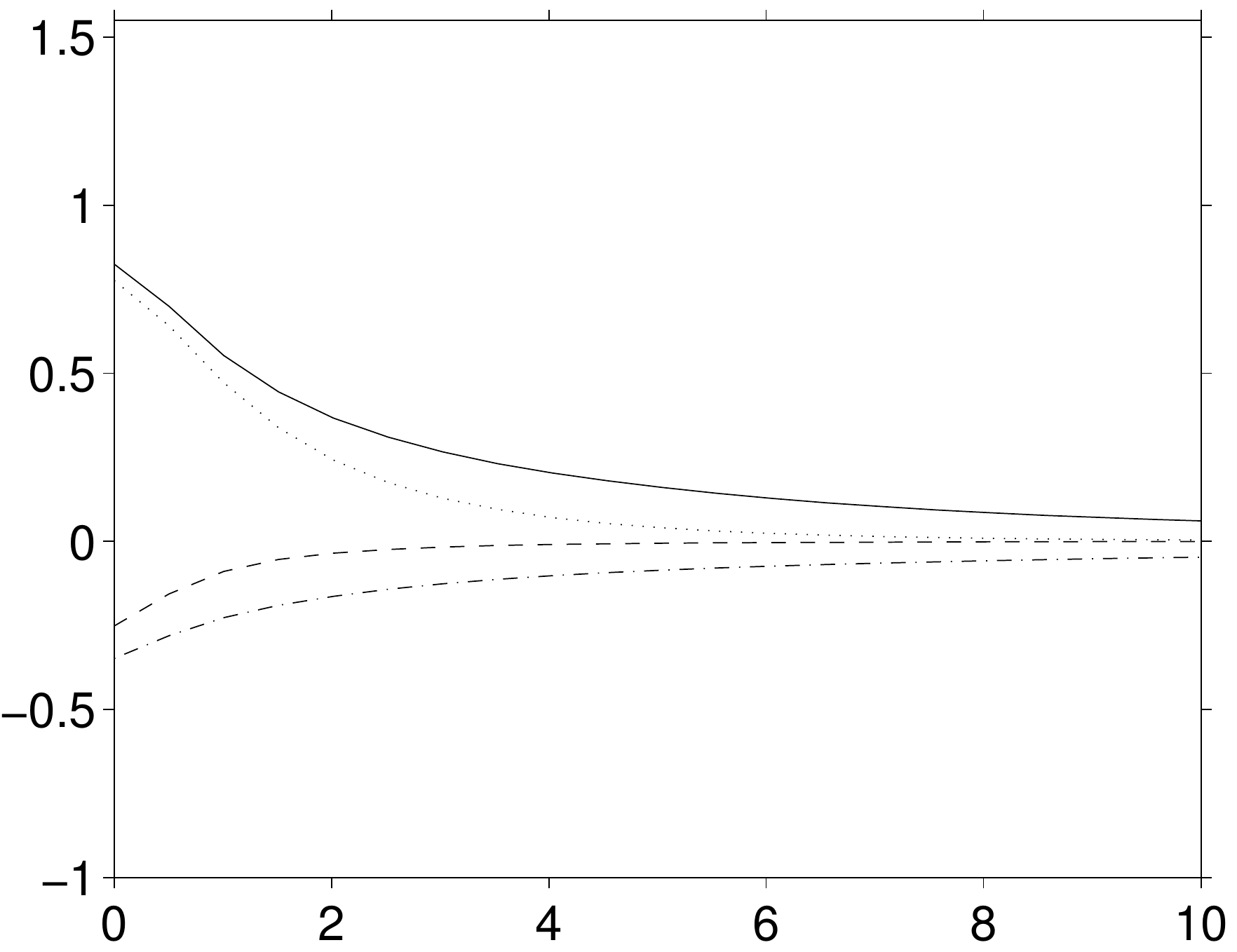}
    \centerline{threshold
    }
  \end{minipage}
  \\[1ex]
  \begin{minipage}{.5\linewidth}
    \centerline{$(b)$}
    \includegraphics[width=\linewidth]
    {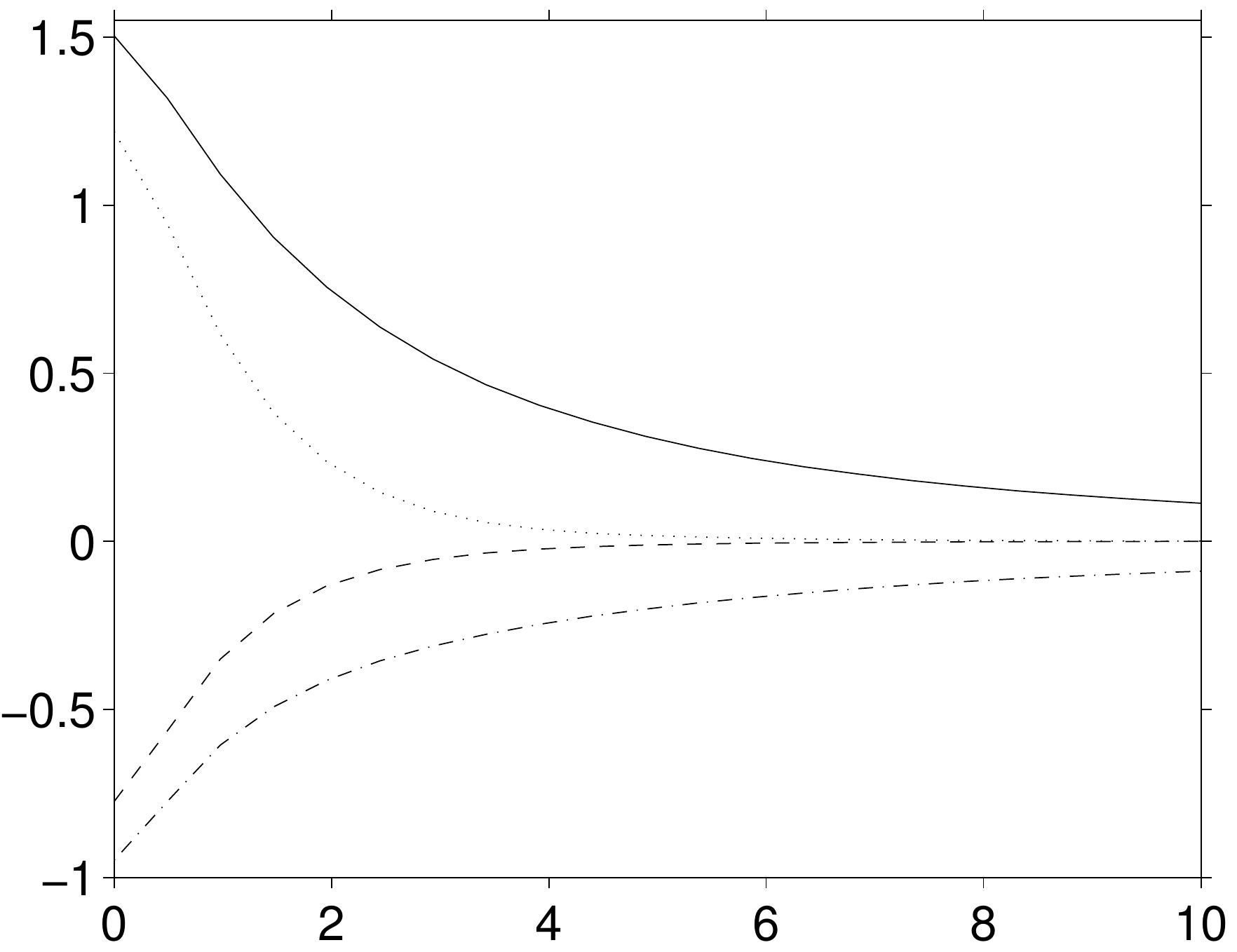}
    \centerline{threshold
    }
  \end{minipage}
  \caption{Fractional contribution from each quadrant to the Reynolds
    shear stress as a function of threshold: \dashed, first; \solid,
    second; \chndot, third; \dotted, fourth. $(a)$ $y^+=20$; $(b)$
    $y^+=74$. The threshold is normalized by the r.m.s.\ intensities
    of each plane: 
    $\sqrt{\langle u_f^\prime u_f^\prime\rangle}\cdot\sqrt{\langle
      v_f^\prime v_f^\prime\rangle}$. } 
  \label{fig-results-quaduv-fluid-frac-contrib}
\end{figure}
\begin{figure}
  \centering
  \begin{minipage}{5ex}
    $\displaystyle\frac{v^\prime_c}{u_b}$
  \end{minipage}
  \begin{minipage}{.5\linewidth}
    \centerline{$(a)$}
    \includegraphics[width=\linewidth]
    {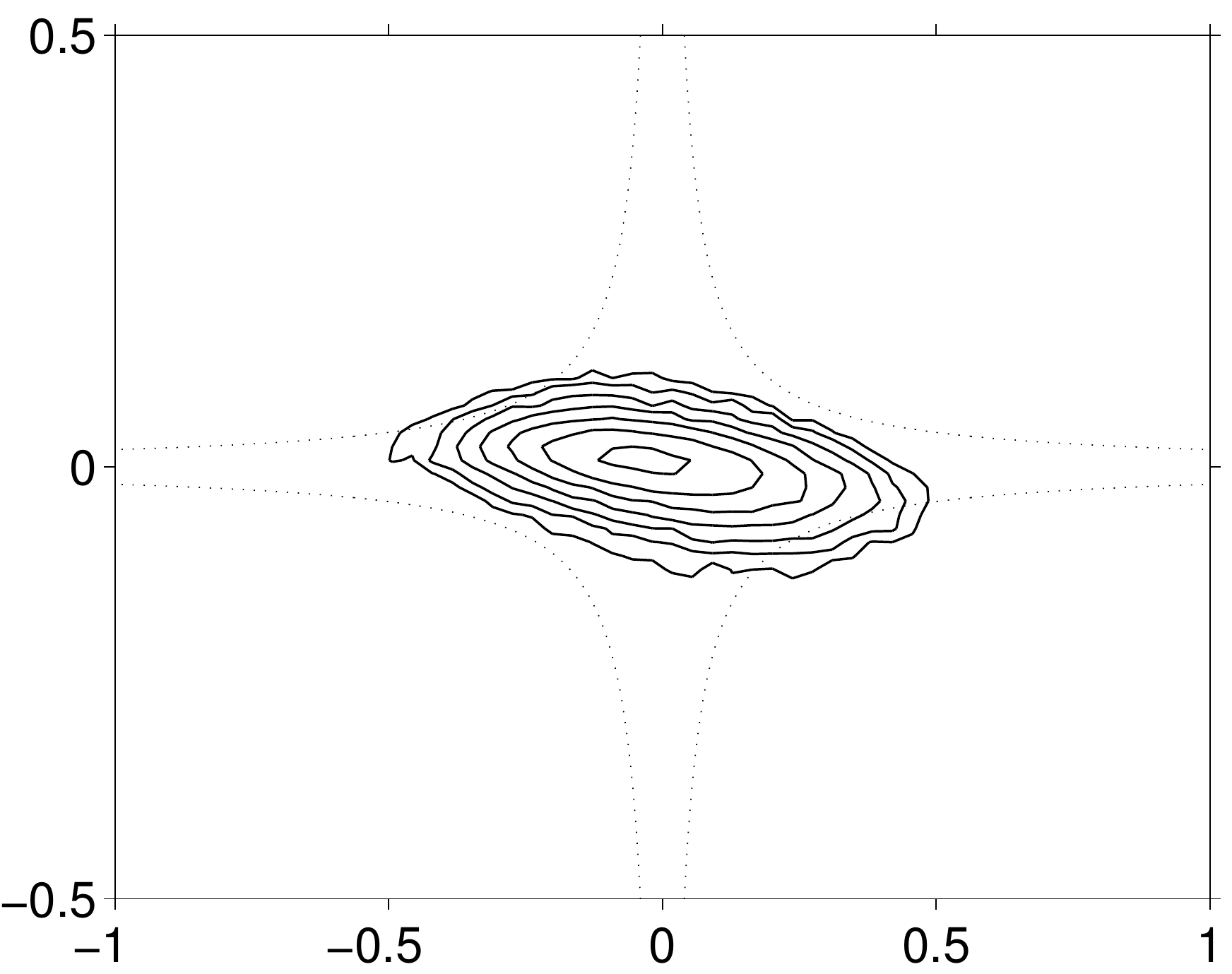}
    \centerline{$u^\prime_c/u_b$}
  \end{minipage}
  \\[1ex]
  \begin{minipage}{5ex}
    $\displaystyle\frac{v^\prime_c}{u_b}$
  \end{minipage}
  \begin{minipage}{.5\linewidth}
    \centerline{$(b)$}
    \includegraphics[width=\linewidth]
    {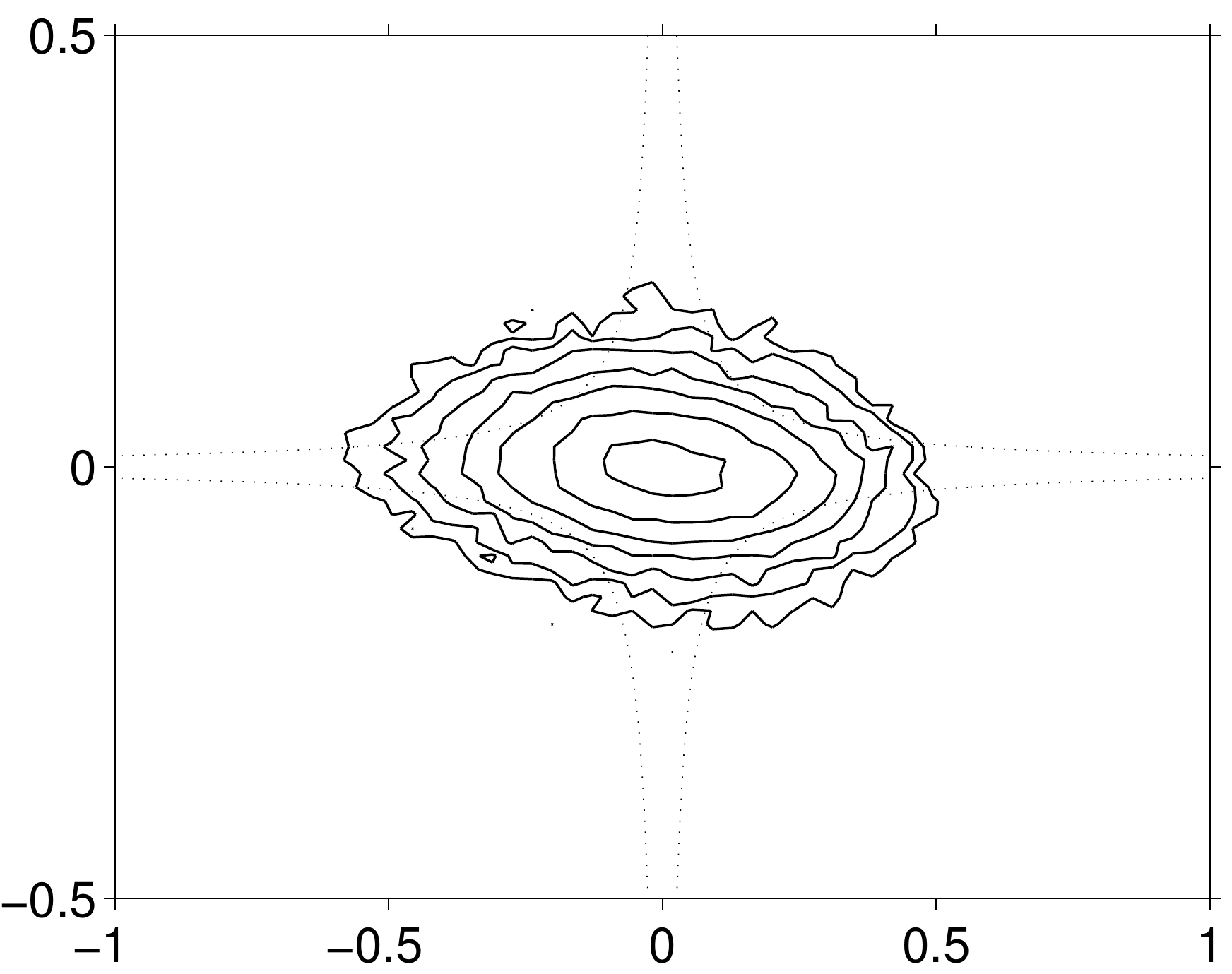}
    \centerline{$u^\prime_c/u_b$}
  \end{minipage}
  \caption{Isocontours of the joint probability density function of
    streamwise and wall-normal particle velocity fluctuations in case B
    (evaluated from 250 instantaneous particle distributions).
    The values for the contours vary between 0.0125 and 0.8 of the
    maximum, increasing by a factor of 2 between each level. 
    $(a)$ wall-normal bin centered at $y^+=17$; $(b)$ $y^+=73$.
    The dotted lines indicate $|u^\prime_c v^\prime_c|=-8\langle
    u^\prime_c v^\prime_c\rangle$.}
  \label{fig-results-quaduv-part}
\end{figure}
\clearpage
\begin{figure}
  \centering
  \begin{minipage}{6ex}
    {$R_{Lp,\alpha}$}
  \end{minipage}
  \begin{minipage}{.5\linewidth}
    \centerline{$(a)$}
    \includegraphics[width=\linewidth]
    {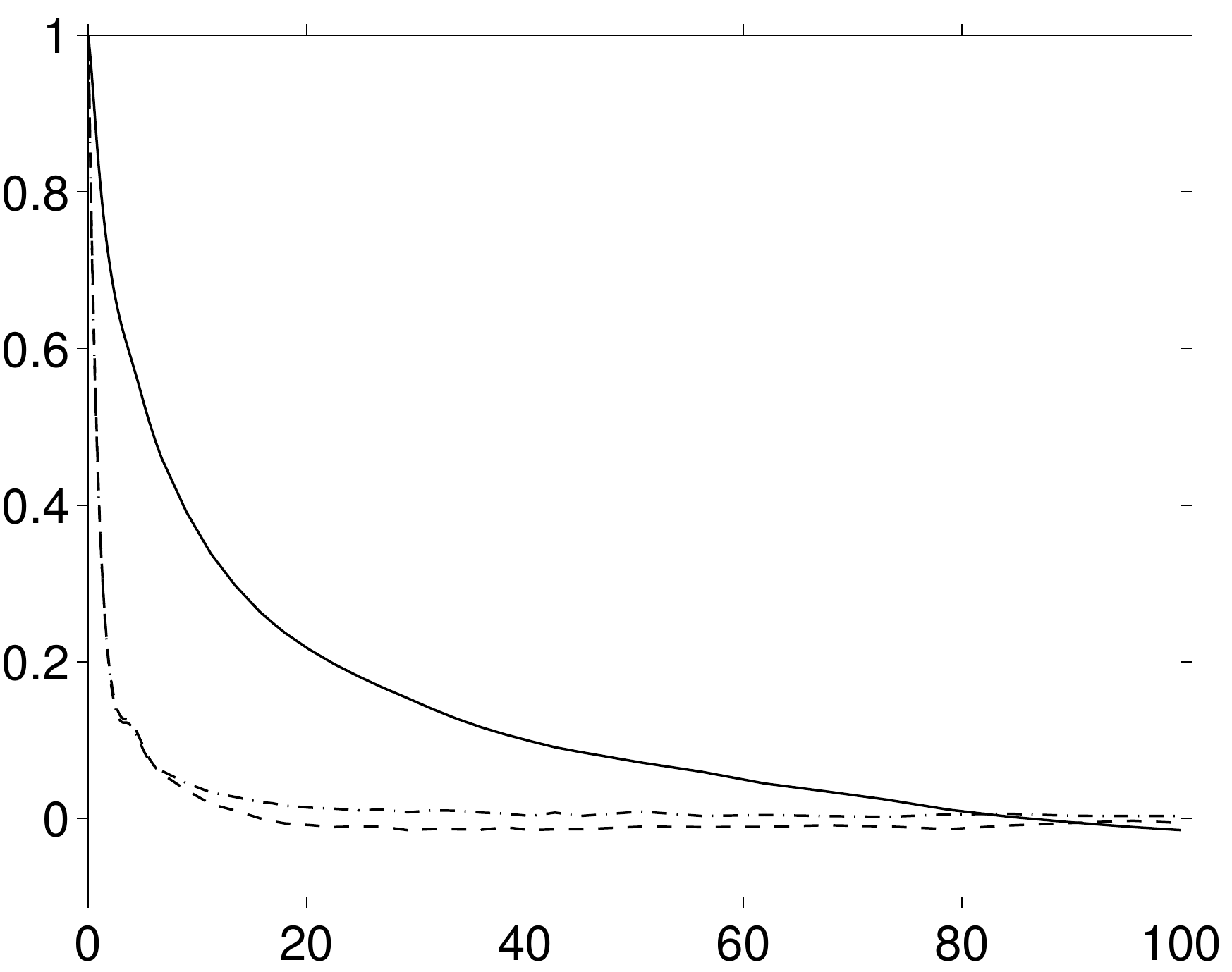}
    \centerline{$\tau u_b/h$}
  \end{minipage}
  \\[1ex]
  \begin{minipage}{6ex}
    {$R_{Lp,\alpha}$}
  \end{minipage}
  \begin{minipage}{.5\linewidth}
    \centerline{$(b)$}
    \includegraphics[width=\linewidth]
    {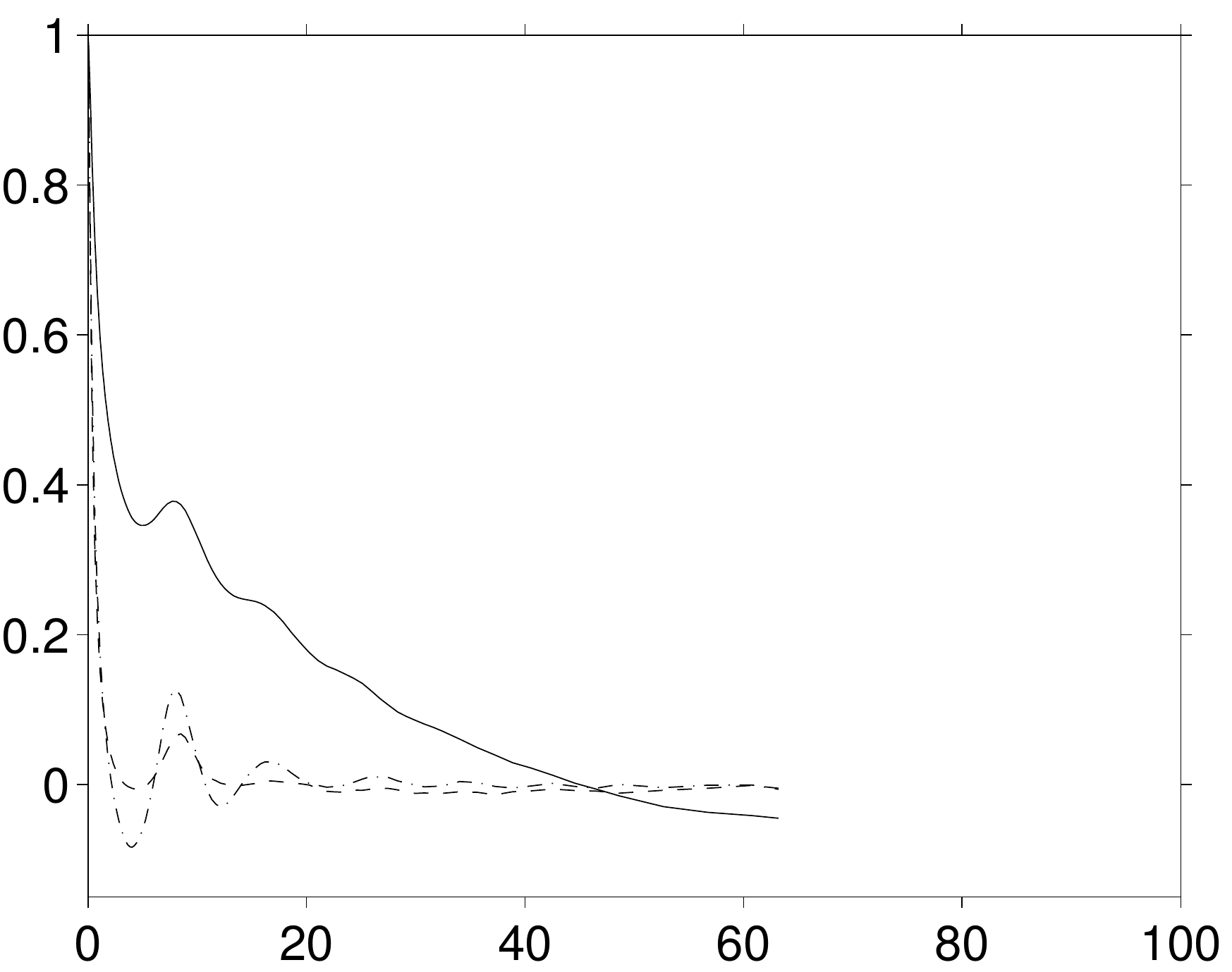}
    \centerline{$\tau u_b/h$}
  \end{minipage}
  \caption{Lagrangian particle velocity auto-correlations as a
    function of the separation time $\tau$: 
    $(a)$ case A; $(b)$ for case B.
    \solid~$\alpha=1$; \dashed~$\alpha=2$; \chndot~$\alpha=3$.
    }
  \label{fig-results-lag-corr}
\end{figure}
\begin{figure}
  \centering
  \begin{minipage}{10ex}
    {$\displaystyle\frac{T_{Lp,\alpha}u_b}{h}$}
  \end{minipage}
  \begin{minipage}{.5\linewidth}
    \includegraphics[width=\linewidth]
    {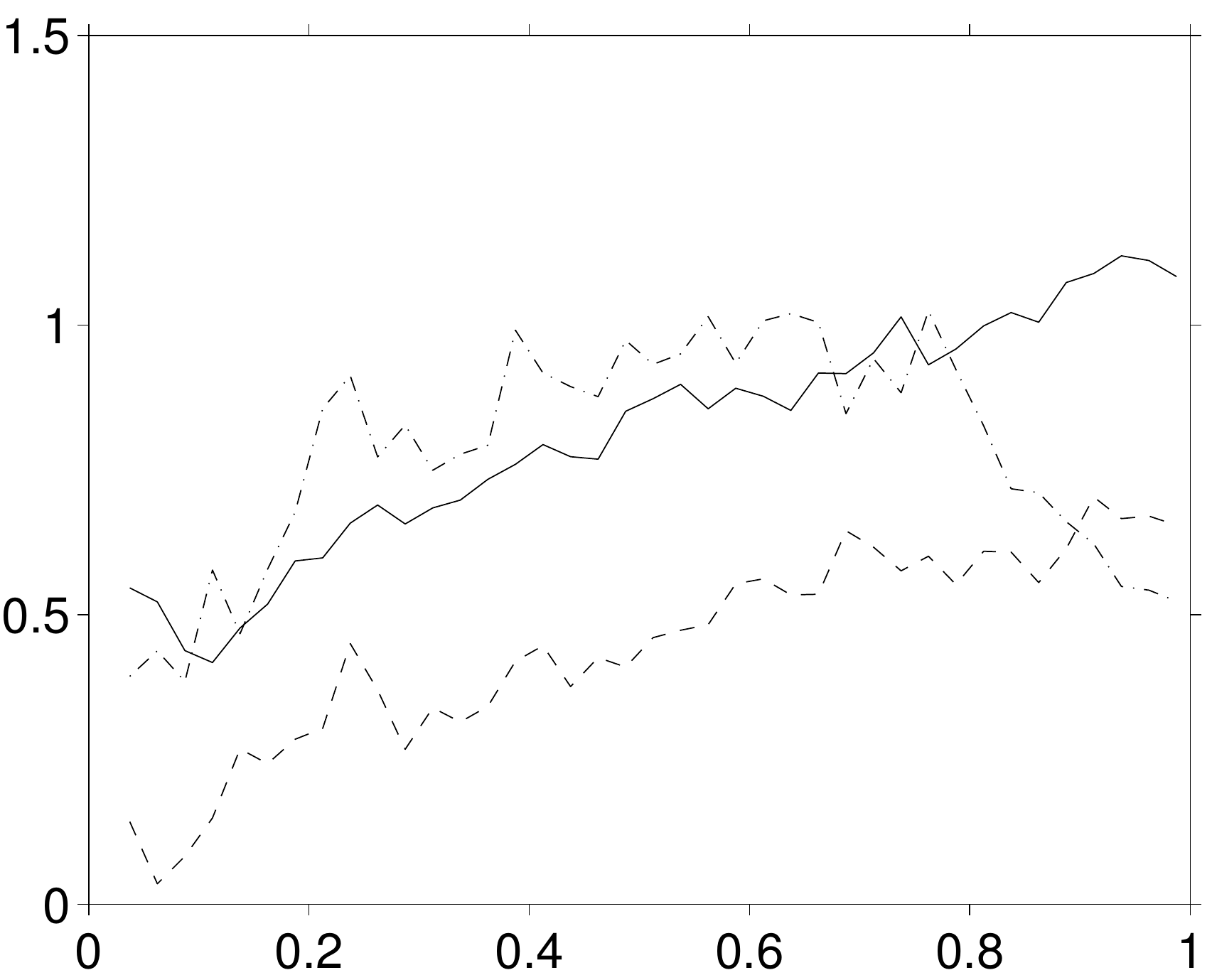}
    \centerline{$y/h$}
  \end{minipage}
  \caption{Integral time scales (in bulk units) computed from the
    Lagrangian particle velocity auto-correlations as a function of
    the particle's wall-distance at the initial time, case B. 
    \solid~$0.1\cdot T_{Lp,1}u_b/h$; 
    \dashed~$T_{Lp,2}u_b/h$;
    \chndot~$T_{Lp,3}u_b/h$. 
    Please note the scaling factor applied to the data of the streamwise
    component. 
    }
  \label{fig-results-lag-corr-integral-scale-y}
\end{figure}
\clearpage
\begin{figure}
  \centering
  \begin{minipage}{5ex}
    $\displaystyle\frac{{d}_{min}}{{d}_{min}^{hom}}$
  \end{minipage}
  \begin{minipage}{.5\linewidth}
    \includegraphics[width=\linewidth]{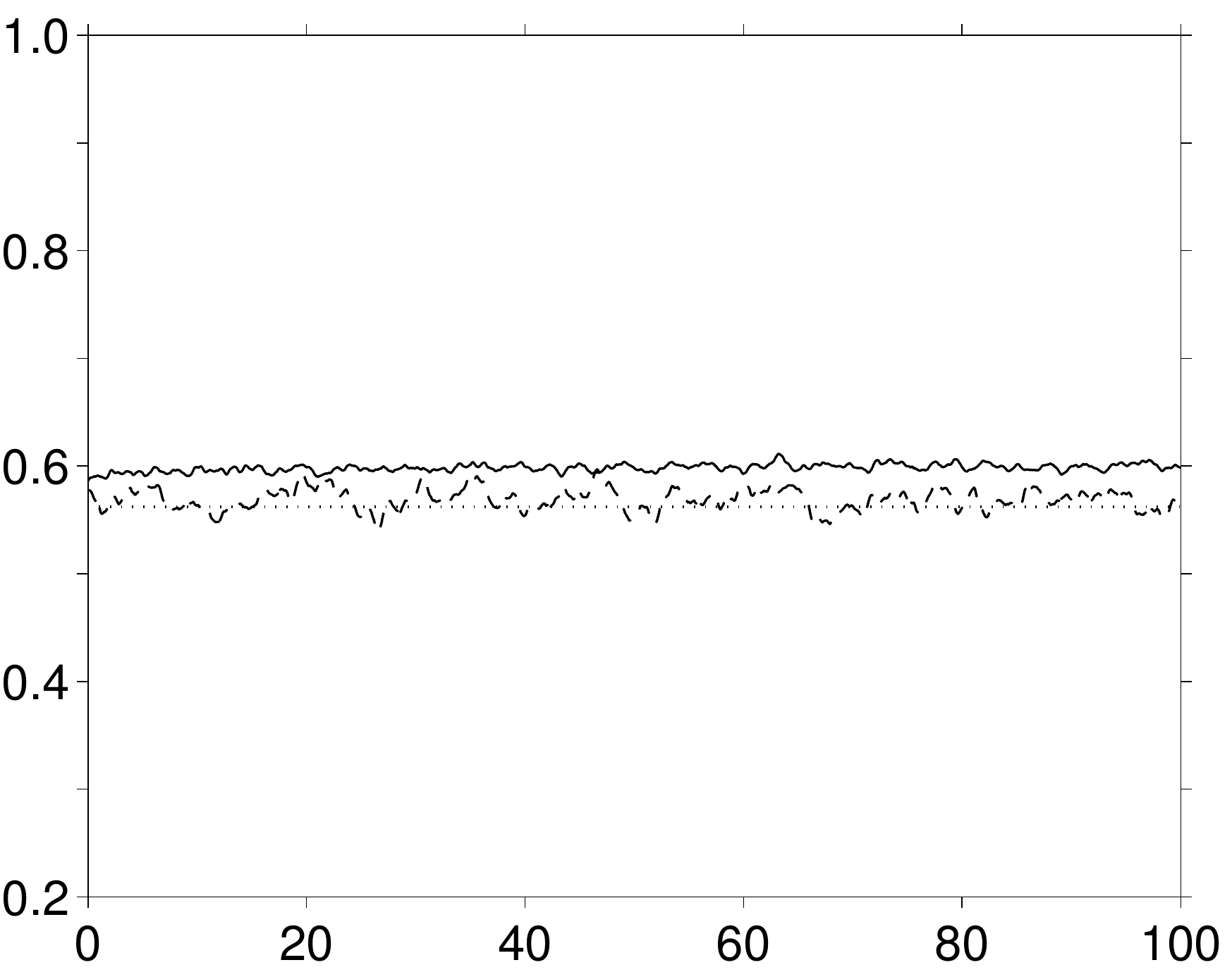}
    \centerline{$(t-t_1)u_b/h$}
  \end{minipage}
  \caption{Time evolution of the average distance to the nearest
    neighbor, normalized by the value for a homogeneous distribution. 
    Symbols as in figure~\ref{fig-results-up}; $t_1$ marks the
    beginning of the observation interval. 
    The dashed line corresponds to the value for a random distribution
    with the same solid volume fraction (computed for 524288 particles
    and correspondingly streamwise extended domain).}
  \label{fig-results-dist}
\end{figure}
\begin{figure}
  \centering
  \rotatebox{90}{$pdf$}
  \begin{minipage}{.5\linewidth}
    \centerline{$(a)$}
    \includegraphics[width=\linewidth]{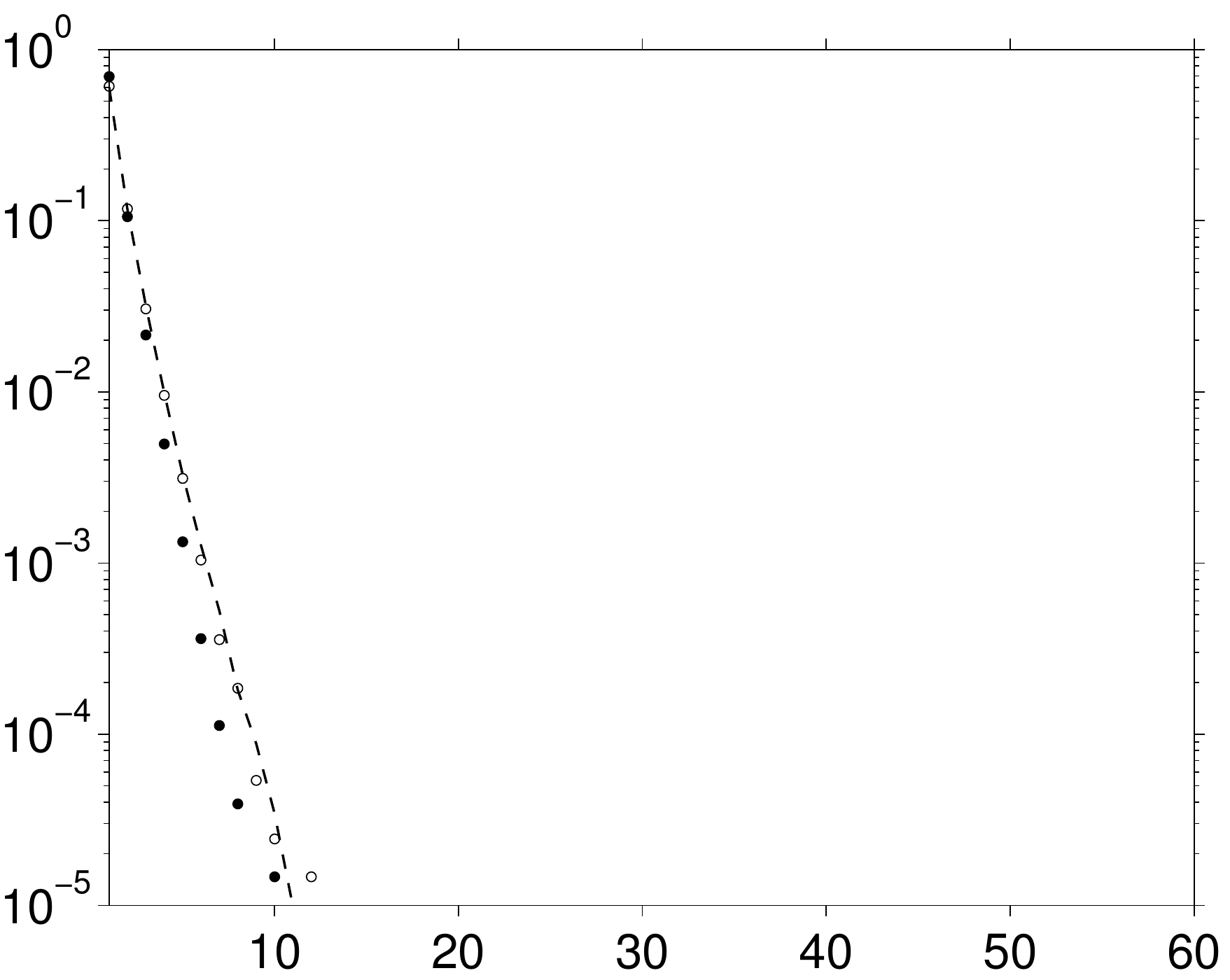}
    \centerline{$n_c$}
  \end{minipage}
  \\[1ex]
  \rotatebox{90}{$pdf$}
  \begin{minipage}{.5\linewidth}
    \centerline{$(b)$}
    \includegraphics[width=\linewidth]{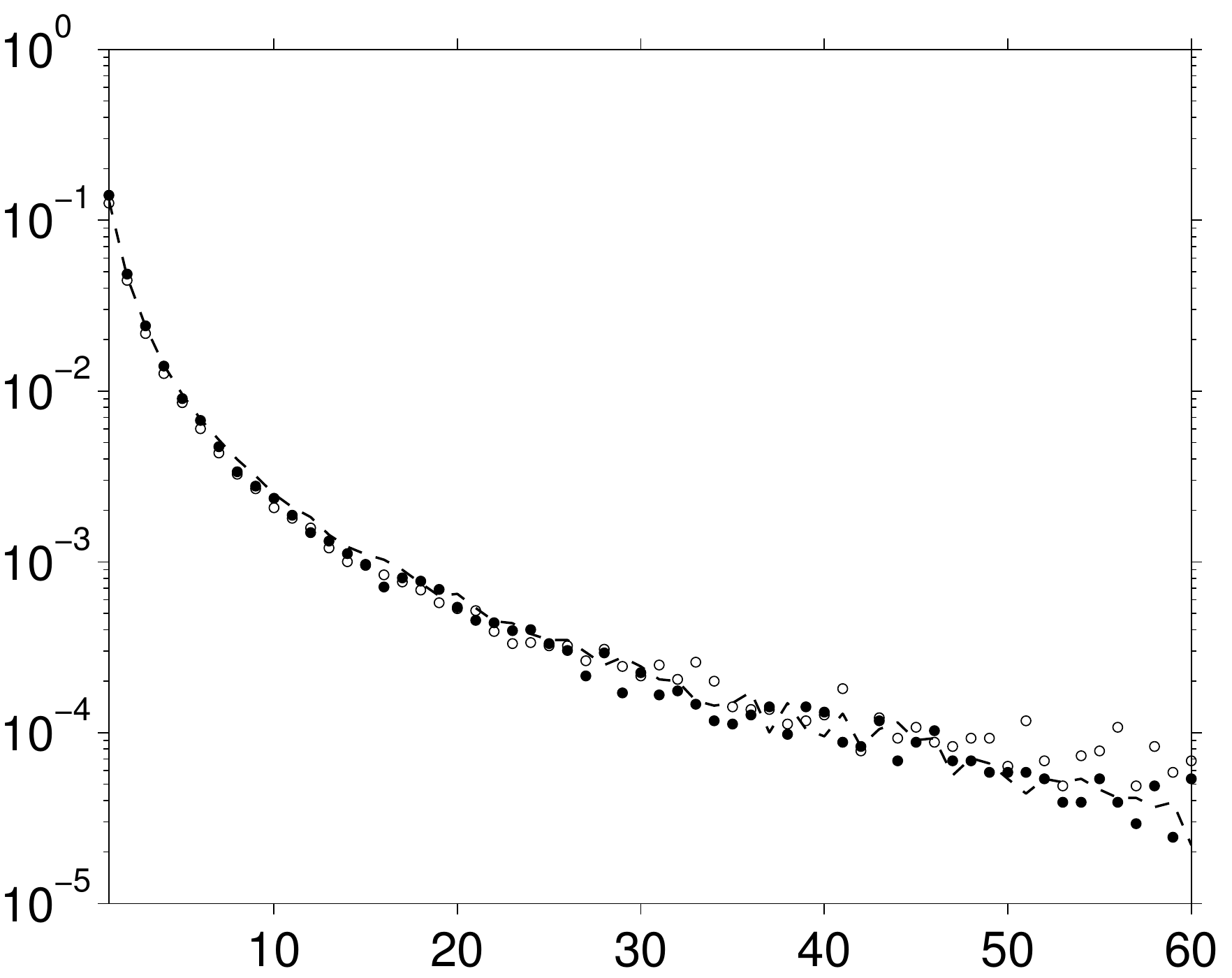}
    \centerline{$n_c$}
  \end{minipage}
  \\[1ex]
  \rotatebox{90}{$pdf$}
  \begin{minipage}{.5\linewidth}
    \centerline{$(c)$}
    \includegraphics[width=\linewidth]{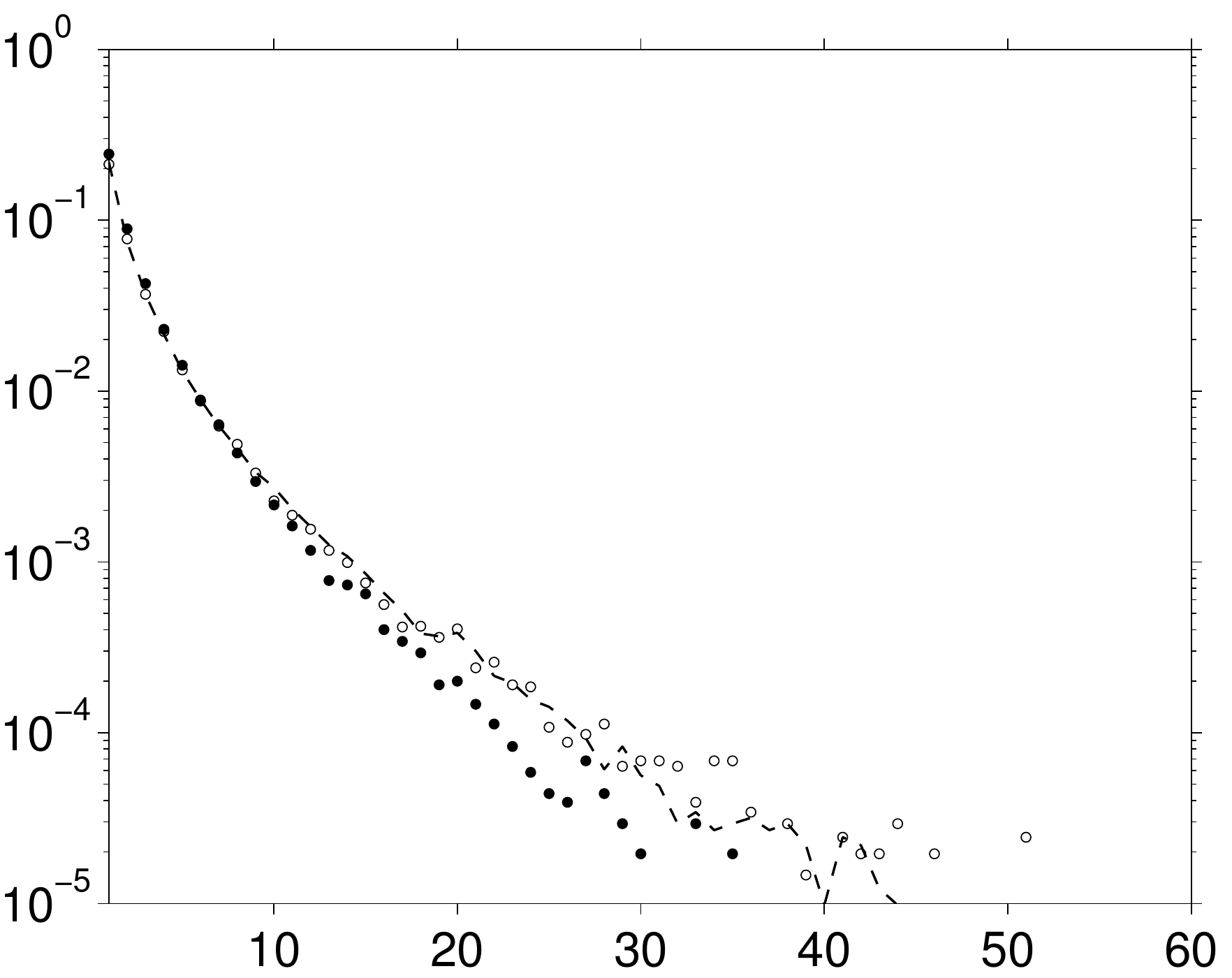}
    \centerline{$n_c$}
  \end{minipage}
  \caption{
    Probability density function of the occurrence of particle
    clusters 
    with $n_c$ members in cases A ($\circ$) and B ($\bullet$). The
    graphs correspond to different cut-off lengths: $(a)$ spherical
    cut-off $l_c=2.5D$; $(b)$ spherical cut-off $l_c=4D$; 
    $(c)$ elliptical cut-off with axes of length $l_{cx}=7.5D$,
    $l_{cy}=l_{cz}=2.5D$. The probability was evaluated from 400 (50)
    instantaneous particle distributions in case A (B), spanning the
    respective observation intervals. The dashed lines correspond to
    the probability for a random distribution of particles with the
    same solid volume fraction, evaluated from 100 realizations.}
  \label{fig-results-clusters}
\end{figure}
\begin{figure}
  \centering
  \begin{minipage}{4ex}
    {$pdf$}
  \end{minipage}
  \begin{minipage}{.5\linewidth}
    \centerline{$(a)$}
    \includegraphics[width=\linewidth]{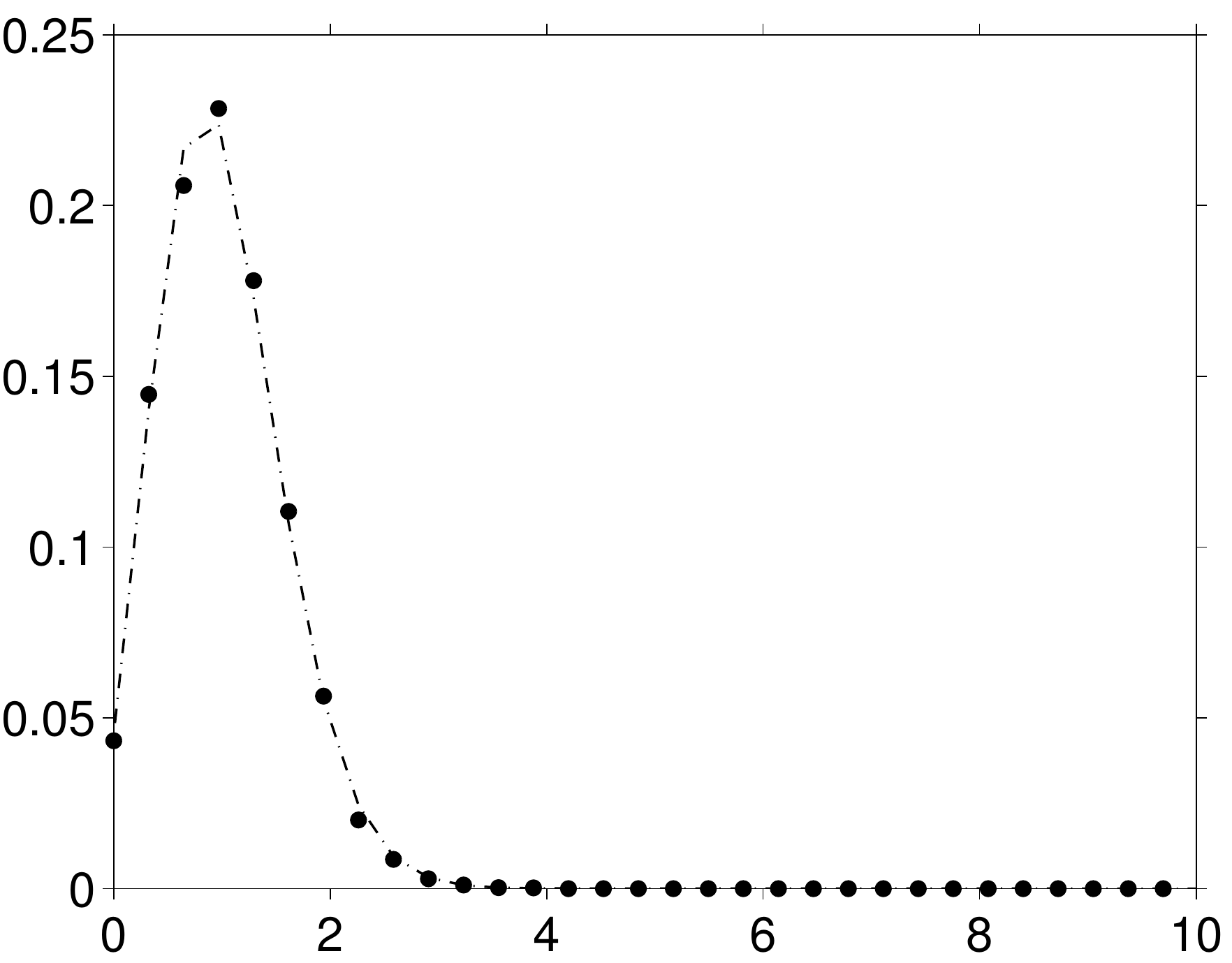}
    \centerline{$C/C_0$}
  \end{minipage}
  \\[1ex]
  \begin{minipage}{4ex}
    {$D_2$}
  \end{minipage}
  \begin{minipage}{.5\linewidth}
    \centerline{$(b)$}
    \includegraphics[width=\linewidth]
    {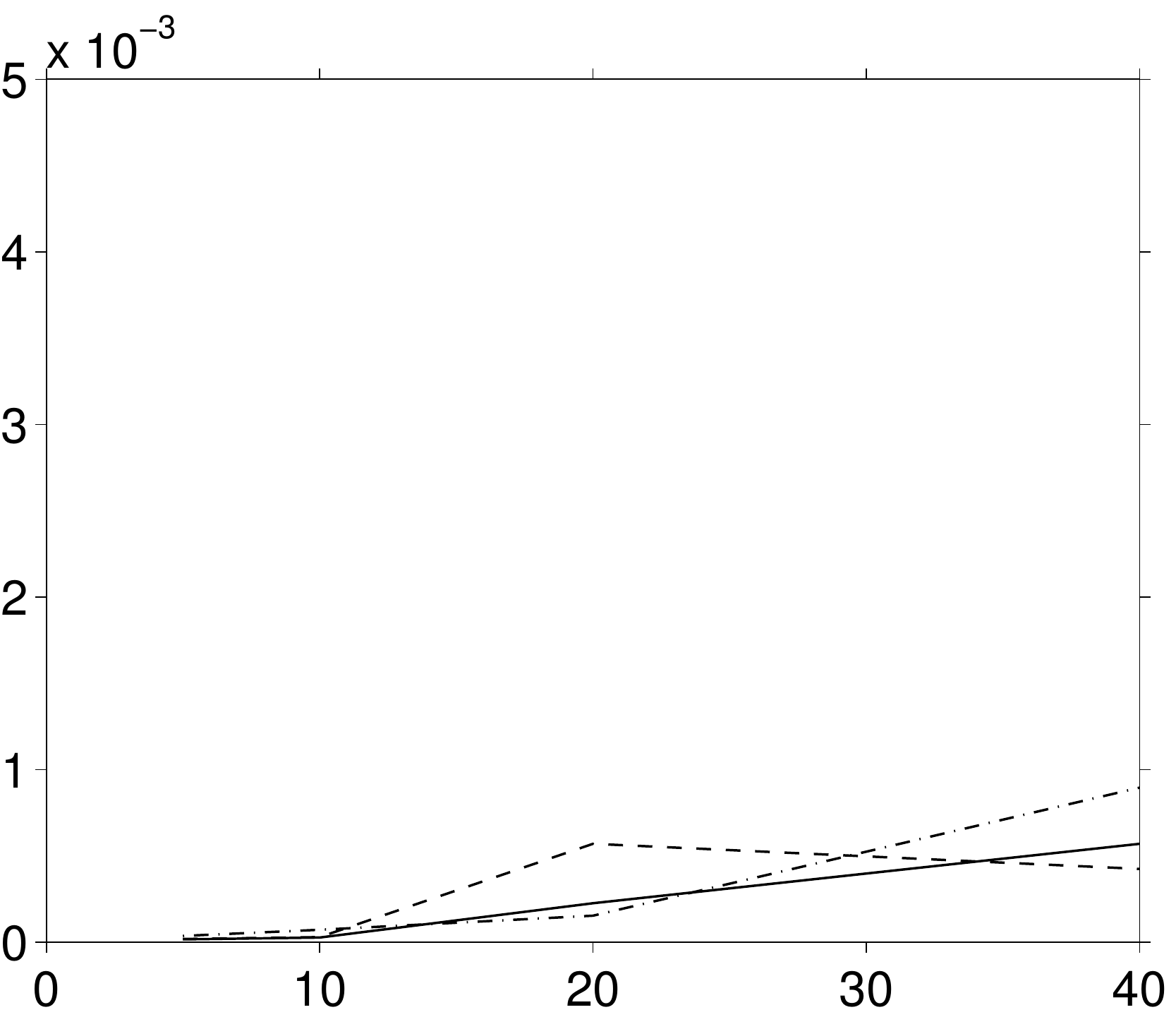}
    \centerline{$\mbox{box size}/D$}
  \end{minipage}
  \caption{Local concentration of particles in case B. $(a)$
    Probability density function of the number of 
    particles per box of size $20D\times D\times 20D$ at a wall
    distance of $y^+=28$: DNS results
    ($\bullet$) versus Poisson distribution (\chndot); $C_0$ is the
    average number of particles per box in the current y-slab. 
    $(b)$ The deviation between the pdf of the DNS data and
    the Poisson distribution according to the definition
    (\ref{equ-results-disp-diff-pdf}), 
    plotted as a function of the linear box 
    dimension (in the directions of the $x$ and $z$ coordinate): 
    \solid~$y^+=28$; \dashed~$y^+=73$; \chndot~$y^+=220$.
  }
  \label{fig-results-local-concentr}
\end{figure}
\clearpage
\begin{figure}
  \centering
  \begin{minipage}{3ex}
    $\displaystyle\frac{x}{h}$
  \end{minipage}
  \begin{minipage}{.47\linewidth}
    \centerline{$(a)$}
    \ifpdf
    \includegraphics*[width=\linewidth,viewport=260 235 905 1500]
    {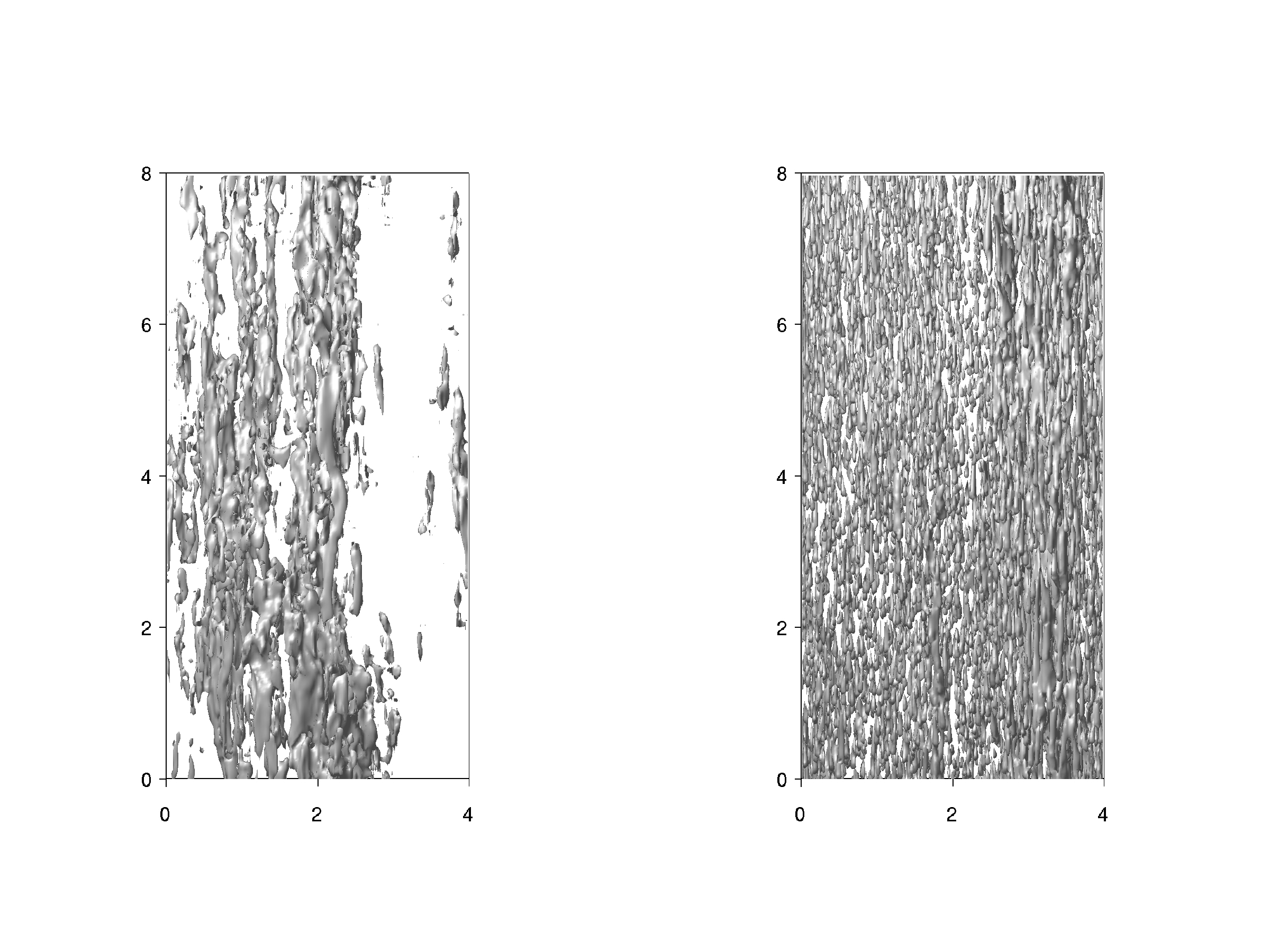}
    \fi
    \centerline{$z/h$}
  \end{minipage}\hfill
  \begin{minipage}{.47\linewidth}
    \centerline{$(b)$}
    \ifpdf
    \includegraphics*[width=\linewidth,viewport=1455 235 2100 1500]
    {matlab/channelp12_uflucx_xz_full_12_3_6utau.jpg}
    \fi
    \centerline{$z/h$}
  \end{minipage}
  \\[1ex]
  \begin{minipage}{3ex}
    $\displaystyle\frac{y}{h}$
  \end{minipage}
  \begin{minipage}{.47\linewidth}
    \centerline{$(c)$}
    \ifpdf  
    \hfill
    \includegraphics*[width=.93\linewidth,viewport=155 320 2270 1480]
    {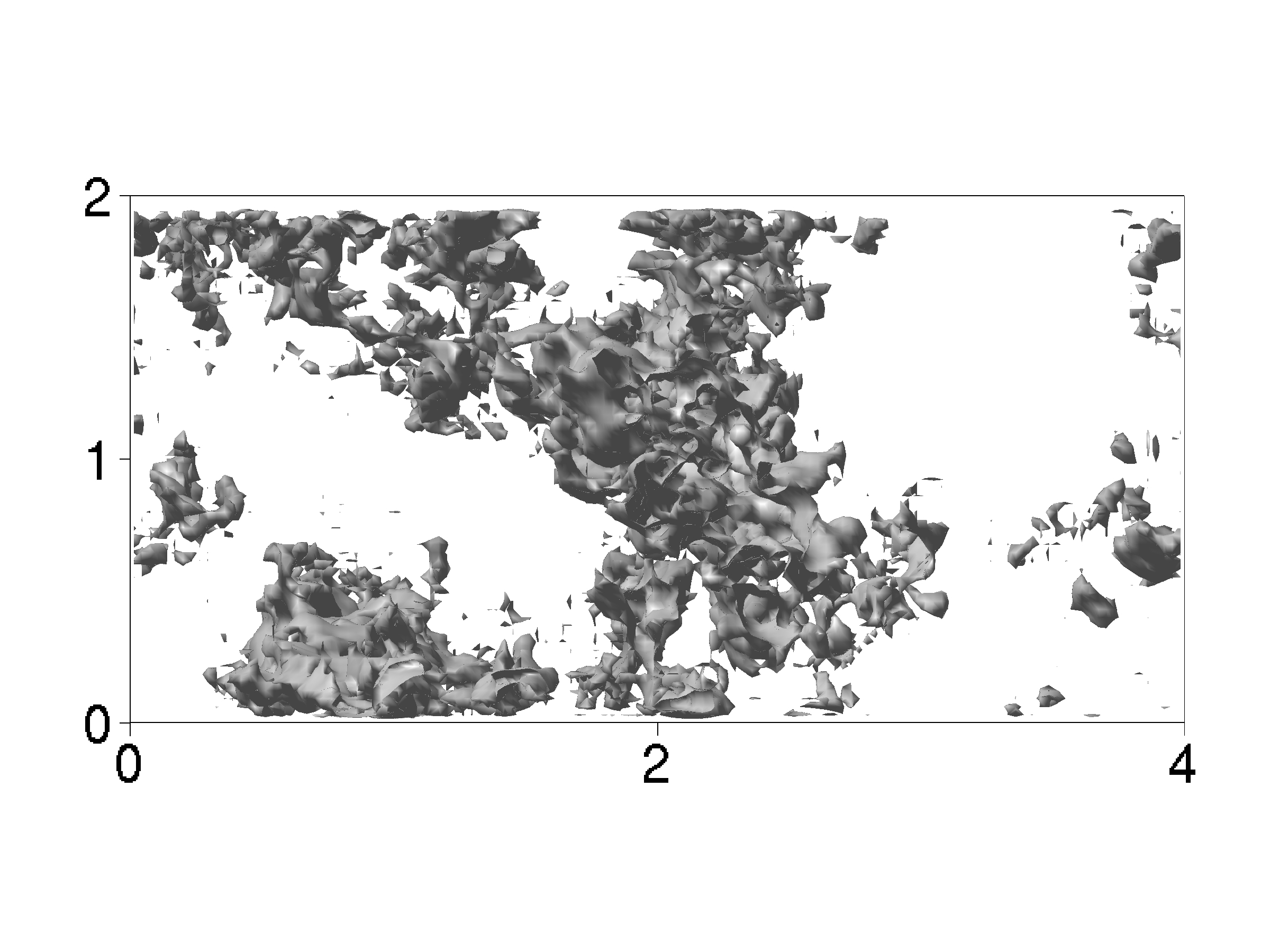}
    \fi
    \centerline{$z/h$}
  \end{minipage}
  \begin{minipage}{.47\linewidth}
    \centerline{$(d)$}
    \ifpdf
    \hfill
    \includegraphics*[width=.93\linewidth,viewport=155 320 2270 1480]
    {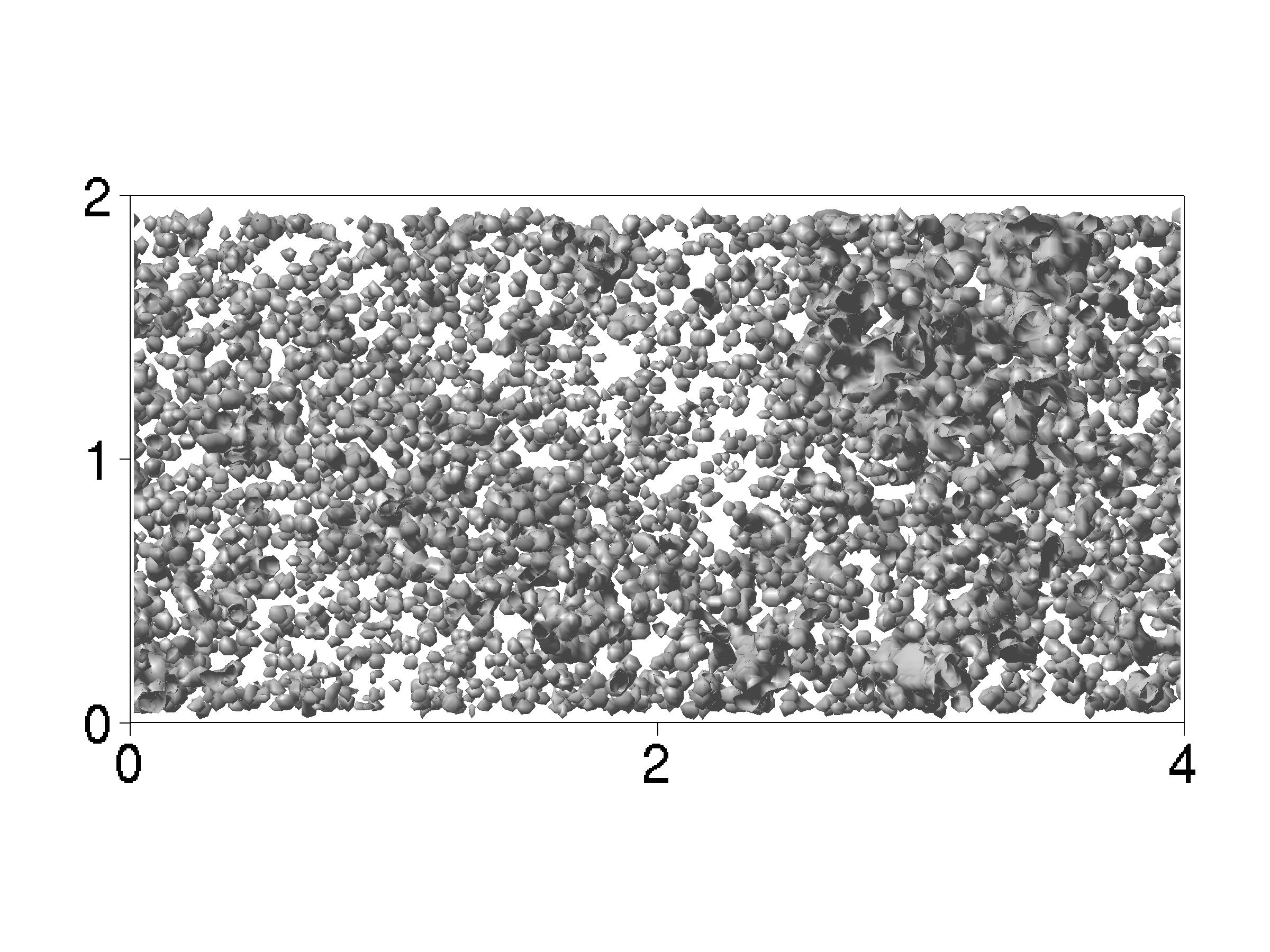}
    \fi
    \centerline{$z/h$}
  \end{minipage}
  \caption{Instantaneous three-dimensional isosurfaces of streamwise
    velocity fluctuations  $u^\prime=\pm3.6u_\tau$ (equivalent to
    $\pm0.3u_b$) of case B. The
    graphs $(a,c)$ show surfaces corresponding to positive values,
    $(b,d)$ show negative-valued surfaces. The view in $(a,b)$ is into
    the wall, in $(c,d)$ it is in the mean flow direction. 
    Please note that only every eighth grid point in each direction
    was used. 
    {Here and in figures~\ref{fig-results-5-fluid-sequence},
      \ref{fig-results-5-particle-snapshot},
      \ref{fig-results-5-particle-snapshot-neutral} the full
      computational domain is shown.}
  } 
  \label{fig-results-5-fluid-snapshot}
\end{figure}
\begin{figure}
  \centering
  \begin{minipage}{3ex}
    $\displaystyle\frac{x}{h}$
  \end{minipage}
  \begin{minipage}{.47\linewidth}
    \centerline{$(a)$}
    \ifpdf
    \includegraphics*[width=\linewidth,viewport=1300 400 2230 1340]
    {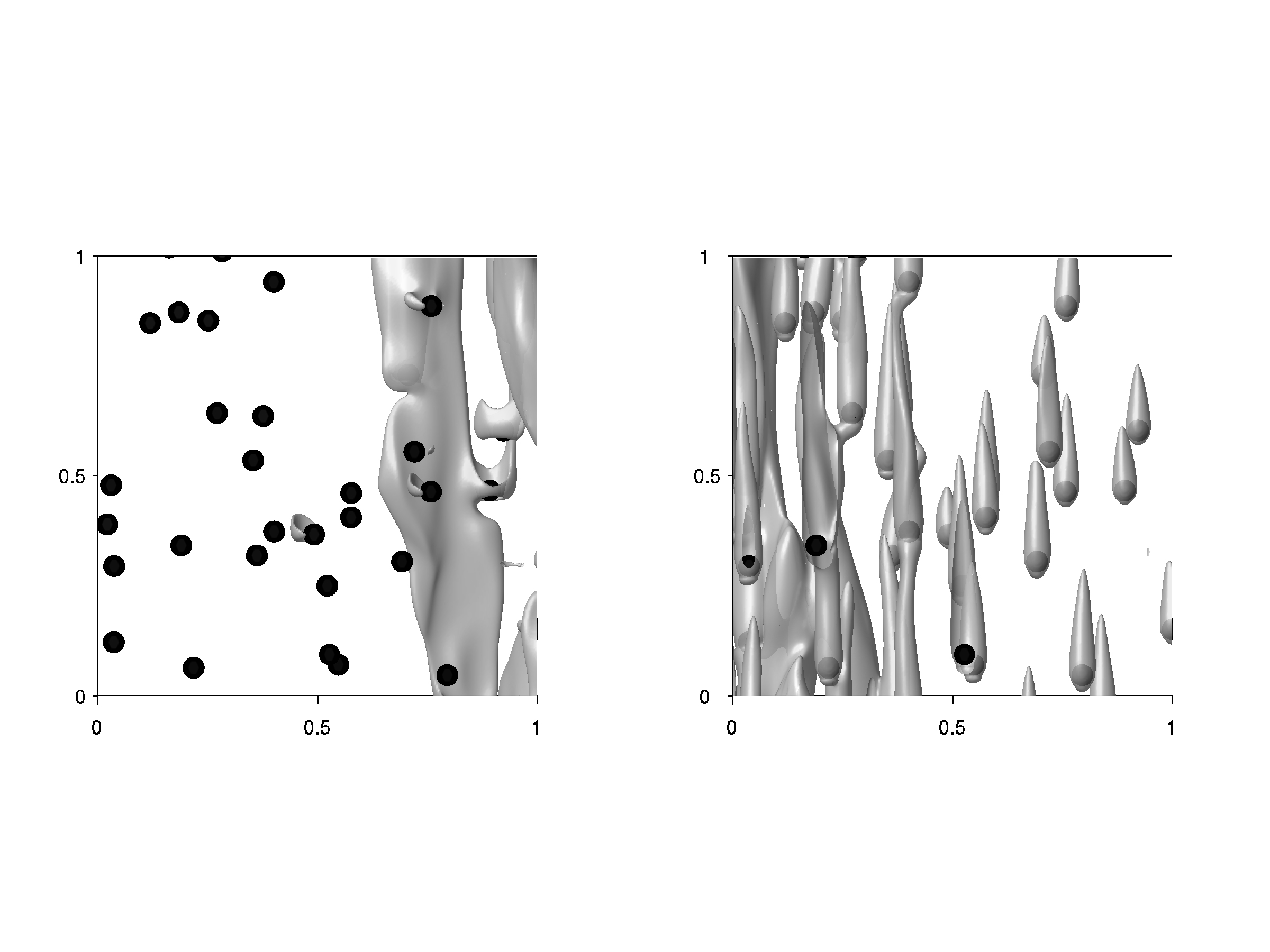}
    \fi
    \centerline{$z/h$}
  \end{minipage}\\
  \begin{minipage}{3ex}
    $\displaystyle\frac{y}{h}$
  \end{minipage}
  \begin{minipage}{.47\linewidth}
    \centerline{$(b)$}
    \ifpdf
    \includegraphics*[width=\linewidth,viewport=500 500 1870 1270]
    {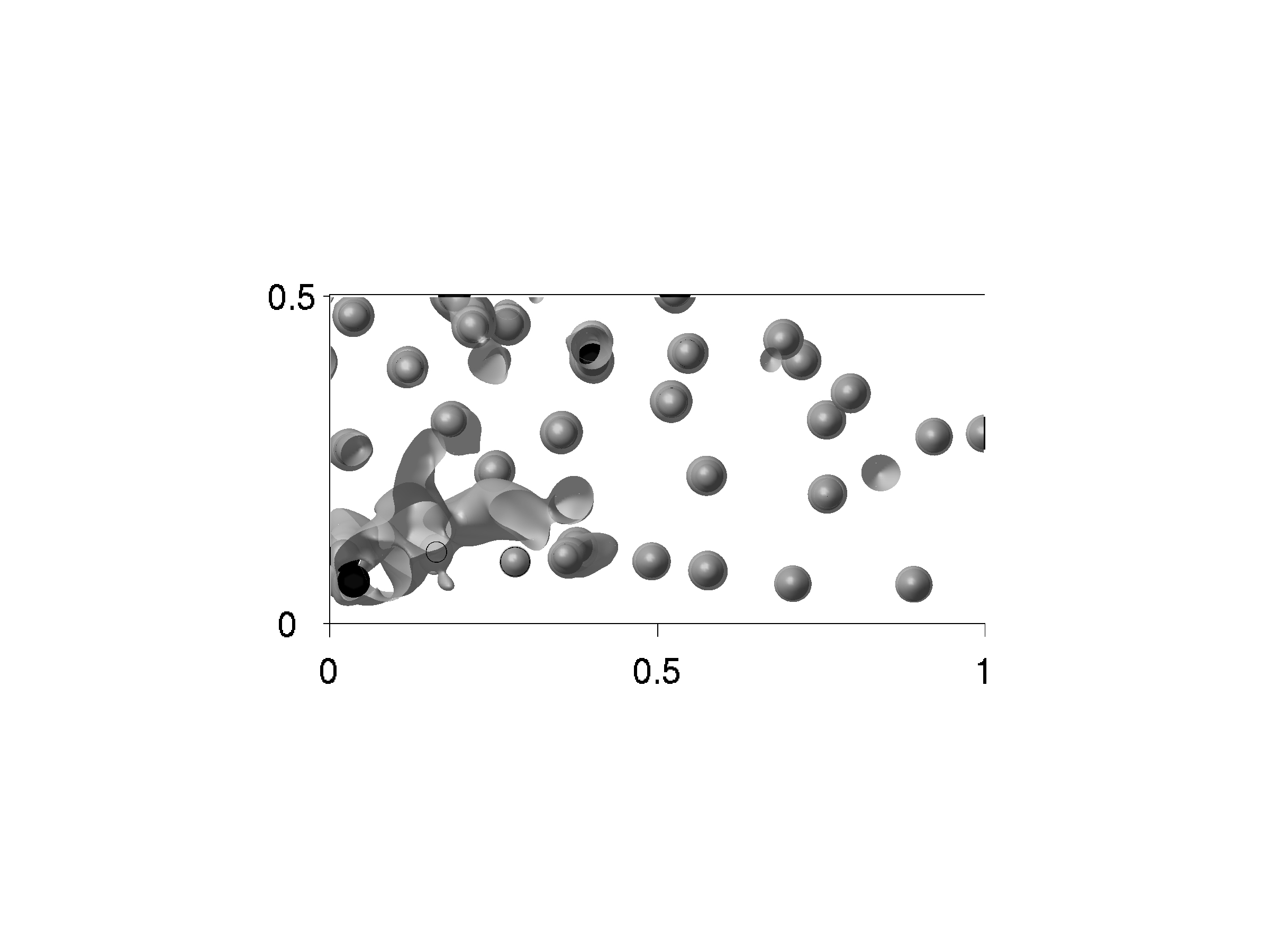}
    \fi
    \centerline{$z/h$}
  \end{minipage}
  \caption{
    {The same negative-valued isosurfaces as in
      figure~\ref{fig-results-5-fluid-snapshot}$(b,d)$, but shown in 
      the sub-domain $[0,h]\times[0,0.5h]\times[0,h]$. In addition,
      the particles are indicated by black spheres.}
  } 
  \label{fig-results-5-fluid-snapshot-zoom}
\end{figure}
\begin{figure}
  \centering
  \begin{minipage}{3ex}
    $\displaystyle\frac{y}{h}$
  \end{minipage}
  \begin{minipage}{.47\linewidth}
    \centerline{$(a)$}
    \ifpdf  
    \hfill
    \includegraphics*[width=.93\linewidth,viewport=155 385 2270 1480]
    {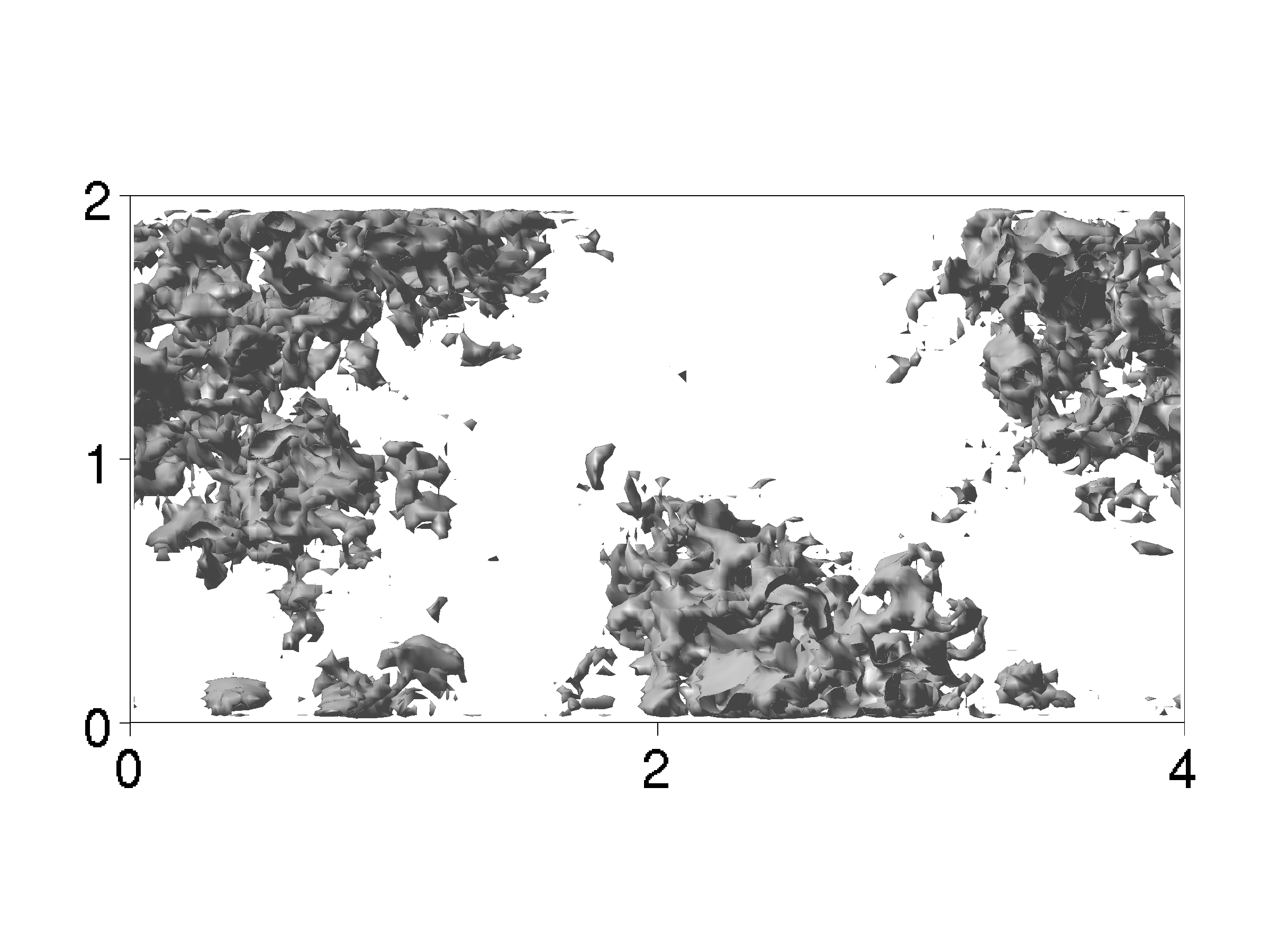}
    \fi
  \end{minipage}
  \begin{minipage}{.47\linewidth}
    \centerline{$(b)$}
    \ifpdf
    \hfill
    \includegraphics*[width=.93\linewidth,viewport=155 385 2270 1480]
    {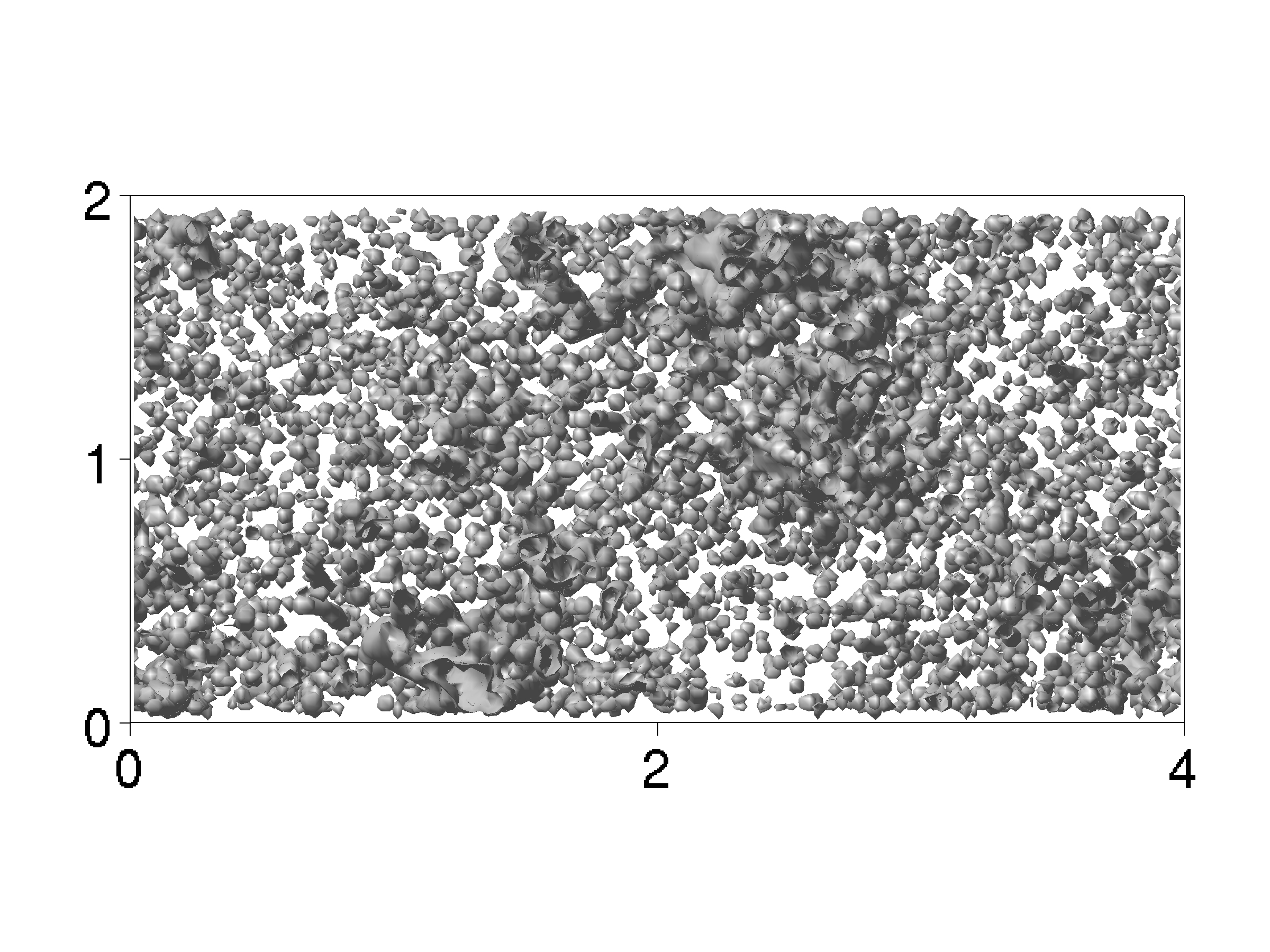}
    \fi
  \end{minipage}
  \\
  \begin{minipage}{3ex}
    $\displaystyle\frac{y}{h}$
  \end{minipage}
  \begin{minipage}{.47\linewidth}
    \ifpdf  
    \hfill
    \includegraphics*[width=.93\linewidth,viewport=155 385 2270 1480]
    {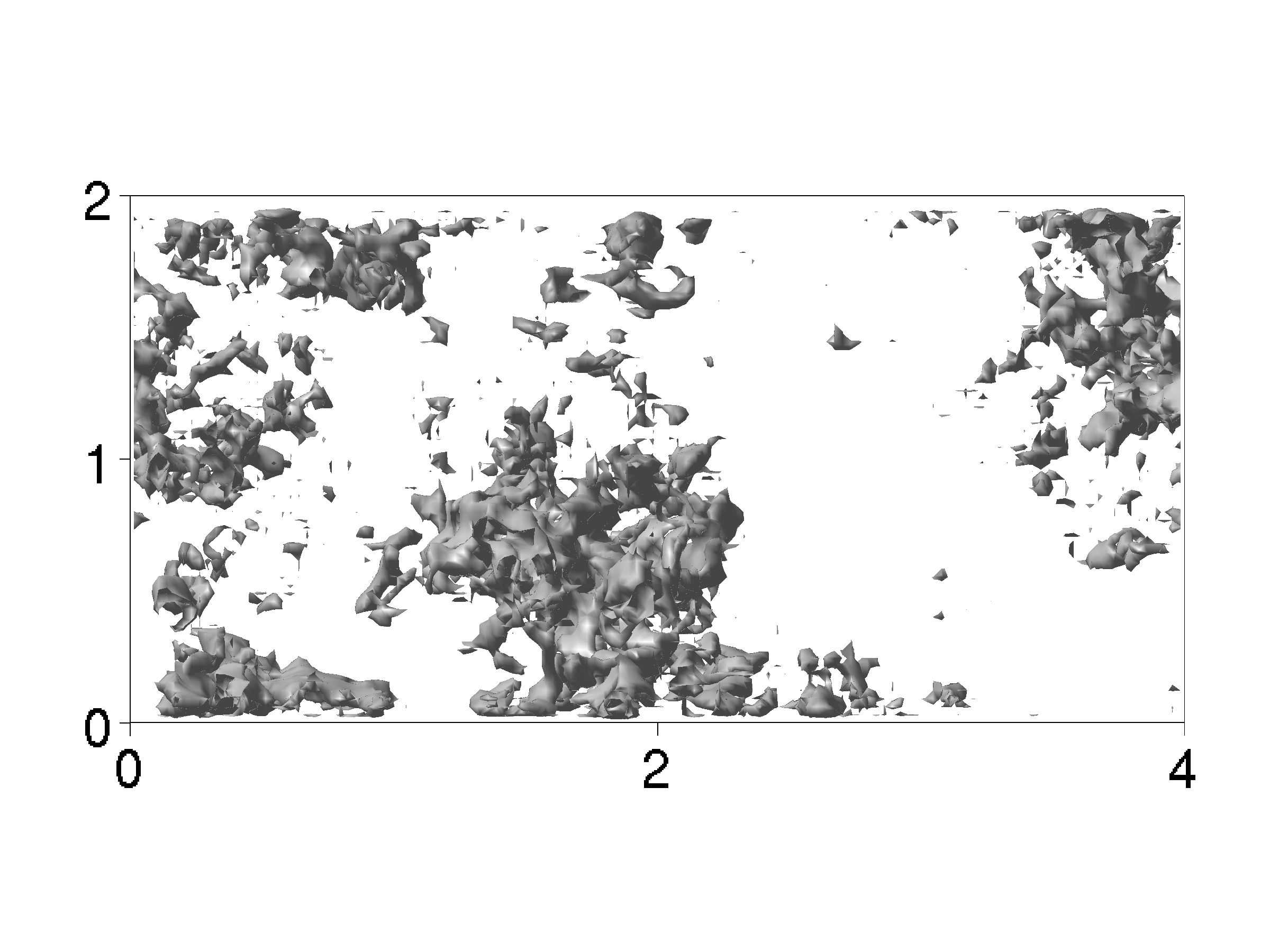}
    \fi
  \end{minipage}
  \begin{minipage}{.47\linewidth}
    \ifpdf
    \hfill
    \includegraphics*[width=.93\linewidth,viewport=155 385 2270 1480]
    {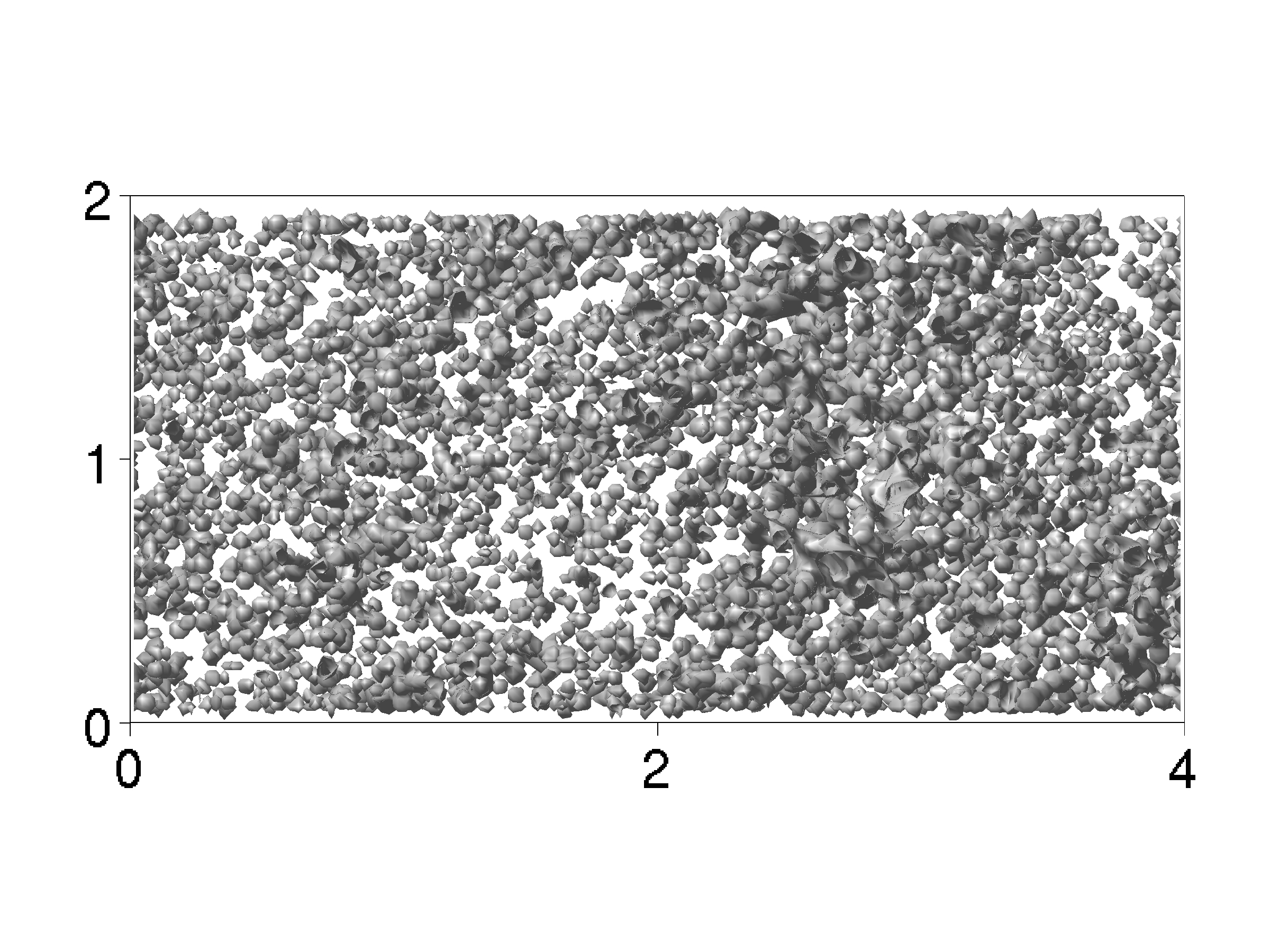}
    \fi
  \end{minipage}
  \\
  \begin{minipage}{3ex}
    $\displaystyle\frac{y}{h}$
  \end{minipage}
  \begin{minipage}{.47\linewidth}
    \ifpdf  
    \hfill
    \includegraphics*[width=.93\linewidth,viewport=155 385 2270 1480]
    {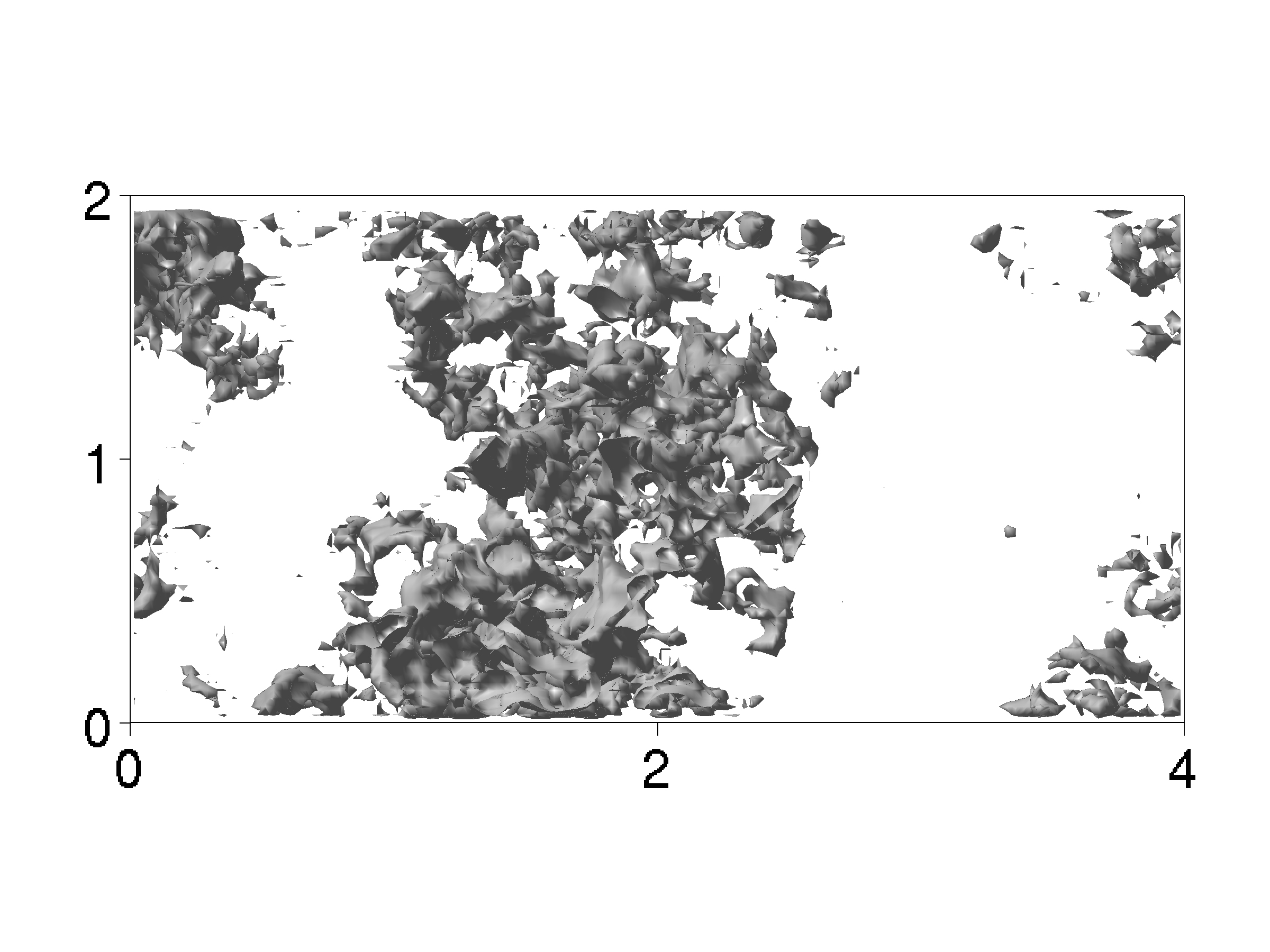}
    \fi
  \end{minipage}
  \begin{minipage}{.47\linewidth}
    \ifpdf
    \hfill
    \includegraphics*[width=.93\linewidth,viewport=155 385 2270 1480]
    {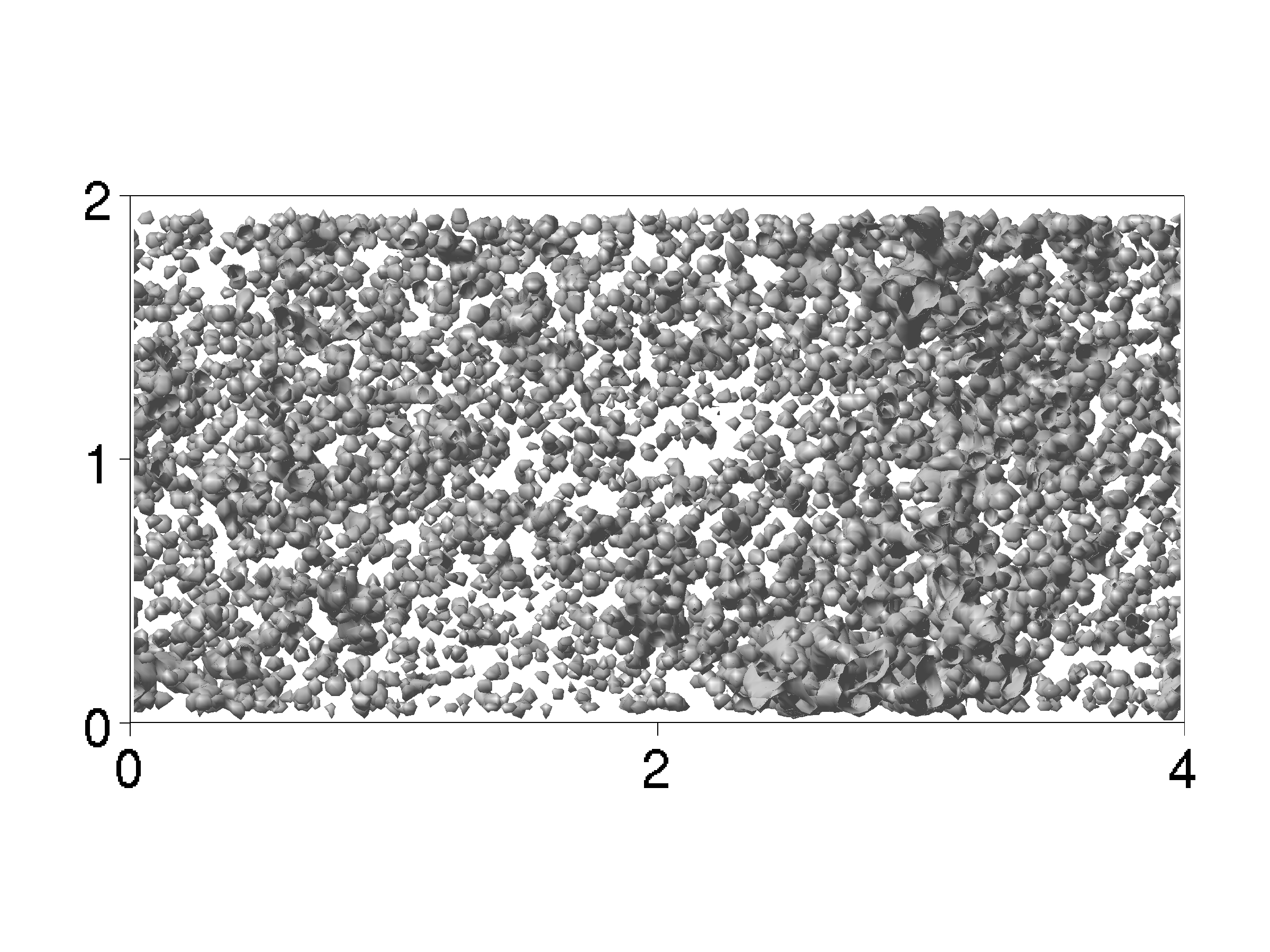}
    \fi
  \end{minipage}
  \\
  \begin{minipage}{3ex}
    $\displaystyle\frac{y}{h}$
  \end{minipage}
  \begin{minipage}{.47\linewidth}
    \ifpdf  
    \hfill
    \includegraphics*[width=.93\linewidth,viewport=155 310 2270 1480]
    {matlab/channelp12_uflucx_yz_pos_full_12_3_6utau.jpg}
    \fi
    \centerline{$z/h$}
  \end{minipage}
  \begin{minipage}{.47\linewidth}
    \ifpdf
    \hfill
    \includegraphics*[width=.93\linewidth,viewport=155 310 2270 1480]
    {matlab/channelp12_uflucx_yz_neg_full_12_3_6utau.jpg}
    \fi
    \centerline{$z/h$}
  \end{minipage}
  \caption{Isosurfaces of $u^\prime=\pm3.6u_\tau$ in the cross-stream
    plane for successive snapshots with time increasing from top to
    bottom by increments of approximately $14.9h/u_b$. 
    Positive-valued surfaces are shown in $(a)$, negative ones in
    $(b)$. 
    The plots in the last row are identical to those in
    figure~\ref{fig-results-5-fluid-snapshot}$(c,d)$.} 
  \label{fig-results-5-fluid-sequence}
\end{figure}
\begin{figure}
  \centering
  \begin{minipage}{3ex}
    $\displaystyle\frac{x}{h}$
  \end{minipage}
  \begin{minipage}{.47\linewidth}
    \centerline{$(a)$}
    \ifpdf
    \includegraphics*[width=\linewidth,viewport=240 200 920 1530]
    {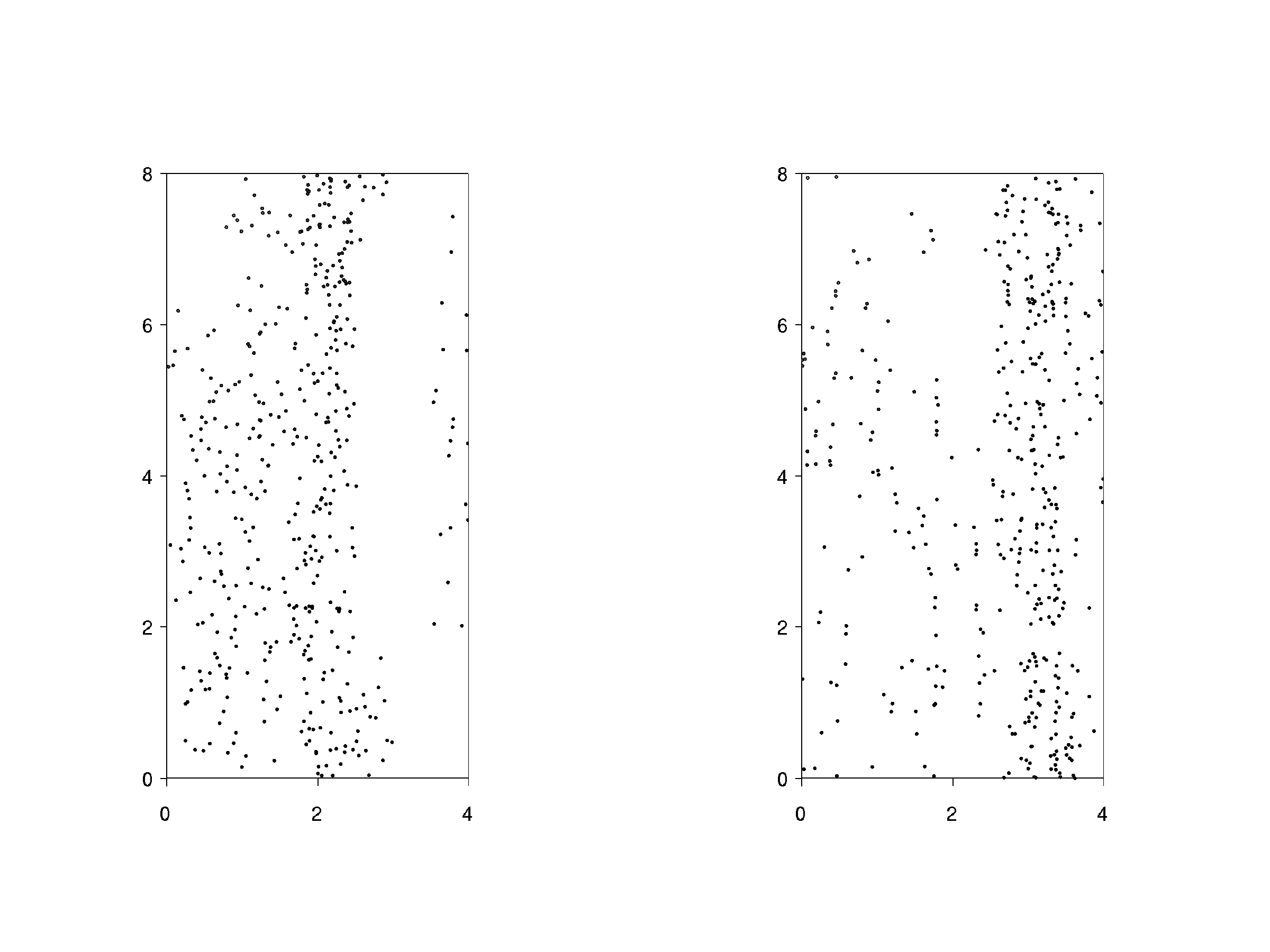}
    \fi
    \centerline{$z/h$}
  \end{minipage}\hfill
  \begin{minipage}{.47\linewidth}
    \centerline{$(b)$}
    \ifpdf
    \includegraphics*[width=\linewidth,viewport=1435 200 2115 1530]
    {matlab/channelp12_particles_xz_full_12.jpg}
    \fi
    \centerline{$z/h$}
  \end{minipage}
  \\[1ex]    
  \hspace*{.5ex}
  \begin{minipage}{3ex}
    $\displaystyle\frac{y}{h}$
  \end{minipage}
  \begin{minipage}{.47\linewidth}
    \centerline{$(c)$}
    \ifpdf
    \includegraphics*[width=.9\linewidth,viewport=155 320 2265 1470]
    {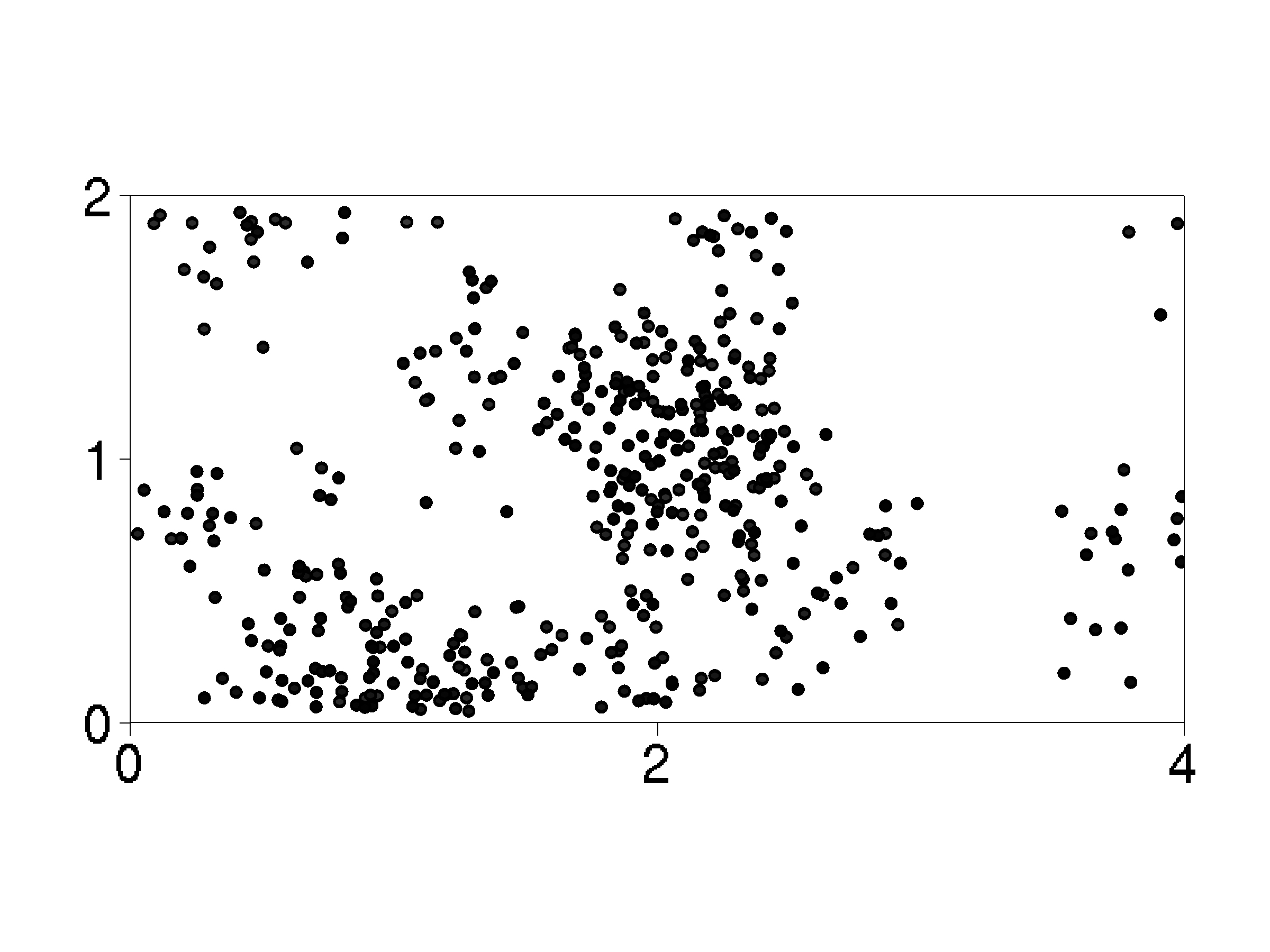}
    \fi
    \centerline{$z/h$}
  \end{minipage}
  \begin{minipage}{.47\linewidth}
    \centerline{$(d)$}
    \ifpdf
    \hfill
    \includegraphics*[width=.9\linewidth,viewport=155 320 2265 1470]
    {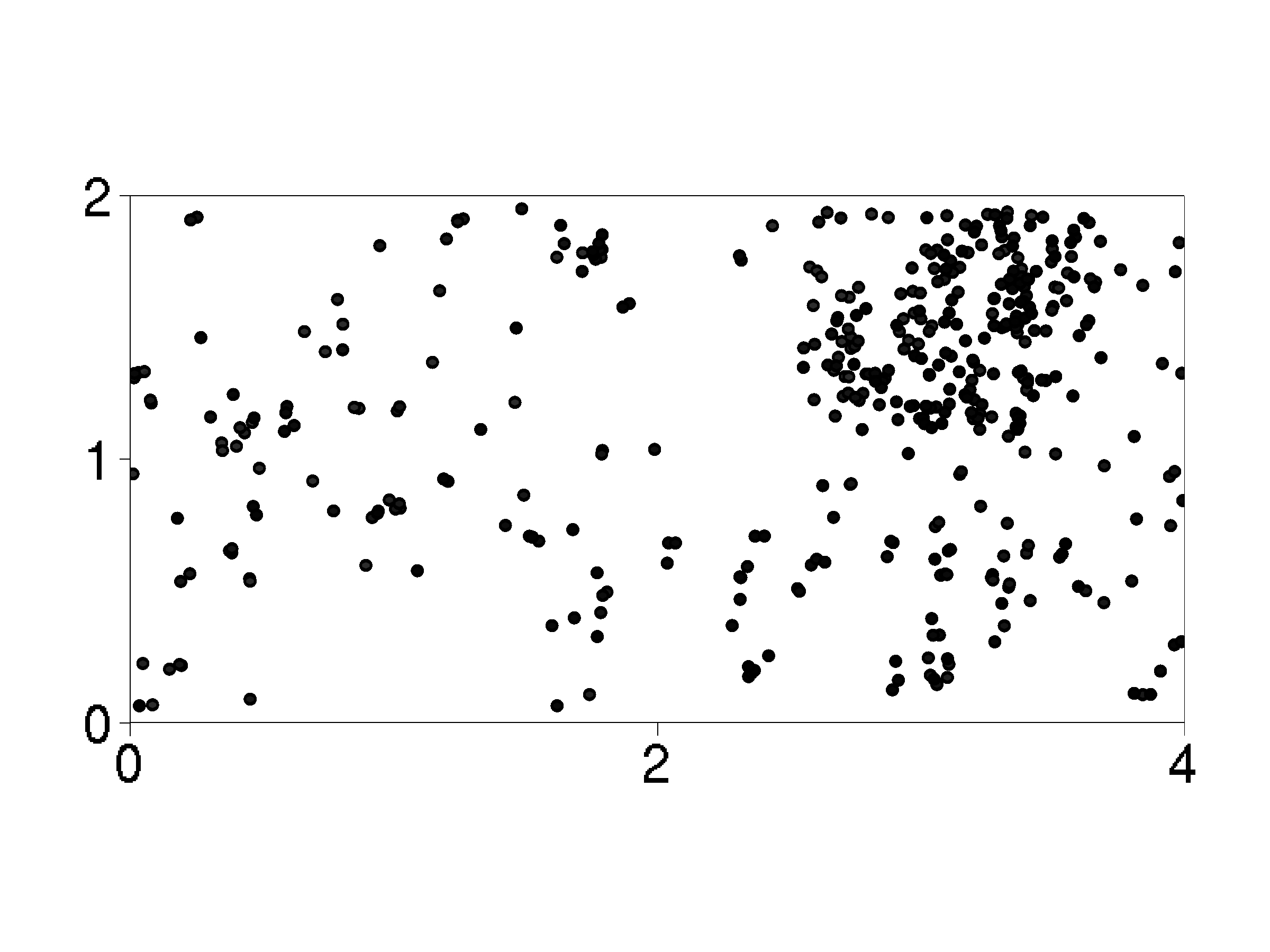}
    \fi
    \centerline{$z/h$}
  \end{minipage}
  \caption{Instantaneous positions of particles sorted according to
    their streamwise velocity fluctuation: 
    $(a)$ $u_c^\prime\geq +0.2u_b$; 
    $(b)$ $u_c^\prime\leq -0.2u_b$. The field corresponds to the one
    shown in figure~\ref{fig-results-5-fluid-snapshot}. 
  } 
  \label{fig-results-5-particle-snapshot}
\end{figure}
\begin{figure}
  \centering
  \begin{minipage}{3ex}
    $\displaystyle\frac{x}{h}$
  \end{minipage}
  \begin{minipage}{.47\linewidth}
    \centerline{$(a)$}
    \ifpdf
    \includegraphics*[width=\linewidth,viewport=240 200 920 1530]
    {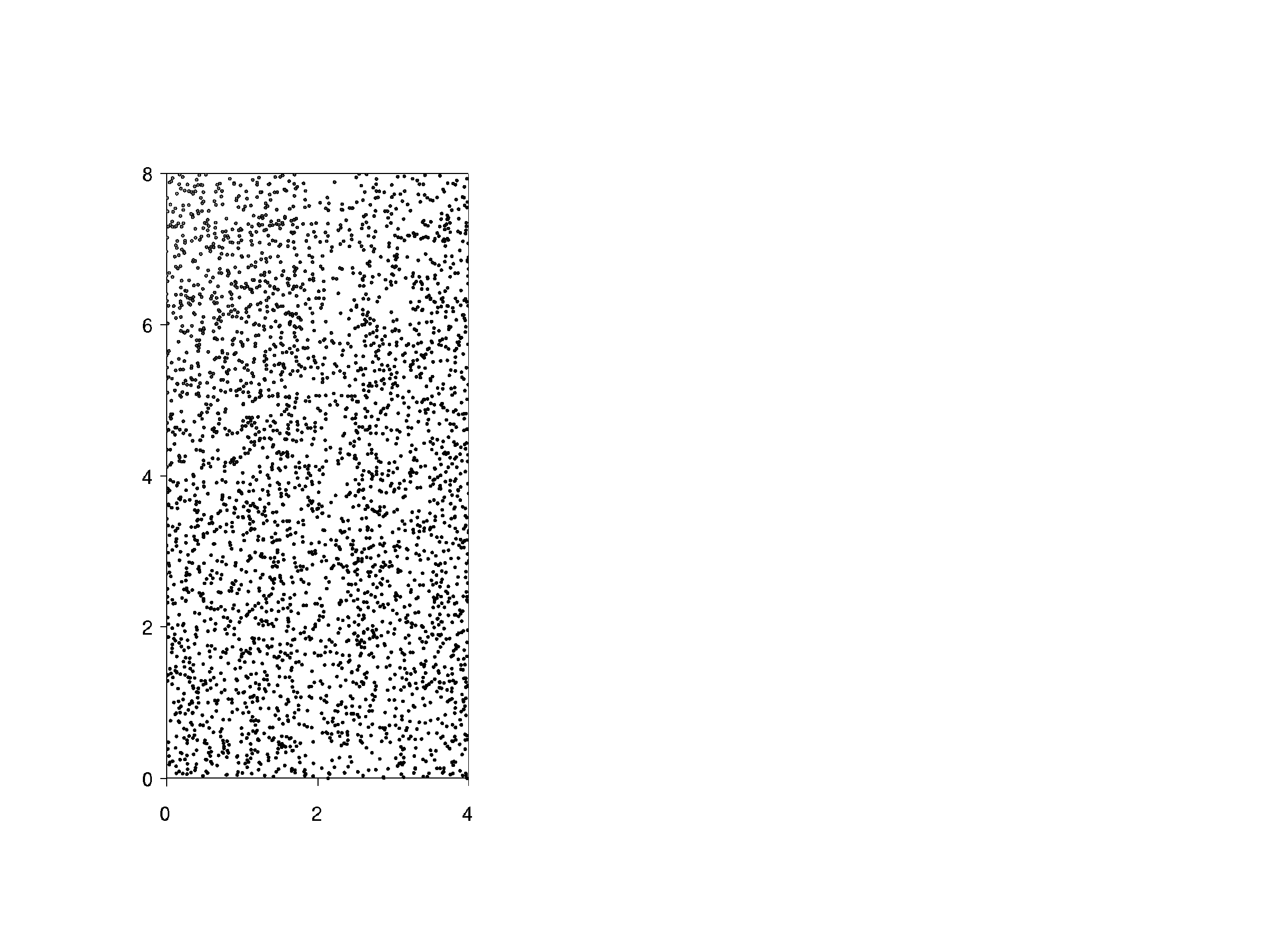}
    \fi
    \centerline{$z/h$}
  \end{minipage}
  \\[1ex]
  \begin{minipage}{3ex}
    $\displaystyle\frac{y}{h}$
  \end{minipage}
  \hspace*{4ex}
  \begin{minipage}{.47\linewidth}
    \centerline{$(b)$}
    \ifpdf
    \includegraphics*[width=.9\linewidth,viewport=155 320 2265 1470]
    {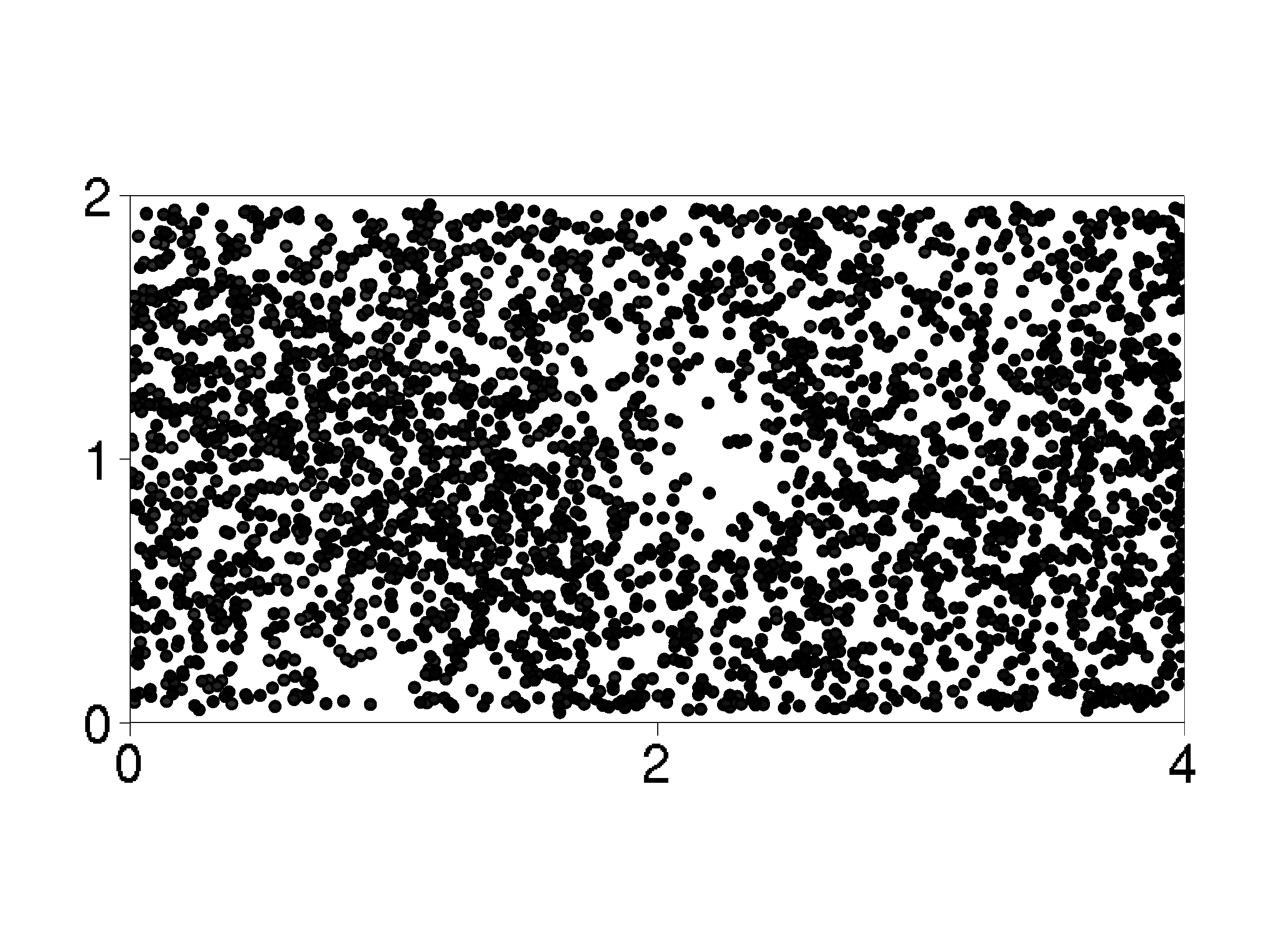}
    \fi
    \centerline{$z/h$}
  \end{minipage}
  \caption{As figure~\ref{fig-results-5-particle-snapshot}, but
    showing particles with a streamwise velocity close to the average
    value, i.e.\ for which $-0.2u_b<u_c^\prime<+0.2u_b$.
  } 
  \label{fig-results-5-particle-snapshot-neutral}
\end{figure}
\begin{figure}
  \centering
  \begin{minipage}{4ex}
    {$R_{uu}$}
  \end{minipage}
  \begin{minipage}{.5\linewidth}
    \centerline{$(a)$}
    \includegraphics[width=\linewidth]
    {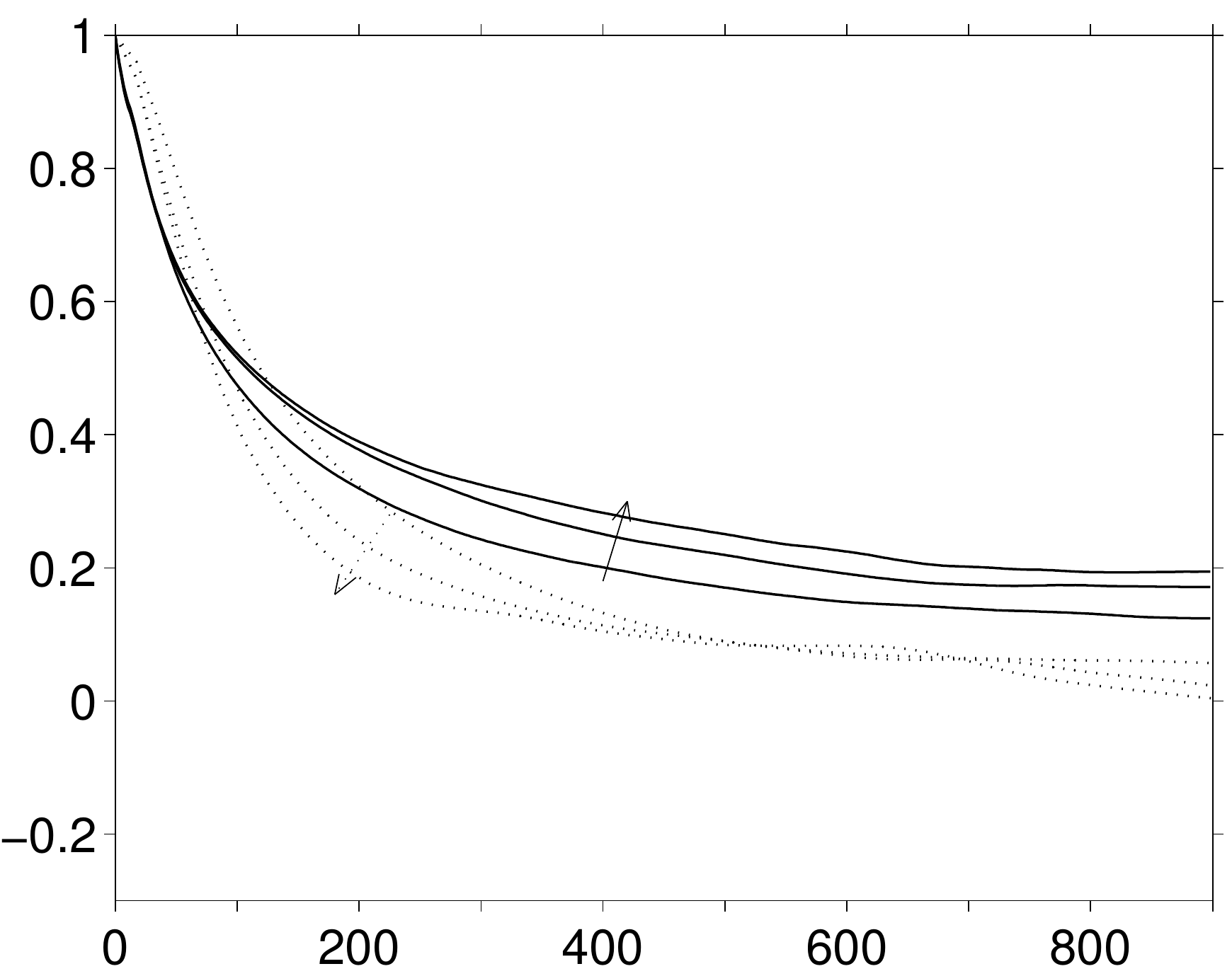}
    \centerline{$r_x^+$}
  \end{minipage}
  \\[1ex]
  \begin{minipage}{4ex}
    {$R_{vv}$}
  \end{minipage}
  \begin{minipage}{.5\linewidth}
    \centerline{$(b)$}
    \includegraphics[width=\linewidth]
    {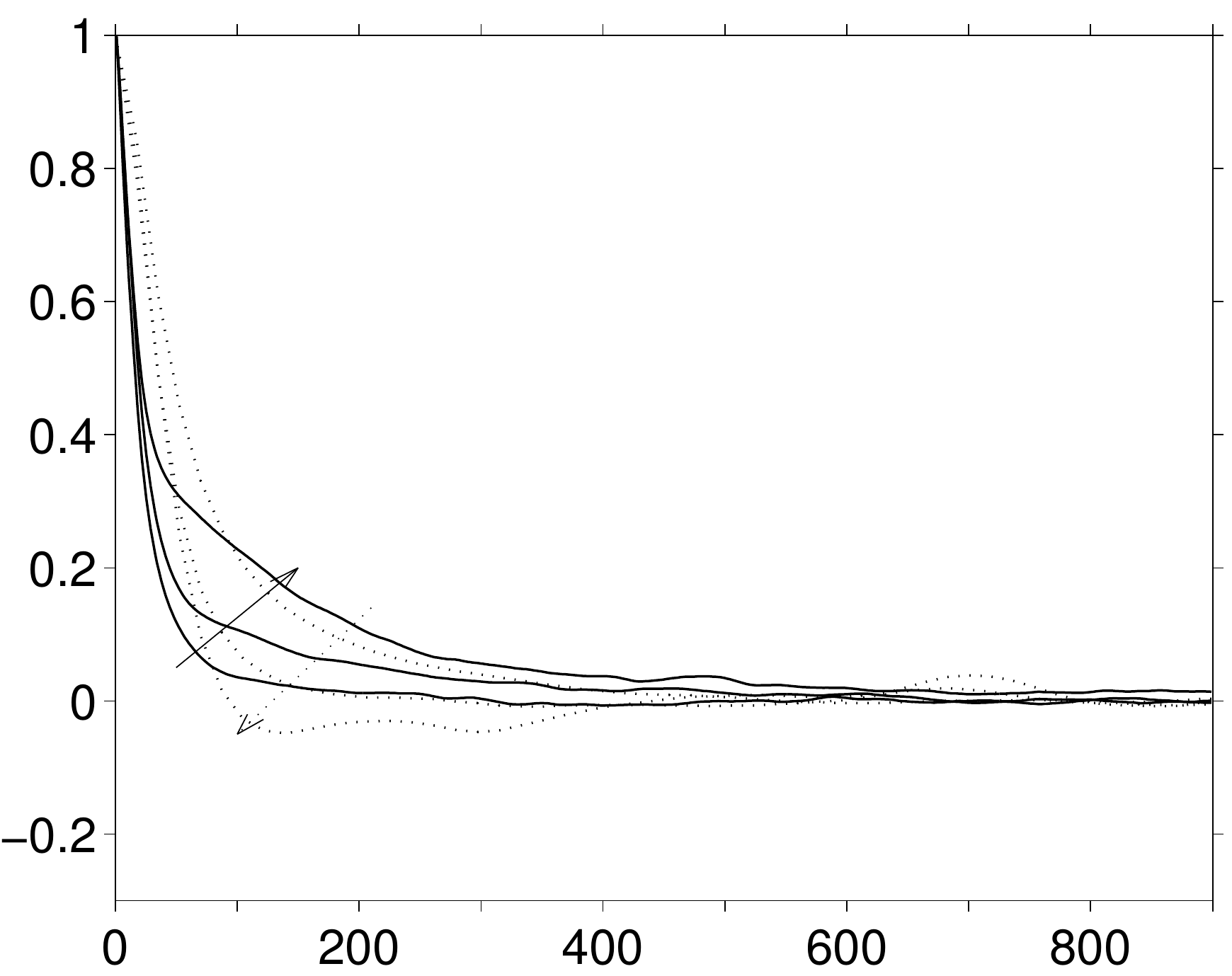}
    \centerline{$r_x^+$}
  \end{minipage}
  \\[1ex]
  \begin{minipage}{4ex}
    {$R_{ww}$}
  \end{minipage}
  \begin{minipage}{.5\linewidth}
    \centerline{$(c)$}
    \includegraphics[width=\linewidth]
    {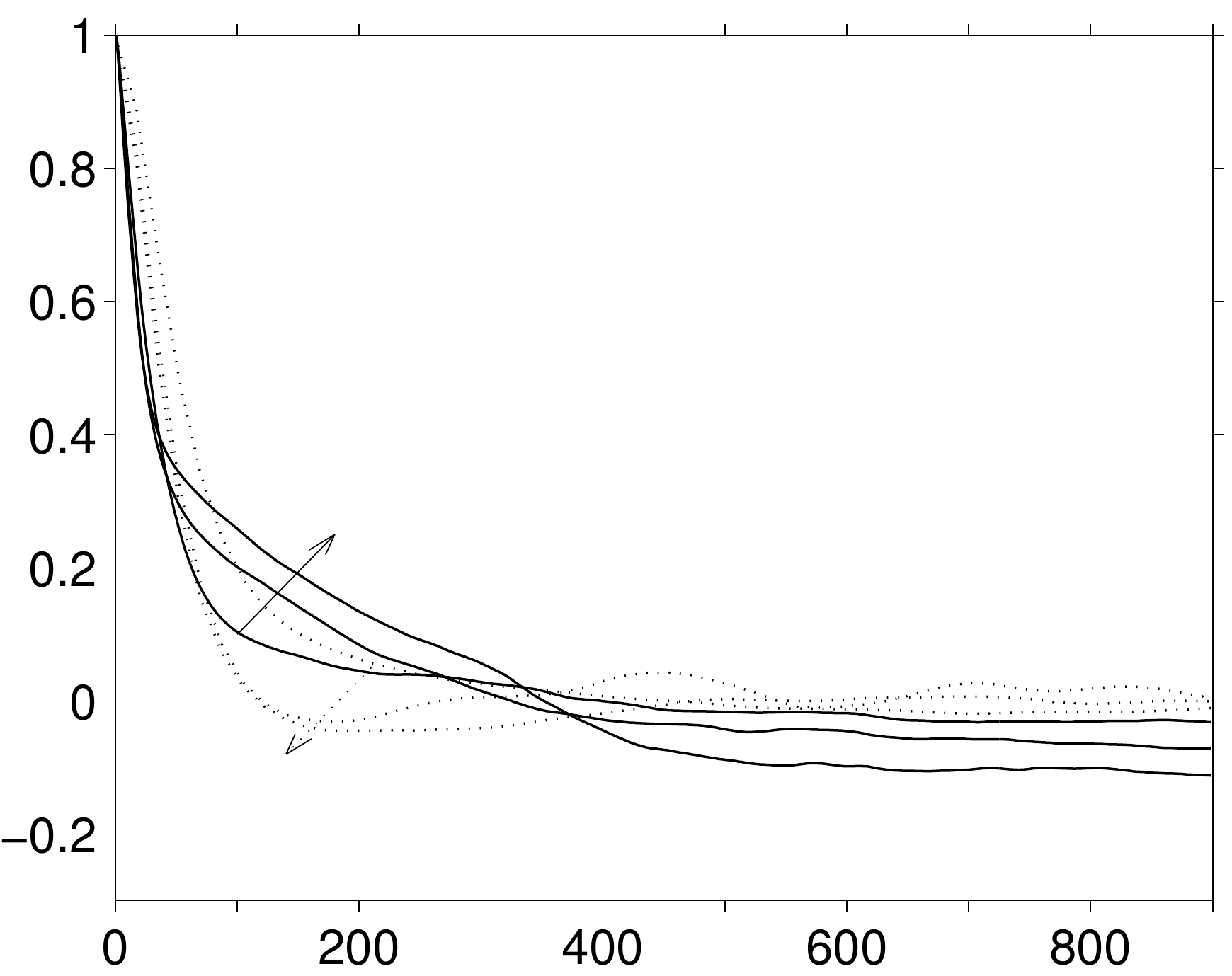}
    \centerline{$r_x^+$}
  \end{minipage}
  \caption{Two-point autocorrelations of fluid velocity fluctuations
    for streamwise separations, case B (evaluated from 12
    instantaneous flow fields, taking into account only points in the
    fluid domain).
    The graphs show the following components of velocity: $(a)$
    streamwise, $(b)$ wall-normal, $(c)$ spanwise. 
    The lines correspond to $y^+=\{21,87,225\}$, increasing along the
    arrows. 
    The single-phase results of \cite{kim:87} are indicated by the
    dotted lines. 
  } 
  \label{fig-results-5-autocorrelations-dx}
\end{figure}
\begin{figure}
  \centering
  \begin{minipage}{4ex}
    {$R_{uu}$}
  \end{minipage}
  \begin{minipage}{.5\linewidth}
    \centerline{$(a)$}
    \includegraphics[width=\linewidth]
    {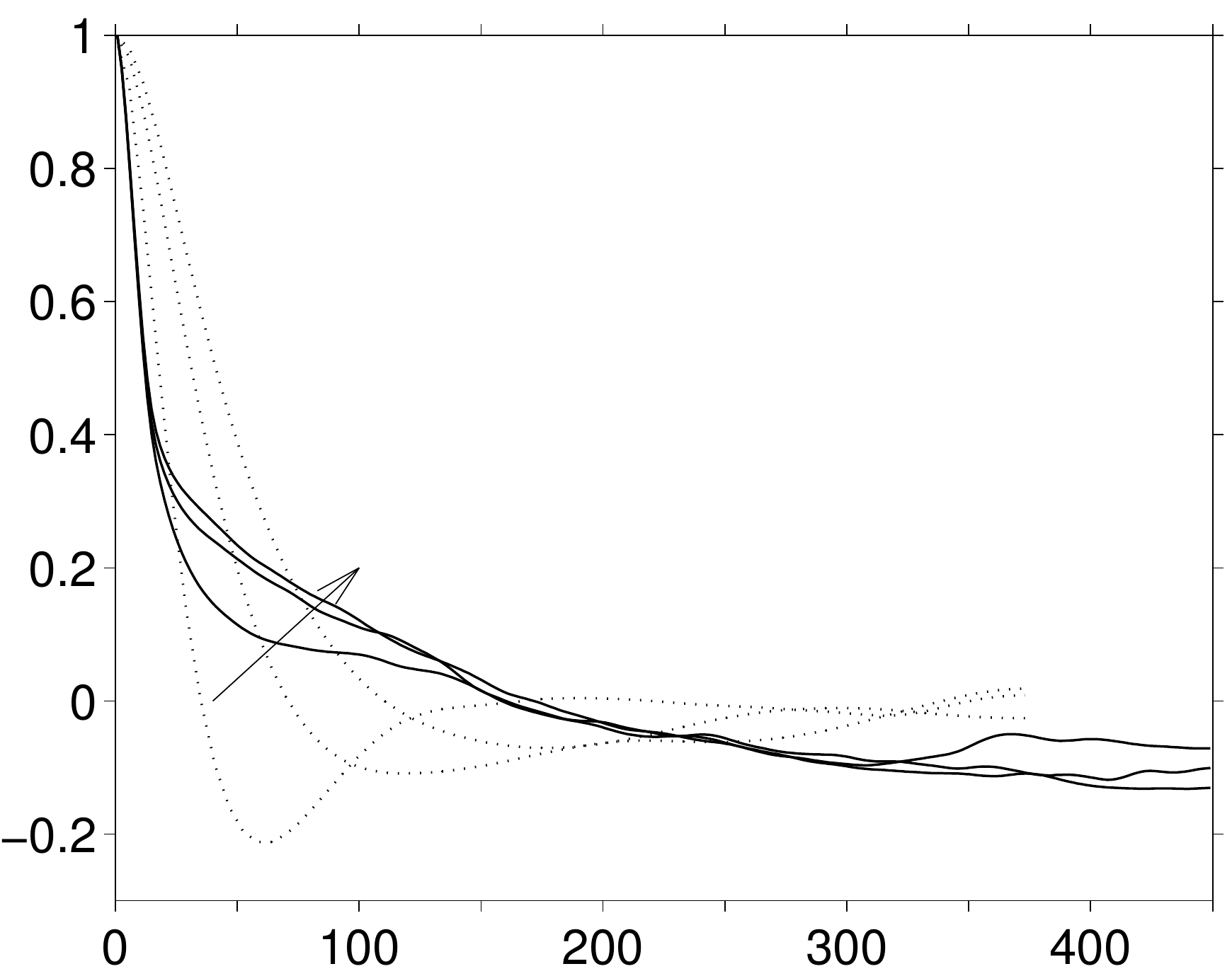}
    \centerline{$r_z^+$}
  \end{minipage}
  \\[1ex]
  \begin{minipage}{4ex}
    {$R_{vv}$}
  \end{minipage}
  \begin{minipage}{.5\linewidth}
    \centerline{$(b)$}
    \includegraphics[width=\linewidth]
    {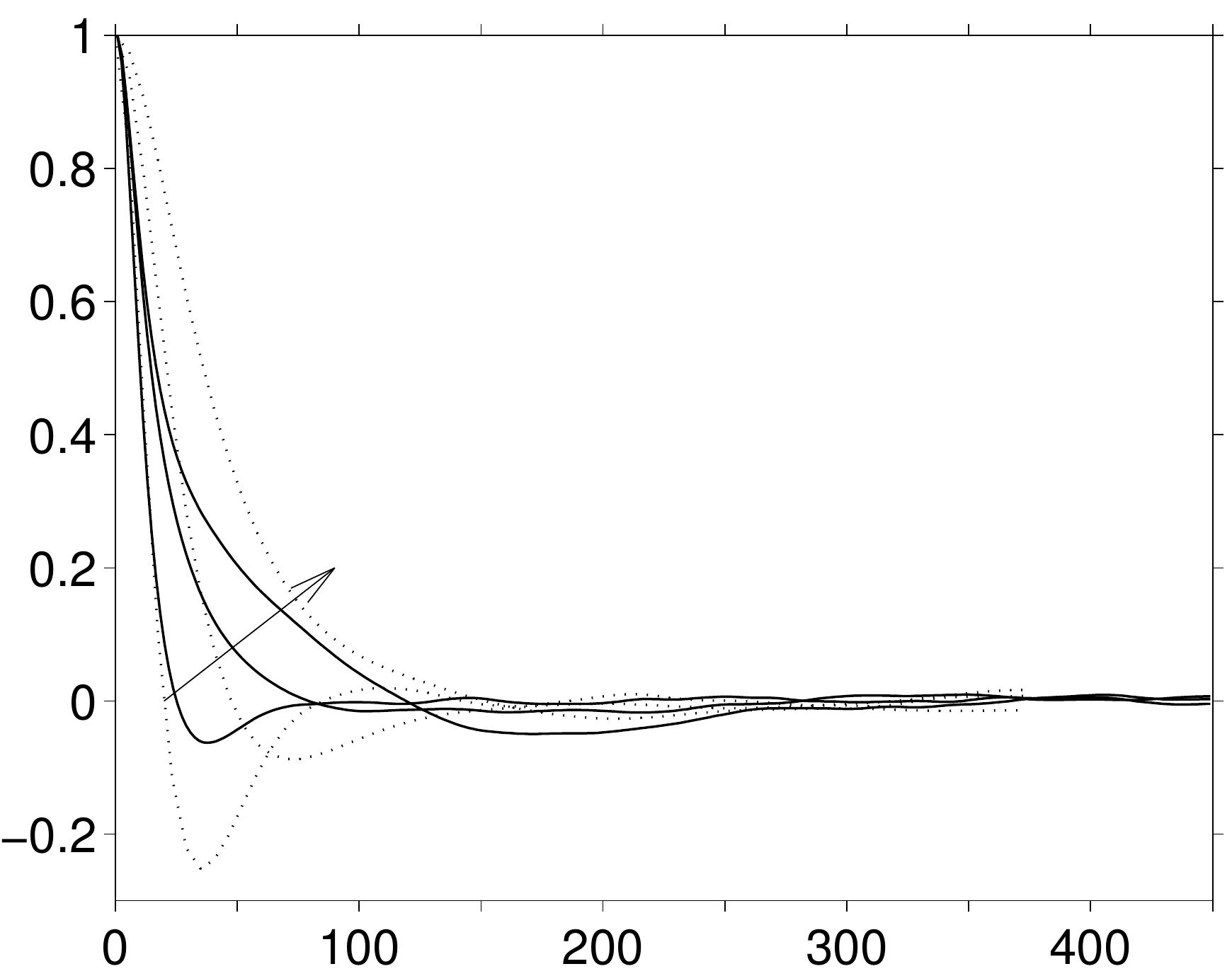}
    \centerline{$r_z^+$}
  \end{minipage}
  \\[1ex]
  \begin{minipage}{4ex}
    {$R_{ww}$}
  \end{minipage}
  \begin{minipage}{.5\linewidth}
    \centerline{$(c)$}
    \includegraphics[width=\linewidth]
    {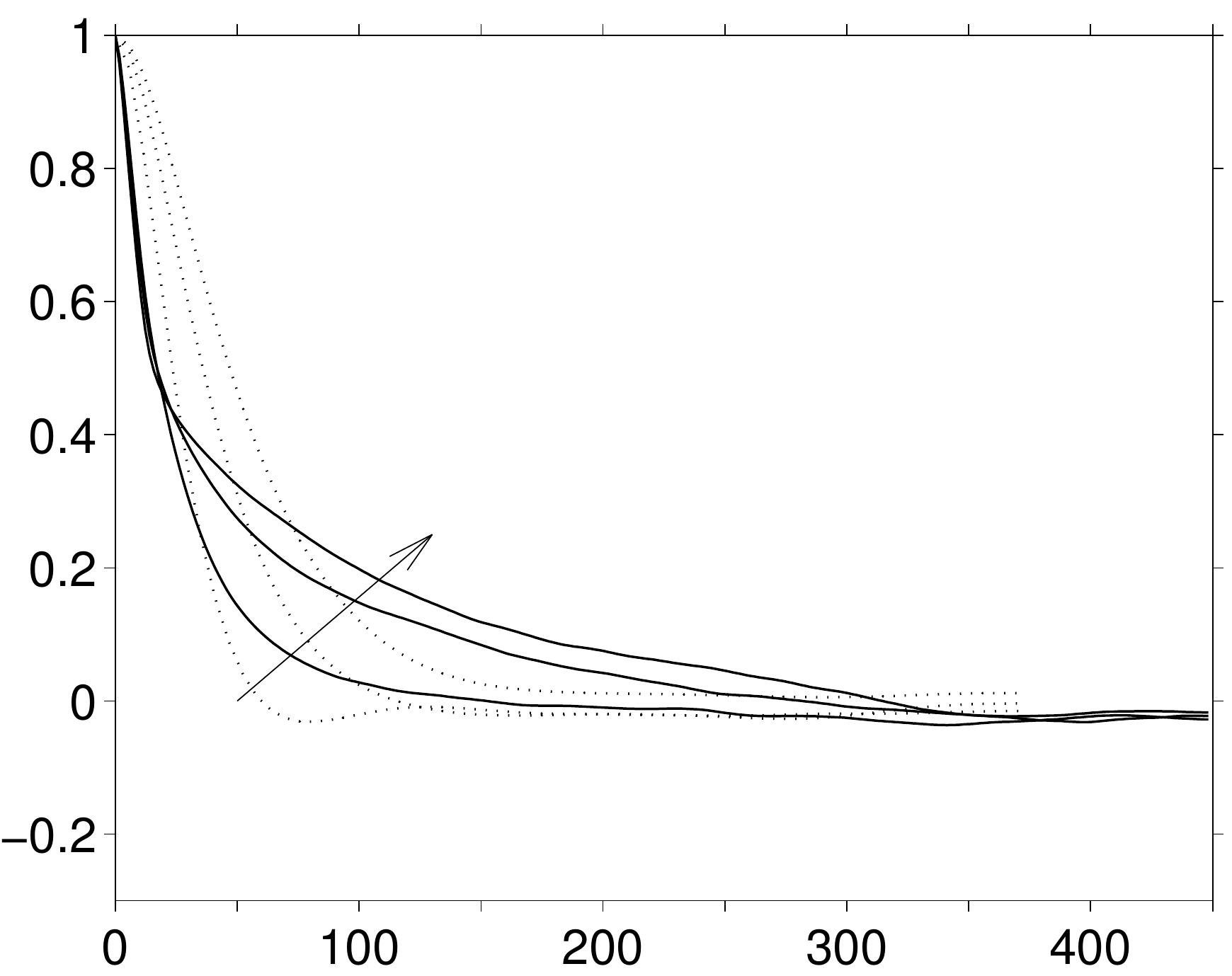}
    \centerline{$r_z^+$}
  \end{minipage}
  \caption{As figure~\ref{fig-results-5-autocorrelations-dx}, but for
    spanwise separations.} 
  \label{fig-results-5-autocorrelations-dz}
\end{figure}
\clearpage
\begin{figure}
  \centering
  \begin{minipage}{9ex}
    {$\displaystyle\frac{\langle u\rangle}{u_b}$}
  \end{minipage}
  \begin{minipage}{.5\linewidth}
    \centerline{$(a)$}
    \includegraphics[width=.98\linewidth]{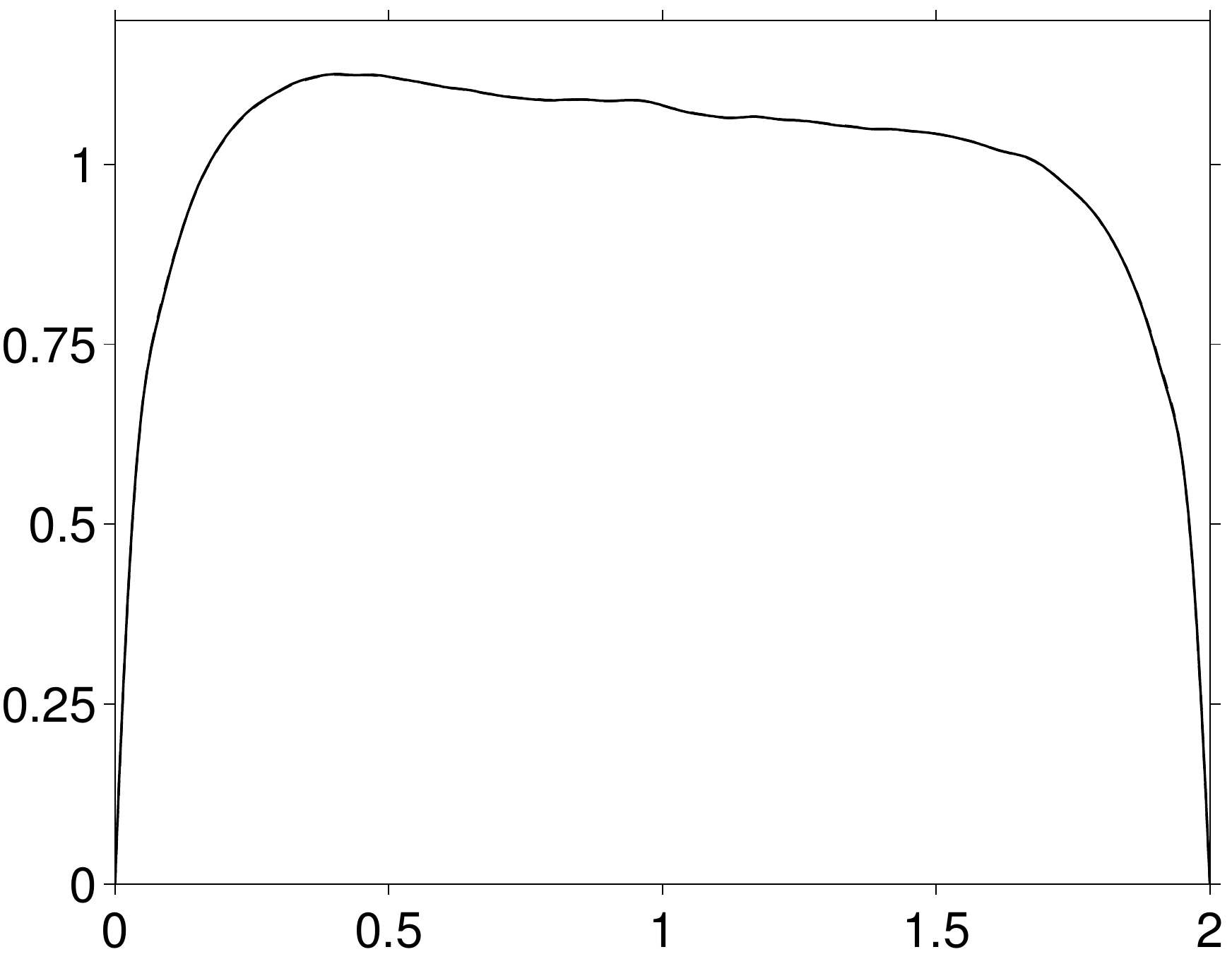}
    \centerline{$y/h$}
  \end{minipage}
  \\[1ex]
  \begin{minipage}{11ex}
    $\sqrt{\langle u_\alpha^\prime u_\alpha^\prime\rangle}^+$
  \end{minipage}
  \begin{minipage}{.5\linewidth}
    \centerline{$(b)$}
    \includegraphics[width=.98\linewidth]{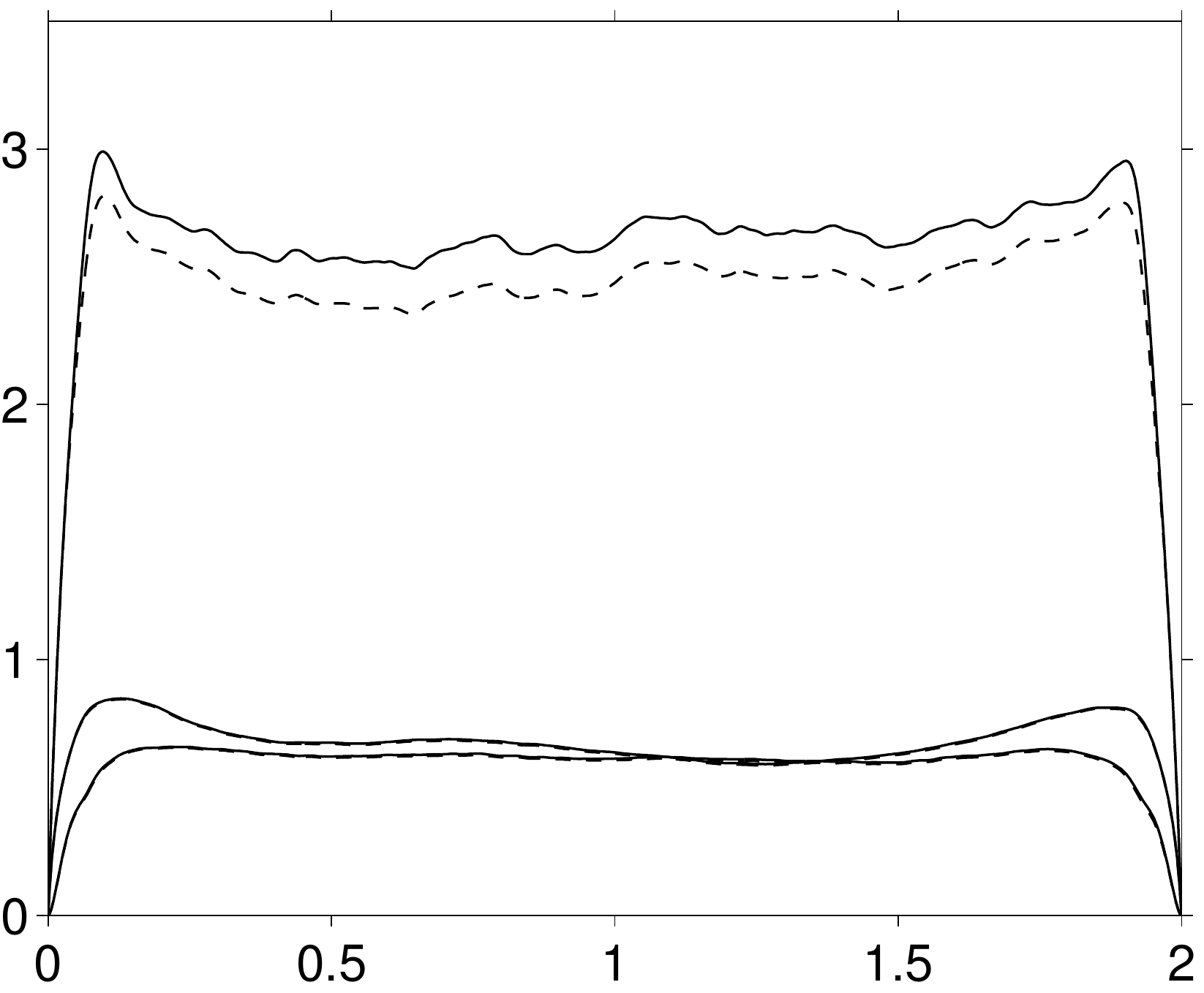}
    \centerline{$y/h$}
  \end{minipage}
  \\[1ex]
  \begin{minipage}{9ex}
    {$\displaystyle
      \langle u^\prime v^\prime\rangle^+$}
  \end{minipage}
  \begin{minipage}{.5\linewidth}
    \centerline{$(c)$}
    \includegraphics[width=.98\linewidth]{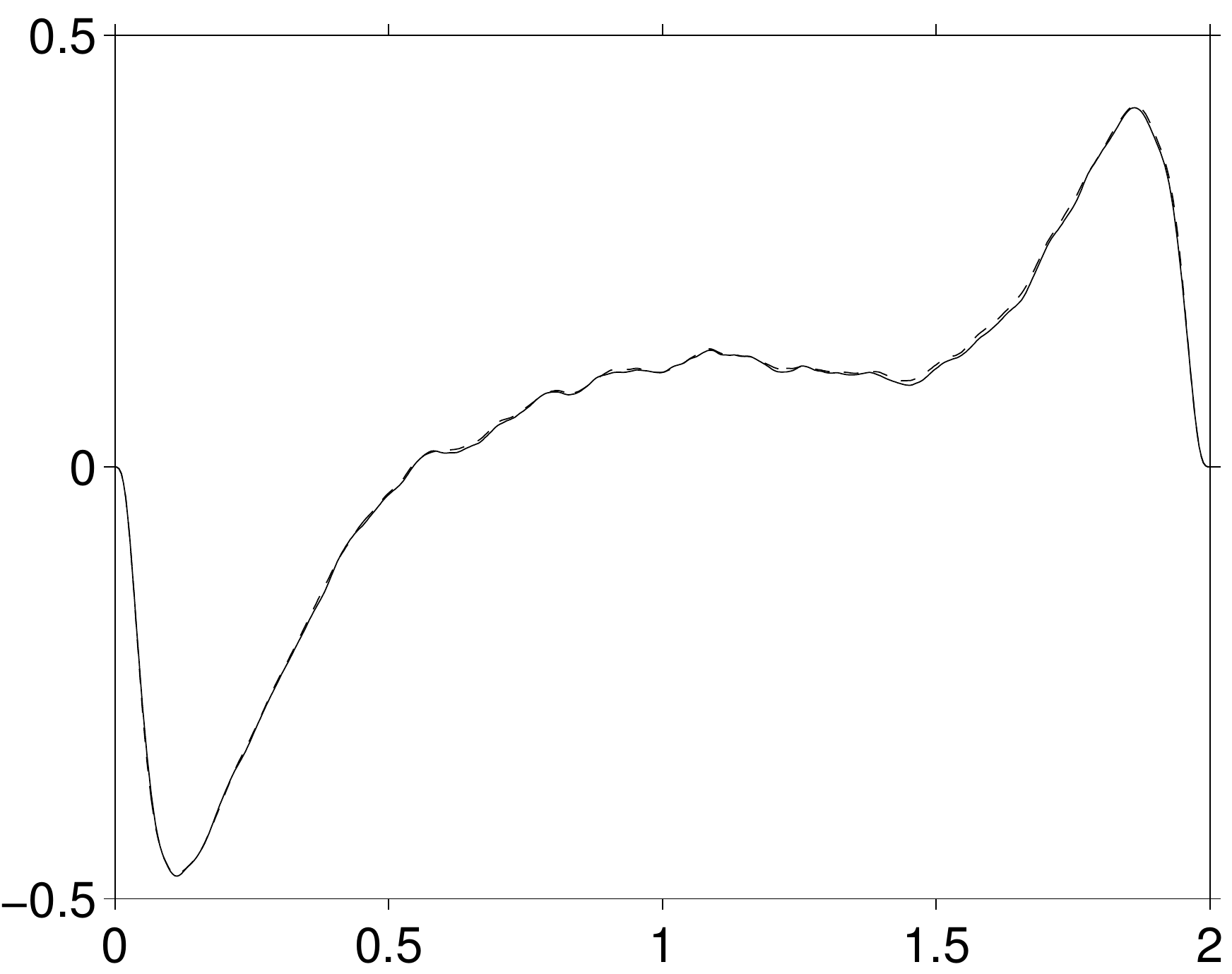}
    \centerline{$y/h$}
  \end{minipage}
  \caption{Eulerian statistics of case B, computed from 12 snapshots
    of the flow field by using two different methods: 
    \solid~averaging over the composite field
    (including fluid and solid nodes);
    \dashed, averaging over the fluid nodes only. $(a)$ The
    wall-normal profile of the mean 
    velocity, $(b)$ the r.m.s.\ velocity fluctuations, $(c)$ the
    Reynolds shear stress.}  
\label{fig-app-fluid-only}
\end{figure}
\begin{figure}
  \centering
  \rotatebox{90}{
    $({\langle u\rangle}-{\langle u\rangle_f})/u_b$
  }
  \begin{minipage}{.5\linewidth}
    \centerline{$(a)$}
    {\includegraphics[width=.98\linewidth]{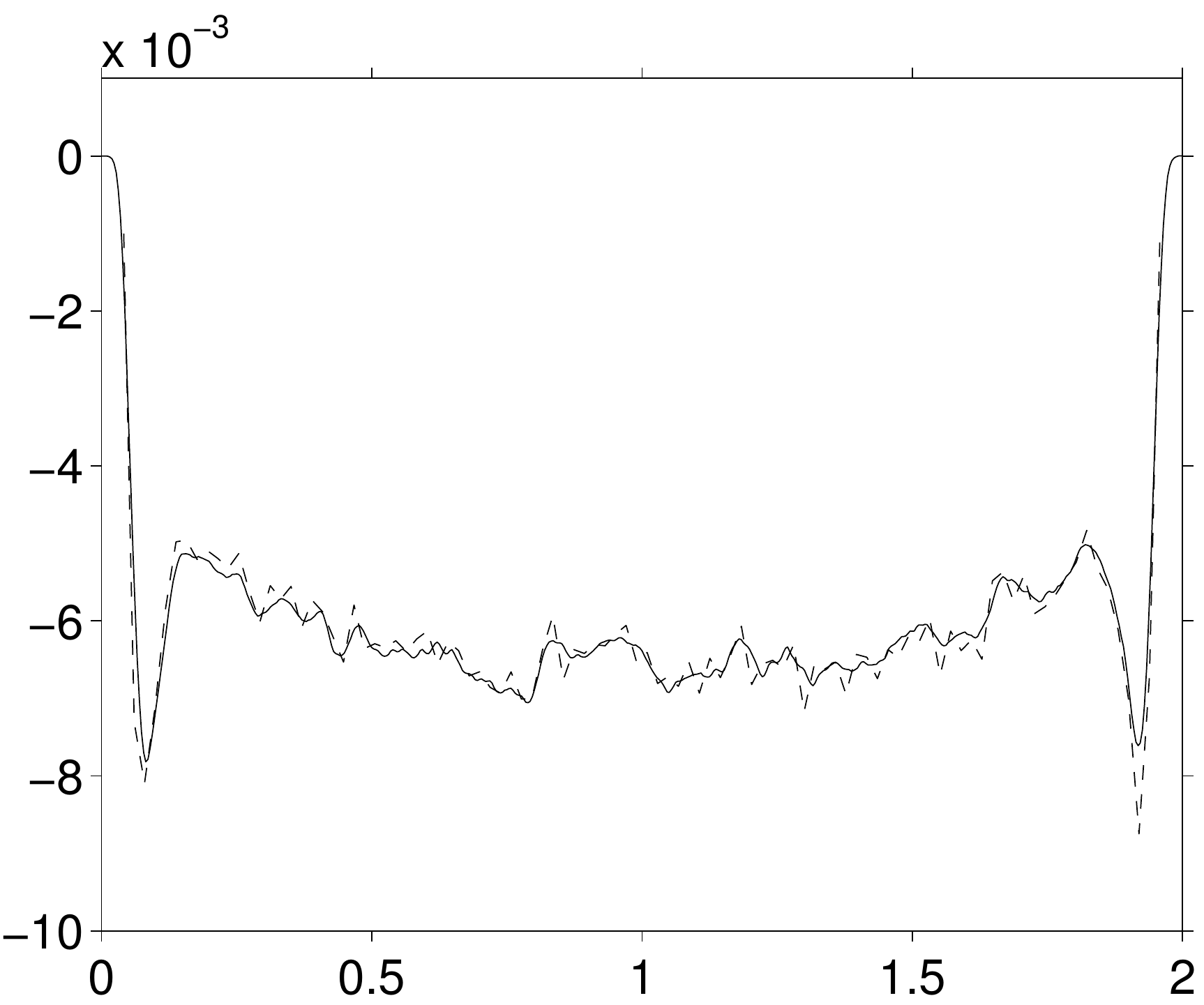}}
    \centerline{$y/h$}
  \end{minipage}
  \\[1ex]
  \raisebox{-5ex}{
    \rotatebox{90}{
      $({\langle u^{\prime}u^{\prime}\rangle}-
      {\langle u^{\prime}u^{\prime}\rangle_f})/u_b^2$
    }}
  \begin{minipage}{.5\linewidth}
    \centerline{$(b)$}
  {\includegraphics[width=.98\linewidth]{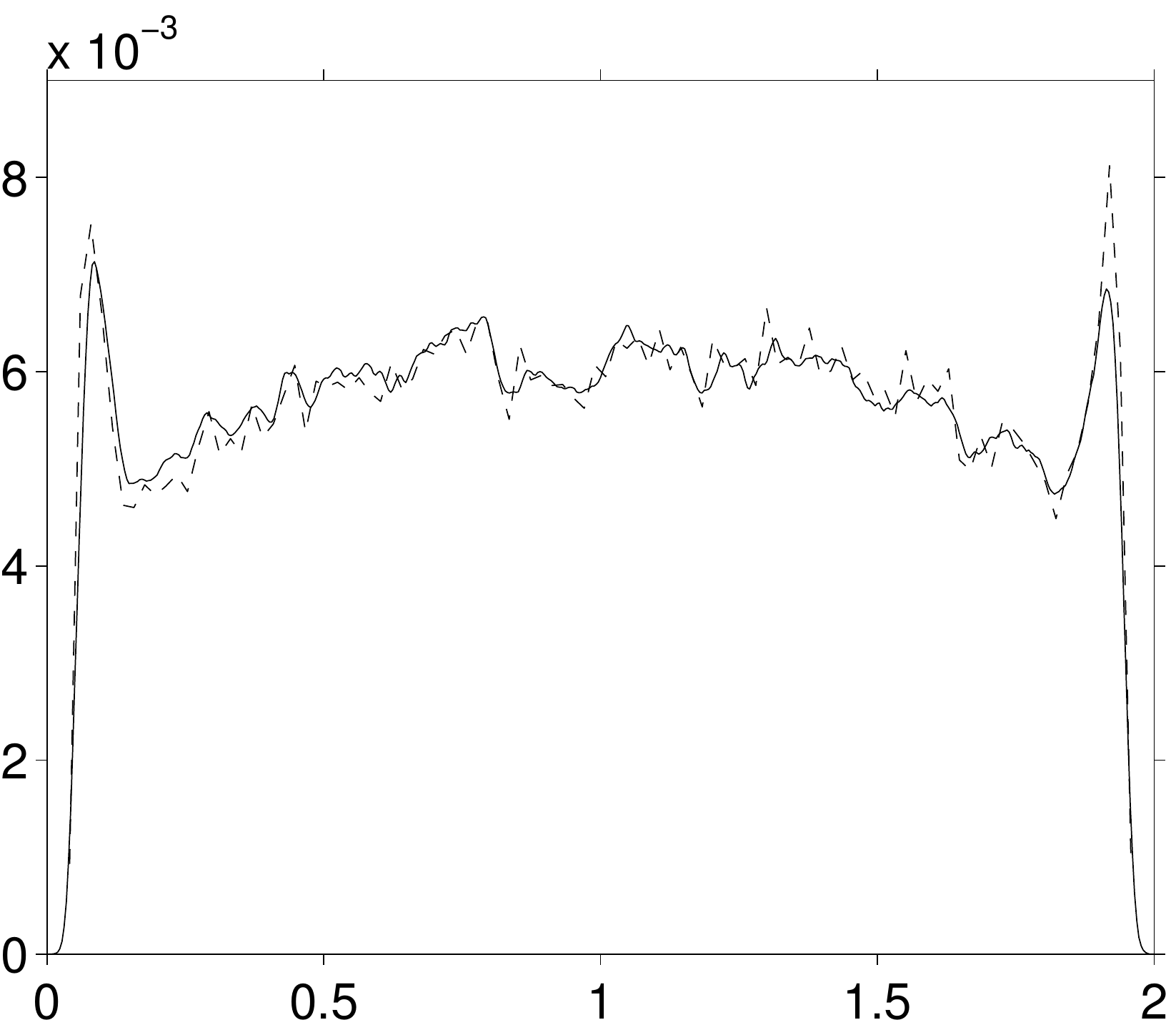}}
    \centerline{$y/h$}
  \end{minipage}
  \caption{
    {The difference between the two procedures of accumulating
      the statistics shown in figure~\ref{fig-app-fluid-only}: 
      $(a)$ the mean velocity difference; 
      $(b)$ the streamwise normal stress component.
      The dashed lines indicate the proportionality with the profile of
      the solid volume fraction $C\cdot\langle\phi_s\rangle(y)$, where 
      $C=\{-1.4682,1.3634\}$ in $(a)$ and $(b)$, respectively, was
      obtained by a least-squares fit.}
  }
  \label{fig-app-fluid-only-diff}
\end{figure}
\begin{figure}
  \centering
  \begin{minipage}{20.ex}
    $\langle f_x\rangle\frac{h}{u_\tau^2}$,\\
    $\left(\frac{\rho_p}{\rho_f}-1\right)g_x\frac{h}{u_\tau^2}\,
    \langle\phi_s\rangle$
  \end{minipage}
  \begin{minipage}{.5\linewidth}
    \includegraphics[width=\linewidth]{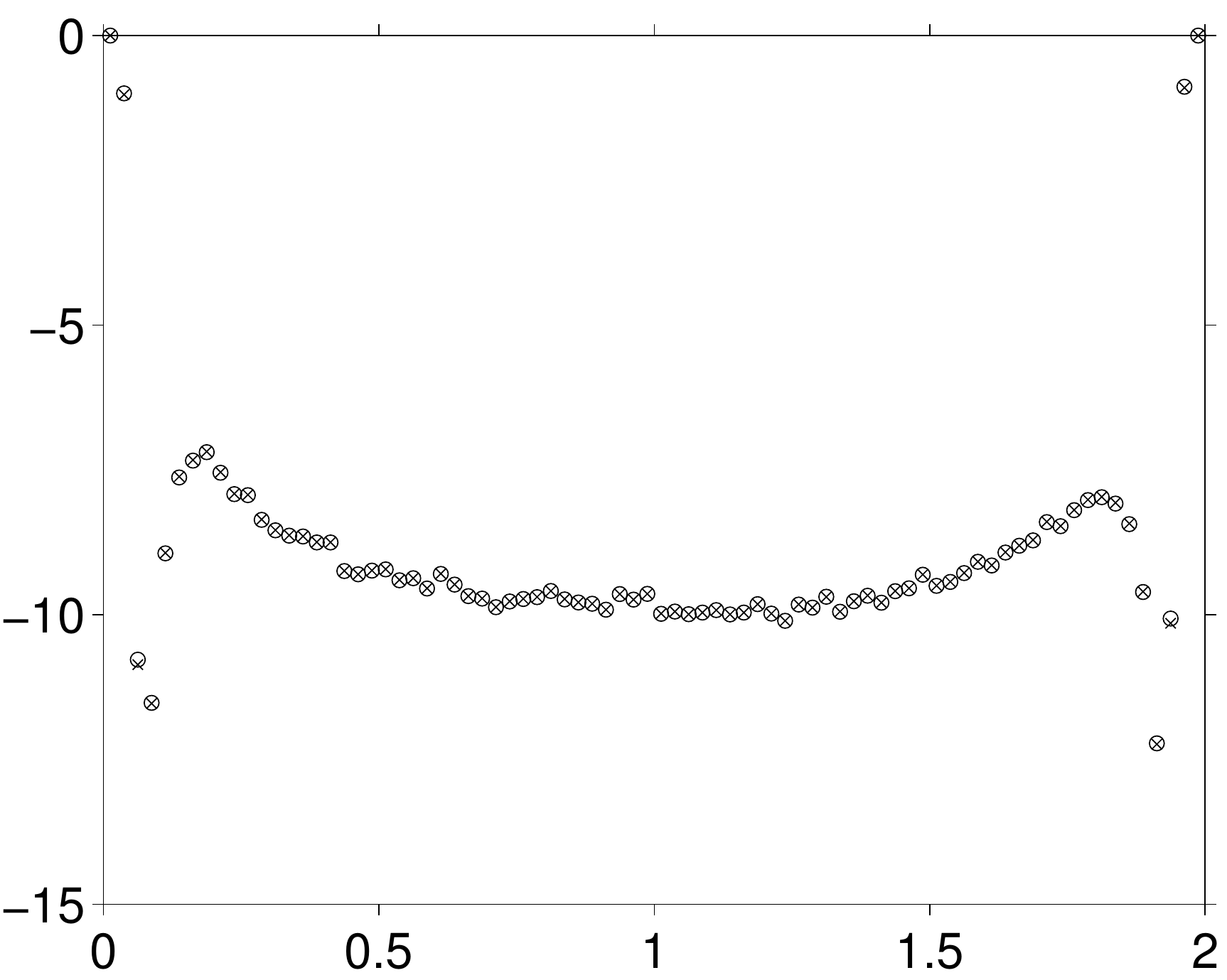}
    \centerline{$y/h$}
  \end{minipage}
  \caption{Wall-normal profiles of the mean streamwise force density
    ($\circ$) and the mean buoyancy term ($\times$) for case B.}
  \label{fig-results-converge-newton}
\end{figure}

\end{document}